\documentclass[twocolumn,superscriptaddress]{revtex4-2}
\usepackage[utf8]{inputenc}
\usepackage[english]{babel}
\usepackage{amsmath}
\usepackage{braket}
\usepackage{bm}
\usepackage{chemformula}
\usepackage{siunitx}
\usepackage{amsbsy}

\usepackage{graphicx}
\usepackage{amssymb}
\usepackage{amsfonts,dsfont}
\usepackage{subfigure}

\usepackage[normalem]{ulem}
\usepackage[bbgreekl]{mathbbol}
\usepackage{mathrsfs,empheq}
\usepackage{relsize,scalerel}

\usepackage[toc,page]{appendix}
\usepackage{hyperref}
\usepackage[savepos]{zref}

\newcommand{\eq}{\sss{eq}}

\newcommand{\Rscr}{\mathscr{R}}

\newcommand{\Pcal}{\mathcal{P}}
\newcommand{\Rcal}{\mathcal{R}}
\newcommand{\Hcal}{\mathcal{H}}
\newcommand{\Fcal}{\mathcal{F}}

\newcommand{\Zcal}{\mathcal{Z}}
\newcommand{\Wcal}{\mathcal{W}}

\newcommand{\ft}{\text{f}}

\newcommand{\bbf}{\mathbb{f}}
\newcommand{\bnot}{\boldsymbol{0}}

\newcommand{\bZ}{\boldsymbol{Z}}

\newcommand{\bR}{\boldsymbol{R}}

\newcommand{\bT}{\boldsymbol{T}}
\newcommand{\bD}{\boldsymbol{D}}

\newcommand{\bM}{\boldsymbol{M}}

\newcommand{\bG}{\boldsymbol{G}}
\newcommand{\bX}{\boldsymbol{X}}
\newcommand{\bY}{\boldsymbol{Y}}
\newcommand{\bPi}{\boldsymbol{\Pi}}
\newcommand{\bu}{\boldsymbol{u}}
\newcommand{\bl}{\boldsymbol{l}}
\newcommand{\bq}{\boldsymbol{q}}
\newcommand{\bk}{\boldsymbol{k}}
\newcommand{\btau}{\boldsymbol{\tau}}
\newcommand{\vareps}{\varepsilon}
\newcommand{\bft}{f}
\newcommand{\bfBO}{\textbf{f}^{\sss{\,(BO)}}}
\newcommand{\bvareps}{\boldsymbol{\varepsilon}}

\newcommand{\bPhi}{\boldsymbol{\Phi}}
\newcommand{\bPsi}{\boldsymbol{\Psi}}
\newcommand{\bvarPhi}{\boldsymbol{\varPhi}}

\newcommand{\bvarPsi}{\boldsymbol{\varPsi}}
\newcommand{\bRcal}{\boldsymbol{\Rcal}}

\newcommand{\bLambda}{\boldsymbol{\Lambda}}
\newcommand{\epsel}{\epsilon^{\scriptscriptstyle{\infty}}}
\newcommand{\bepsel}{\boldsymbol{\epsilon}^{\scriptscriptstyle{\infty}}}
\newcommand{\bbOmega}{\mathbb{\Omega}}

\newcommand{\sss}[1]{\scriptscriptstyle{\text{#1}}}
\newcommand{\hs}{\sss{hs}}

\newcommand{\rscha}{\bRcal}
\newcommand{\rschatrial}{\bRcal}
\newcommand{\phischatrial}{\bvarPhi}
\newcommand{\psischatrial}{\bvarPsi}
\newcommand{\schatrial}{\rscha,\phischatrial}
\newcommand{\schatrialeq}{\bRcaleq,\bPhieq}
\newcommand{\Hschatrial}{\Hcal_{\schatrial}}
\newcommand{\Hschatrialeq}{\Hcal_{\schatrialeq}}
\newcommand{\HschaR}{\Hcal_{\rschatrial, \phischaR}}
\newcommand{\rhoschatrial}{{\tilde \rho}_{\scriptscriptstyle{\rscha},\scriptscriptstyle{\phischatrial}}}
\newcommand{\rhoschastart}{{\tilde \rho}_{\scriptscriptstyle{\rscha^{(0)}},\scriptscriptstyle{\phischatrial^{(0)}}}}

\newcommand{\rhoschatrialfixed}{{\tilde \rho}_{\scriptscriptstyle{\rscha},\scriptscriptstyle{\phischatrial}(\rscha)}}

\newcommand{\fBO}{\text{f}^{\sss{\,(BO)}}}
\newcommand{\fscha}{\text{f}^{\Hschatrial}}
\newcommand{\bfscha}{\textbf{f}^{\Hschatrial}}
\newcommand{\bfschaR}{\textbf{f}^{\HschaR}}

\newcommand{\Tr}[1]{\textup{Tr}\left[#1\right]}
\newcommand{\Avg}[2]{\left\langle #1\right\rangle_{#2}}
\newcommand{\Avgschatrial}[1]{\Avg{#1}{ \rhoschatrial}}

\newcommand{\schaR}{\rscha,\bPhi_{\rscha}}
\newcommand{\phischaR}{\bPhi_{\sssrcal}}
\newcommand{\rhoschaR}{\rho_{\schaR}}
\newcommand{\AvgschaR}[1]{\Avg{#1}{ \!\!\rhoschaR}}
\newcommand{\AvgschaRfixed}[1]{\Avg{#1}{ \rhoschatrialfixed}}

\newcommand{\rhotrial}{\tilde{\rho}}
\newcommand{\rschaeq}{\bRcal_{\eq}}
\newcommand{\rschahs}{\bRcal_{\hs}}

\newcommand{\phischaeq}{\bPhi_{\eq}}
\newcommand{\bPhieq}{\bPhi_{\eq}}

\newcommand{\Hscha}{\Hcal^{\sss{(S)}}}

\newcommand{\inv}{\scriptscriptstyle{-1}}

\newcommand{\bDR}{\bD_{\sssrcal}}
\newcommand{\DR}{D_{\sssrcal}}

\newcommand{\Phin}{\overset{\sss{(n)}}{\Phi}}

\newcommand{\Phithree}{\overset{\sss{(3)}}{\Phi}}
\newcommand{\bPhithree}{\overset{\sss{(3)}}{\bPhi}}
\newcommand{\Phicent}{\overset{\cent}{\Phi}}
\newcommand{\Dthree}{\overset{\sss{(3)}}{D}}
\newcommand{\bDthree}{\overset{\sss{(3)}}{\bD}}
\newcommand{\DthreeR}{\overset{\sss{(3)}}{D}{}_{\sssrcal}}
\newcommand{\bDthreeR}{\overset{\sss{(3)}}{\bD}{}_{\sssrcal}}

\newcommand{\bDthreeeq}{\overset{\sss{(3)}}{\bD}{}_{\eq}}

\newcommand{\Phifour}{\overset{\sss{(4)}}{\Phi}}
\newcommand{\bPhifour}{\overset{\sss{(4)}}{\bPhi}}

\newcommand{\bDfour}{\overset{\sss{(4)}}{\bD}}
\newcommand{\DfourR}{\overset{\sss{(4)}}{D}{}_{\sssrcal}}
\newcommand{\bDfourR}{\overset{\sss{(4)}}{\bD}{}_{\sssrcal}}

\newcommand{\bDfoureq}{\overset{\sss{(4)}}{\bD}{}_{\eq}}

\newcommand{\bLambdaeq}{\bLambda_{\eq}}

\newcommand{\bLambdaR}{\bLambda_{\sssrcal}}
\newcommand{\LambdaR}{\Lambda_{\sssrcal}}

\newcommand{\bDscha}{{\bD}^{\sss{(S)}}}
\newcommand{\Rcaleq}{\Rcal_{\eq}}
\newcommand{\bRcaleq}{\bRcal_{\eq}}
\newcommand{\DF}{D^{\sss{(F)}}}
\newcommand{\bDF}{\bD^{\sss{(F)}}}
\newcommand{\DS}{D^{\sss{(S)}}}
\newcommand{\bDS}{{\bD}^{\sss{(S)}}}
\newcommand{\Eel}{E_{\sss{el}}}
\newcommand{\Fel}{F_{\sss{el}}}
\newcommand{\Sel}{S_{\sss{el}}}
\newcommand{\Sion}{S_{\sss{ion}}}

\newcommand{\Nat}{N_{\sss{a}}}
\newcommand{\Nc}{N_{\sss{c}}}
\newcommand{\Natu}{n_{\sss{a}}}

\newcommand{\doubledot}{\,\textbf{\text{:}}\,}
\newcommand{\singledot}{\boldsymbol{\cdot}}

\newcommand{\sssrcal}{\scriptscriptstyle{\bRcal}}
\renewcommand{\ss}[1]{\scriptscriptstyle{#1}}
\renewcommand{\Im}{\mathrm{Im}}
\renewcommand{\Re}{\mathrm{Re}}
\newcommand{\static}{\sss{(stat)}}
\newcommand{\bPiB}{\overset{\sss{(B)}}{\bPi}}
\newcommand{\PiB}{\overset{\sss{(B)}}{\Pi}}
\newcommand{\be}{\boldsymbol{e}}
\renewcommand{\bf}{\boldsymbol{f}}
\newcommand{\id}{\sss{id}}

\newcommand{\OS}{\sss{(os)}}
\newcommand{\se}{\sss{se}}
\newcommand{\pert}{\sss{(pert)}}
\newcommand{\cent}{\sss{(cent)}}

\newcommand{\Rlat}{\Rscr_{\sss{lat}}}
\newcommand{\RSlat}{\Rscr^{\sss{(S)}}_{\sss{lat}}}

\newcommand{\ols}{\overline{s}}
\newcommand{\olbl}{\overline{\bl}}
\newcommand{\Vol}{\Omega_{\sss{Vol}}}

\newcommand{\stress}{P}
\newcommand{\bstress}{\boldsymbol{P}}
\newcommand{\stressbo}{\stress^{\text{(BO)}}}
\newcommand{\bstressbo}{\bstress^{\text{(BO)}}}

\newcommand{\columnlabel}[1]{\phantomsection\label{#1}\zsaveposx{#1}}
\newcommand{\columnref}[1]{%
  \hyperref[#1]{%
    \pageref*{#1}-%
    \ifdim\zposx{#1}sp<.5\pdfpagewidth
      1% column 1
    \else
      2% column 2
    \fi
  }
}

\newcommand{\eqname}{Eq.}
\newcommand{\bpm}{\begin{pmatrix}}
\newcommand{\epm}{\end{pmatrix}}

\begin{document}

\title{The Stochastic Self-Consistent Harmonic Approximation: Calculating Vibrational Properties of Materials with Full Quantum and Anharmonic Effects}

\author{Lorenzo Monacelli}
\email{lorenzo.monacelli@roma1.infn.it}
\affiliation{Dipartimento di Fisica, Università di Roma Sapienza, Piazzale Aldo Moro 5, 00185 Roma, Italy}

\author{Raffaello Bianco}
\email{raffaello.bianco@ehu.eus}
\affiliation{Centro de F\'isica de Materiales (CSIC-UPV/EHU), Manuel de Lardizabal pasealekua 5, 20018 Donostia/San Sebasti\'an, Spain}

\author{Marco Cherubini}
\affiliation{Dipartimento di Fisica, Università di Roma Sapienza, Piazzale Aldo Moro 5, 00185 Roma, Italy}
\affiliation{Center for Life NanoScience, Istituto Italiano di Tecnologia, viale ReginaElena 291, 00161 Rome, Italy}

\author{Matteo Calandra}
\affiliation{Sorbonne Universit\'e, CNRS, Institut des Nanosciences de Paris, UMR7588, F-75252 Paris, France}
\affiliation{Dipartimento di Fisica, Università di Trento, Via Sommarive 14, 38123 Povo, Italy.}

\author{Ion Errea}
\email{ion.errea@ehu.eus}
\affiliation{Centro de F\'isica de Materiales (CSIC-UPV/EHU), Manuel de Lardizabal pasealekua 5, 20018 Donostia/San Sebasti\'an, Spain}
\affiliation{Fisika Aplikatua Saila, Gipuzkoako Ingeniaritza Eskola, University of the Basque Country (UPV/EHU),
Europa Plaza 1, 20018 Donostia/San Sebasti\'an, Spain}
\affiliation{Donostia International Physics Center (DIPC), Manuel Lardizabal pasealekua 4, 20018 Donostia/San Sebasti\'an, Spain}

\author{Francesco Mauri}
\email{francesco.mauri@uniroma1.it}
\affiliation{Dipartimento di Fisica, Università di Roma Sapienza, Piazzale Aldo Moro 5, 00185 Roma, Italy}

\date{\today}

\begin{abstract}
The efficient and accurate calculation of how ionic quantum and thermal fluctuations impact the free energy of a crystal, its atomic structure, and phonon spectrum is one of the main challenges of solid state physics, especially when strong anharmonicy invalidates any perturbative approach. To tackle this problem, we present the implementation on a modular Python code of the stochastic self-consistent harmonic approximation method. This technique rigorously describes the full thermodyamics of crystals accounting for nuclear quantum and thermal anharmonic fluctuations. The approach requires the evaluation of the Born-Oppenheimer energy, as well as its derivatives with respect to ionic positions (forces) and cell parameters (stress tensor) in supercells, which can be provided, for instance, by first principles density-functional-theory codes. The method performs crystal geometry relaxation on the quantum free energy landscape, optimizing the free energy with respect to all degrees of freedom of the crystal structure. It can be used to determine the phase diagram of any crystal at finite temperature. It enables the calculation of phase boundaries for both first-order and second-order phase transitions from the Hessian of the free energy. Finally, the code can also compute the anharmonic phonon spectra, including the phonon linewidths, as well as phonon spectral functions. We review the theoretical framework of the stochastic self-consistent harmonic approximation and its dynamical extension, making particular emphasis on the physical interpretation of the variables present in the theory that can enlighten the comparison with any other anharmonic theory. A modular and flexible Python environment is used for the implementation, which allows for a clean interaction with other packages. We briefly present a toy-model calculation to illustrate the potential of the code. Several applications of the method in superconducting hydrides, charge-density-wave materials, and thermoelectric compounds are also reviewed.   
\end{abstract}

\maketitle

\section{Introduction} 

Ions fluctuate at any temperature in matter, also at zero kelvin due to the quantum zero-point motion. Even if the energy of ionic fluctuations is considerably smaller than the electronic one, many physical and chemical properties of materials and molecules cannot be understood without considering ionic vibrations. Since ionic vibrations are excited at much lower temperatures than electrons, ionic fluctuations are mainly responsible for the temperature dependence of thermodynamic properties of materials. They also determine heat and electrical transport through the electron-phonon and/or phonon-phonon interactions, as well as spectroscopic signatures detected in infrared, Raman, and inelastic x-ray or neutron scattering experiments. The large computational power available today has paved the way to material design and characterization, but advanced and reliable methods that accurately calculate vibrational properties of materials in the limit of strong quantum anharmonicity and that are easily interfaced with modern {\it ab initio} codes are required for accurately describing materials' properties {\it in silico}. 

Since electrons are faster than ions, the ionic motion is assumed to be described by the Born-Oppenheimer potential $V(\bR)$, which, at an ionic configuration $\bR$, is given by the electronic ground state energy. In the standard harmonic approximation $V(\bR)$ is Taylor-expanded up to second-order around the $\bR_0$ ionic positions that minimize $V(\bR)$. The resulting Hamiltonian is exactly diagonalizable in terms of phonons, the quanta of vibrations. Harmonic phonons are well-defined quasiparticles whith an infinite lifetime, which energies do not depend on temperature. These two features are intrinsic failures of this approximation: phonons acquire a finite lifetime due to their anharmonic interaction with other phonons (also because of other types of interactions such as the electron-phonon coupling), and phonon energies do depend on temperature experimentally. When higher-order anharmonic terms are small compared to harmonic ones, anharmonicity can be treated within perturbation theory\cite{PhysRev.128.2589,0034-4885-31-1-303,Calandra200738}. Even if within perturbative approaches phonons' temperature dependence and lifetimes can be understood, whenever anharmonic terms of the $V(\bR)$ potential are similar or larger than the harmonic terms in the range sampled by the ionic fluctuations, perturbative approaches collapse and are not valid\cite{ErreaRev2016}. This is often the case when light ions are present, as well as when the system is close to melting or a displacive phase transition, such as a ferroeletric or charge-density wave (CDW) instability.  

In order to calculate from first principles vibrational properties of solids beyond perturbation theory and overcome these difficulties, several methods have been developed in the last years\cite{PhysRevLett.55.2471,PhysRevB.42.11276,PhysRevLett.110.105503,PhysRevB.87.174110,PhysRevLett.112.058501,PhysRevB.84.180301,PhysRevB.87.104111,PhysRevB.88.144301,RevModPhys.67.279,PhysRevB.52.6301,0953-8984-25-30-305401,PhysRevLett.106.165501,PhysRevLett.113.185501,PhysRevB.92.054301,Tadano_2014,PhysRevLett.113.185501,PhysRevLett.112.165501,PhysRevLett.111.177002,PhysRevB.89.064302,PhysRevB.96.014111,PhysRevB.98.024106,PhysRevLett.100.095901,PhysRevB.82.184301,PhysRevB.86.054119,PhysRevB.98.054305,doi:10.1002/adts.201800184,roekeghem2020quantum}. Many of them are based on extracting renormalized phonon frequencies from {\it ab initio} molecular dynamics (AIMD) through velocity autocorrelation functions\cite{PhysRevB.42.11276,PhysRevLett.110.105503,PhysRevB.87.174110,PhysRevLett.112.058501} or by extracting effective force constants from the AIMD trajectory\cite{PhysRevB.84.180301,PhysRevB.87.104111,PhysRevB.88.144301}. In order to include quantum effects on the ionic motion, which are neglected on AIMD, the AIMD trajectory may be substituted by a path-integral molecular dynamics (PIMD) one\cite{RevModPhys.67.279}. Other methods are based on variational principles\cite{PhysRevLett.106.165501,PhysRevB.92.054301,PhysRevB.89.064302,PhysRevB.96.014111,PhysRevB.98.024106,Needs}, which are mainly inspired on the self-consistent harmonic approximation\cite{hooton422} or vibrational self-consistent
field~\cite{:/content/aip/journal/jcp/68/2/10.1063/1.435782} theories, and yield free energies and/or phonon frequencies corrected by anharmonicity non-perturbatively.
    
Even if these methods have often successfully incorporated the effect of anharmonicity beyond perturbation theory in different materials, they usually lack a consistent procedure that prevents them from capturing properly both quantum effects and anharmonicity in the compound. For instance, many of them simply correct the free energy and/or the phonon frequencies assuming that the ions remain fixed at the $\bR_0$ classical positions. However, as it has been  shown recently in several compounds\cite{Errea81,0953-8984-28-49-494001,Errea66,monacelli2019black}, the ionic positions can be strongly altered by quntum effects and anharmonicity even at zero Kelvin. The structural changes are important for both internal degrees of freedom (the Wyckoff positions), and the lattice parameters themselves. Moreover, in many of the aforementioned methods, it is not clear what the meaning of the renormalized phonon frequencies is, i.e., whether they are auxiliary phonon frequencies intrinsic to the devised theoretical framework or if they really represent the physical vibrational excitations probed experimentally.

The stochastic self-consistent harmonic approximation (SSCHA)\cite{PhysRevB.89.064302,PhysRevB.96.014111,PhysRevB.98.024106} is a unique method that provides a full and complete way of incorporating ionic quantum and anharmonic effects on materials' properties without approximating the $V(\bR)$ potential. The SSCHA is defined from a rigorous variational method that directly yields the anharmonic free energy. It can optimize completely the crystal structure, including both internal and lattice degrees of freedom, accounting for the quantum nature of the ions at any target pressure or temperature. It computes thermal expansion even in highly anharmonic crystals. Furthermore, the SSCHA provides a well-defined approach to estimate at which thermodynamic conditions displacive second-order phase transitions occur. This is particularly challenging in \emph{ab initio} molecular dynamics simulations, both for the dynamical slowing down that may hamper the thermalization close to the critical point, and for the difficulties in resolving the two distinct phases that continuously transform one into the other.   Also, the rigorous theoretical approach of the SSCHA yields a clear distinction between auxiliary phonons of the theory and the phonon spectra probed experimentally, which can be accessed from a rigorous dynamical extension of the theory\cite{PhysRevB.96.014111,monacelli2020time,lihm2020gaussian}. Lastly, the code provides non-perturbative third- and fourth-order phonon-phonon scattering matrices that can be fed in any external thermal transport code to compute thermal conductivity and lattice transport properties. Here, we present an implementation of the full SSCHA theory in a modular Python software that can be easily and efficiently interfaced with any total-energy-force engine, e.g., density-functional-theory (DFT) first-principles codes. 

This paper is organized to introduce the reader to the SSCHA algorithm and to review the recent developments in the SSCHA theory that lead to the SSCHA code, following the typical usage of the final user. In Sec.~\ref{sec:overview} we give a simple overview of the method, presenting a simple picture of how it works with a model calculation on a highly anharmonic system with one particle in one dimension. Then, we review the full theory of the SSCHA in details, starting from the free energy calculation and structure optimization in Sec.~\ref{sec:relax}. Then, we describe, in Sec.~\ref{sec:postprocessing}, the post-processing features of the code, which include calculations of the free energy Hessian for second-order phase transitions, as well as phonon spectral function and linewidth calculations. Each section is introduced by an overview of the theory to understand what the code is doing and, then, reports the details of the implementation, together with a guide for setting up a typical run. In Sec.~\ref{sec:python}, specific details of the Python code are provided, including the different execution modes and installation tips. 
As a showcase of the SSCHA, we provide a simple example in a thermoelectric material in Sec.~\ref{sec:model} (\ch{SnTe}), where we fully characterize the thermodynamics of the phase transition between the high-symmetry and low-symmetry phases. This is also a guide on how to correctly analyze the output of the SSCHA calculation and the physical interpretation of the different frequencies. In Sec.~\ref{sec:examples}, we review some important results obtained so far with the SSCHA code. Finally, in Sec. \ref{sec:conclusions}, we summarize the main conclusions.

\section{The variational free energy}
\label{sec:overview}

The SSCHA is a theory that aims at describing the thermodynamics of a crystal, fully accounting for quantum, thermal, and anharmonic effects of nuclei within the Born-Oppenheimer approximation. The basis of all equilibrium thermodynamics is that a system in equilibrium at fixed volume, temperature, and number of particles is at the minimum of the free energy. The free energy is expressed by the sum of the internal energy $E$, which includes the energy of the interaction between the particles (kinetic and potential), and the product between the temperature $T$ and the entropy $S$, which accounts for ``disorder'' and is related to the number of microstates corresponding to the same macrostate of the system:
\begin{equation}
    F = E - TS.
\end{equation}

In a classical picture, the free energy can be thus expressed in terms of the microscopic states of the system, which are determined by the classical probability distribution of atoms $\rho_\mathrm{cla}(\bR)$. We remind that $\bR$ is a vector of coordinates of all atoms in the system (we will use bold symbols to denote vectors and tensors in component free notation). The same holds for a quantum system, but we need to account also for quantum interference. This is achieved by calculating the free energy with the many-body density matrix.
As the system at equilibrium is at the minimum of the free energy, the Gibbs-Bogoliubov variational principle\cite{PhysRevB.69.064509} states that between all possible trial density matrices $\tilde\rho$, the true free energy of the system is reached at the minimum of the functional $\Fcal[\tilde\rho]$:
\begin{equation}
    \Fcal[\tilde \rho] = E[\tilde\rho] - TS[\tilde \rho] \ge   F,
    \label{eq:bogo}
\end{equation}
where 
\begin{equation}
    E[\tilde\rho] = \braket{K + V(\bR)}_{\tilde\rho} 
    \label{eq:toten}
\end{equation}
is the total energy ($K$ is the kinetic energy operator and $V(\bR)$ the potential energy), and and $S[\tilde \rho]$ the entropy calculated with the trial density matrix. $\braket{\cdot}_{\tilde\rho}=Tr[{\tilde \rho}~ \cdot ]$ indicates the quantum average of the the  operator $\cdot$ taken with $\tilde \rho$. 

If we pick any trial density matrix $\tilde\rho$, $\Fcal[\tilde\rho]$ is an upper bound of the true free energy of the system. The SSCHA follows this principle: we optimize a trial density matrix $\tilde\rho$ to minimize the free energy functional $\Fcal[\tilde\rho]$ of \eqname~\eqref{eq:bogo}.
Performing the optimization on any possible trial density matrix is, however, an unfeasible task due its many-body character that hinders an efficient parametrization. This is true also for a classical system: no exact parametrization of $\rho_\mathrm{cla}(\bR)$ can be obtained in a computer with a finite memory.

The SSCHA solves the problem by imposing a constraint on the density matrix. In particular, the quantum probability distribution function that the SSCHA density matrix defines, \columnlabel{eq:TrialGauss}{$\rhoschatrial(\bR)=\bra{\bR}\rhoschatrial\ket{\bR}$}, is a Gaussian. $\rhoschatrial(\bR)$ is the quantum analogue of $\rho_\mathrm{cla}(\bR)$, and determines the probability to find the atoms in the configuration $\bR$. The trial SSCHA density matrix $\rhoschatrial$ is uniquely identified by the average atomic positions (centroids) $\bRcal$ and the quantum-thermal fluctuations around them $\bvarPhi$ (we have explicitly expressed the dependence of ${\tilde \rho}$ on $\rschatrial$ and $\phischatrial$ by adding them as subindexes), just like any Gaussian is defined by the average and mean square displacements. Within the SSCHA, we optimize $\bRcal$ and $\phischatrial$ to minimize the free energy of the system. In this way, we compress the memory requested to store $\rhoschatrial$, as $\rschatrial$ depends only on $3\Nat$ numbers (the coordinates of the atoms), while the fluctuations $\phischatrial$ are encoded in a symmetric square real matrix of $3\Nat\times 3\Nat$. $\Nat$ is the total number of atoms in the system. The free parameters in $\rschatrial$ and $\phischatrial$ can be further reduced by exploiting translation and point group symmetries of the crystal, resulting in an efficient and compact representation of the density matrix $\rhoschatrial$.

The ``harmonic'' in the SSCHA name comes from the fact that any Gaussian density matrix that describes a physical system is the equilibrium solution of a particular harmonic Hamiltonian. Therefore, there is a one-to-one mapping between the trial density matrix $\rhoschatrial$ and an auxiliary trial harmonic Hamiltonian $\Hschatrial$:
\begin{equation}
    \Hschatrial = K + \frac 12 \sum_{ab}(R_a - \Rcal_a) \varPhi_{ab} (R_b - \Rcal_b).
    \label{eq:HSCHAtrial}
\end{equation} 
Here, $\bRcal$ is a real vector and $\phischatrial$ a real matrix that parametrize the trial Hamiltonian, while $K$ and $\bR$ are quantum operators that measure the kinetic energy and the position of the state. For simplicity, unless otherwise specified, all indices $a$, $b$, etc. run over both atomic and Cartesian coordinates from 1 to $3\Nat$. Let us note here, that, inspired by the harmonic shape of $\Hschatrial$, we will also refer to $\phischatrial$ as the auxiliary force constants. 

This mapping with a harmonic Hamiltonian is very useful, as both $\braket{K}_{\rhoschatrial}$ and $S[\rhoschatrial]$ become simply the kinetic energy and entropy of the auxiliary harmonic system $\Hschatrial$, which are analytic functions of $\phischatrial$. Hence, the only quantity that we really need to compute is the average over the interacting Born-Oppenheimer potential 
\begin{equation}
    \braket{V(\bR)}_{\tilde\rho} = \int d\bR\; V(\bR) \tilde\rho(\bR).
    \label{eq:BO:av}
\end{equation}
The potential $V(\bR)$ is the Born-Oppenheimer energy landscape, and can be easily computed \emph{ab initio} by any DFT code (or by any energy and force engine).

The SSCHA algorithm starts with an initial guess on $\rschatrial$ and $\phischatrial$, and proceeds as follows:
\begin{itemize}
    \item Use the trial Gaussian probability distribution function $\rhoschatrial(\bR)$ to extract an ensemble of random nuclear configurations in a supercell.
    \item For each nuclear configuration in the ensemble, compute total energies and forces with an external code, either \emph{ab initio} or via a force field.
    \item Use total energy and forces on the ensemble to compute the free energy functional and its derivatives with respect to the free parameters of our distribution $\rschatrial$ and $\phischatrial$.
    \item Update $\rschatrial$ and $\phischatrial$ to minimize the free energy.
\end{itemize}
These steps are repeated until the minimum of the free energy is found.

\begin{figure*}[t]
    \centering
    \includegraphics[width=\textwidth]{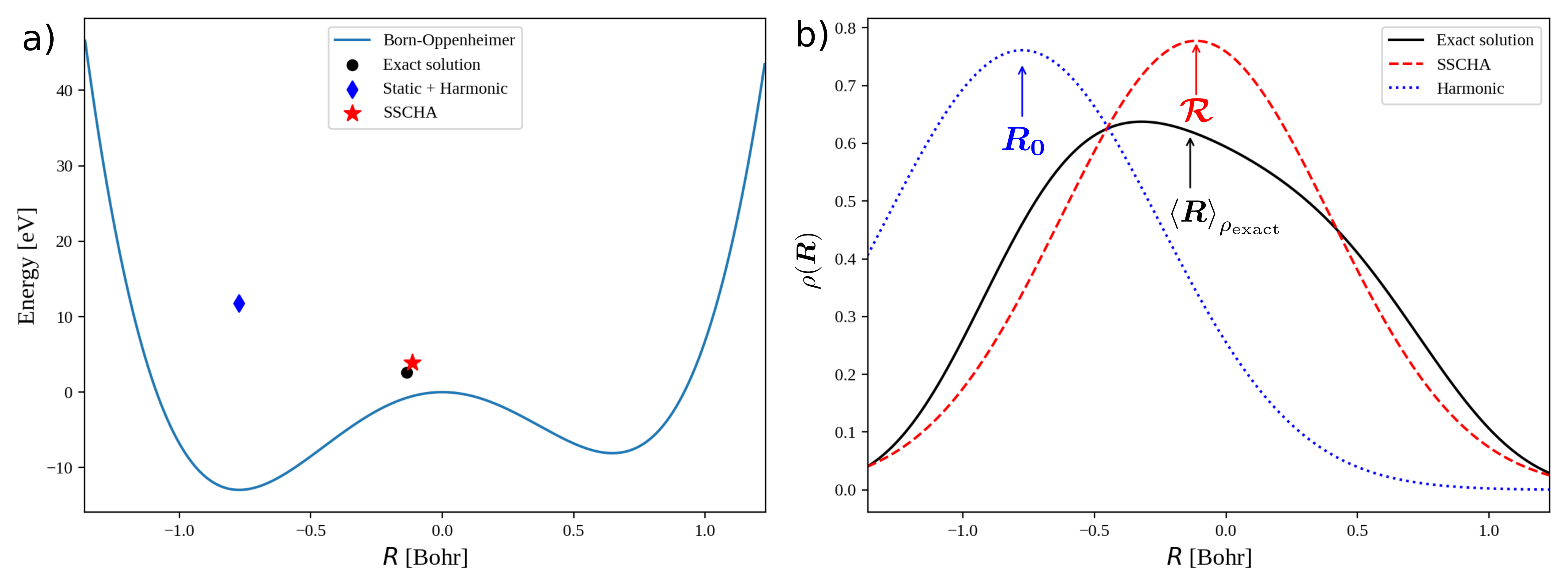}
    \caption{Illustration of the SSCHA method to a one dimensional particle problem at $T = \SI{0}{\kelvin}$. Panel \textbf{a}: The one dimensional Born-Oppenheimer energy landscape $V(R)$ as a function of the particle position $R$. The points represent the solution of the Harmonic approximation, the SSCHA, and the exact solution. The $y$ coordinate of the points are the quantum total energy (including the zero-point motion), while the $x$ axis coordinate is the average position of the particle. The SSCHA outperforms the harmonic approximation and it is very close to the exact solution. Panel \textbf{b}: Representation of the nuclear quantum distribution functions in the different approaches. The arrows point the average position of the particle in each distribution. Both harmonic and the SCHA are Gaussians, while the exact solution is more complex. The harmonic solution is centered around the minimum of the energy landscape $R_0$, while the SSCHA centroid position $\mathcal{R}$ and width are optimized to satisfy the least energy principle. The average position in the exact case is however obtained as $\braket{R}_{\rho_\mathrm{exact}}$.}
    \label{fig:1D_example}
\end{figure*}



To illustrate better the philosophy of the method, we report in \figurename~\ref{fig:1D_example} a simple application of the SSCHA to a one particle in one dimension at $T = \SI{0}{\kelvin}$. In panel \textbf{a}, we plot the very anharmonic ``Born-Oppenheimer'' (BO) energy landscape $V(\bR)$ of our one-dimensional particle (of mass of an electron). In Hartree atomic units it is given by
\begin{equation}
V(R) = 3R^4 + \frac 12 R^3 - 3 R^2.
\label{eq:onedim:pot}
\end{equation}
We first study the classical harmonic result, obtained by Taylor-expanding the potential in~\eqname~\eqref{eq:onedim:pot} to second order around the minimum $R_0$. Then, we use the harmonic solution to build our initial guess for the SSCHA density matrix $\rhoschatrial$
and update the parameters ($\phischatrial$ and $\rschatrial$) until we reach the minimum of the free energy. In \figurename~\ref{fig:1D_example}({a}), we compare the average atomic position and equilibrium free energy obtained with the harmonic approximation, with the SCHA, and the result we obtained with the exact diagonalization of the potential. While the harmonic result clearly overestimates the energy and yields an average atomic position far from the exact result, the SSCHA energy and average position are very close to the exact solution. In \figurename~\ref{fig:1D_example}({b}), we report the probability distribution functions of the particle for the different approximations compared with the exact result. By definition, both the harmonic  and  SSCHA results have Gaussian probability distributions. However, while the harmonic solution is centered in the minimum of the BO energy landscape (and the width is fixed by the harmonic frequencies), the SSCHA distribution is optimized to minimize the free energy. Notice how, even if the exact equilibrium distribution deviates from the Gaussian line-shape, the SSCHA energy and average nuclear position match almost perfectly the exact solution as stated above. The very good result on the free energy reflects that the SSCHA error is variational: the free energy of the exact density matrix is the minimum. This means that the free energy is stationary around the exact solution, assuring that even an approximate density matrix (like the SSCHA solution) describes very well the exact free energy.  
This is an excellent feature of the SCHA, as the free energy and its derivatives fully characterize thermodynamic properties.
Even if this simple calculation is performed at $T = \SI{0}{\kelvin}$, the SSCHA can simulate any finite temperature by mixing quantum and thermal fluctuations on the nuclear distribution.


The previously outlined straightforward implementation of the SSCHA becomes too cumbersome on a real system composed of many particles, especially if {\it ab initio} methods are used to extract $V(\bR)$. The reason is that at any minimization step we need to calculate total energies and forces for many ionic configurations with displaced atoms in a supercell. The bottleneck is the computational cost of the force engine adopted. In the next sections of the paper we will show how the number of force calculations can be minimized and how these issues can be overcome by the code implementation proposed here. The resulting SSCHA code is very efficient, and, in most of the core cases,
much faster than standard AIMD, with the advantage of fully accounting for the quantum nature of nuclei.

\section{Structure relaxation and free energy minimization} 
\label{sec:relax}

In this section we explain the simplest and most common use of the code: the calculation of the free energy and the optimization of a structure by fully accounting for temperature and quantum effects. This enables the simulation of finite temperature and pressure phase-diagrams (with first order boundaries), as well as the calculation of the lattice thermal expansion. We start by briefly reviewing the theory of the SSCHA method. Then, we will explain the details of the implementation, giving tips on how to run a simulation.

\subsection{The SSCHA free energy minimization}
\label{sec:method:free_energy}

\begin{table*}
\begin{ruledtabular}  
\begin{tabular}{lll}
   Symbol  & Meaning     &     First use    \\
\hline
$\bR$  
    & Atomic position (canonical variable) 
    & Eq.~\eqref{eq:toten}\\
$V(\bR)$ 
    & Potential energy 
    & Eq.~\eqref{eq:toten}\\
$\bRcal$ 
    & Trial centroid positions (parameter)
    & Eq.~\eqref{eq:HSCHAtrial}  \\
$\bu$ 
    & Displacement from the average atomic position $\bRcal$ 
    & Eq.~\eqref{eq:psi0}\\    
$\bvarPhi$ 
    &  Trial harmonic matrix (parameter) 
    & Eq.~\eqref{eq:HSCHAtrial}\\
$\bvarPsi$ 
    & Static displacement-displacement correlation matrix relative to $\bvarPhi$ 
    & Eq.~\eqref{eq:psi0}\\    
$\Hschatrial$
    & Trial harmonic Hamiltonian for  given $\bRcal$, $\bvarPhi$ 
    & Eq.~\eqref{eq:HSCHAtrial}\\    
$\rhotrial _{\bRcal,\bvarPhi}$ 
    & Density matrix of $\Hschatrial$ (trial density matrix)
    & Pag.~\columnref{eq:TrialGauss}\\ 
$\rhotrial _{\bRcal,\bvarPhi}(\bR)$ 
    & Gaussian positional probability density for $\rhotrial _{\bRcal,\bvarPhi}$
    & Pag.~\columnref{eq:TrialGauss}\\     
$\Fcal[\bRcal,\bvarPhi]$ 
    & SSCHA Helmholtz free energy functional 
    & Eq.~\eqref{eq:fvariational}\\
$\mathcal{G}[\bRcal,\bvarPhi]$ 
    & SSCHA Gibbs free energy functional 
    & Eq.~\eqref{eq:gvariational}\\    
$\bfscha(\bR)$
    & Forces for the Hamiltonian $\Hschatrial$ acting on the ions when they are in the $\bR$ positions 
    & Eq.~\eqref{eq:gradR}\\    
$\bfBO(\bR)$
    & Born-Oppenheimer forces acting on the ions when they are in the $\bR$ positions
    & Eq.~\eqref{eq:gradR}\\ 
$\bstressbo(\bR)$
    & Born-Oppenheimer stress tensor when the ions are in the $\bR$ positions 
    & Eq.~\eqref{eq:avgstress}\\        
$\bPhi_{\sssrcal}$    
    & 2nd SSCHA force constants for a given $\bRcal$, it is the trial $\bvarPhi$ that minimizes $\Fcal[\bRcal,\bvarPhi]$
    & Pag.~\columnref{phiR}\\  
$\bD_{\sssrcal}$    
    & $\bPhi_{\sssrcal}$ divided by the square root of the masses
    & Eq.~\eqref{eq:2ndD}\\  
$\bPhithree_{\sssrcal},\,\bPhifour_{\sssrcal}$    
    & 3rd and 4th order SSCHA force constants for a given $\bRcal$
    & Eqs.~\eqref{eq:3rdD},~\eqref{eq:4thD}\\     
$\bDthree_{\sssrcal},\bDfour_{\sssrcal}$    
    & $\bPhithree_{\sssrcal},\bPhifour_{\sssrcal}$ divided by the square root of the masses
    & Eqs.~\eqref{eq:3rdD},~\eqref{eq:4thD}\\      
$F(\bRcal)$ 
    & SSCHA positional Helmholtz free energy, given by $\Fcal[\bRcal,\bPhi_{\sssrcal}]$ 
    & Eq.~\eqref{eq:fmin_only_fc}\\      
$\bRcaleq$ 
    & SSCHA equilibrium centroids, trial centroids that minimizes $F(\bRcal)$  
    & Eq.~\eqref{eq:fvariationalmin}\\    
$\phischaeq$ 
    & SSCHA harmonic matrix $\bPhi_{\bRcaleq}$, the trial $\bvarPhi$ at the minimum of the free energy functional 
    & Eq.~\eqref{eq:fvariationalmin}\\   
$\Hscha$ 
    & SSCHA effective harmonic Hamiltonian, given by $\Hschatrialeq$
    & Eq.~\eqref{eq:HSCHA}\\  
$\bDS$    
    & Dynamical matrix of $\Hscha$, given by $\bD_{\bRcaleq}$
    & Pag.~\columnref{DSdef}\\        
$\bDthreeeq,\,\bDfoureq$    
    & Symbols indicating $\bDthree_{\bRcaleq}$ and $\bDfour_{\bRcaleq}$, respectively
    & Eq.~\eqref{eq:static_SE}\\  
$\bDF$    
    & Positional Helmholtz free energy Hessian divided by the square root of the masses 
    & Pag.~\columnref{posHesdivM}\\ 
$\bG(z)$
    & One-phonon Green function
    &  Eq.~\eqref{Eq:Gm1} \\
$\bPi(0),~\bPi(z)$    
    & Static and dynamic SSCHA self-energy
    & Eqs.~\eqref{eq:static_SE},~\eqref{Eq:app_Pi}\\ 
$\bPiB(0),~\bPiB(z)$    
    & Static and dynamic SSCHA bubble self-energy 
    & Eqs.~\eqref{eq:static_bubble_SE},~\eqref{Eq:app_Pi_B}\\ 
$\sigma(\bq,\Omega)$
    & Phonon spectral function (with the reciprocal lattice vector made explicit) 
    & eq.\eqref{eq:spfuq}\\
$\omega_{\mu}(\bq)$
    & Frequency of the $({\mu},\bq)$ SSCHA auxiliary phonon
    & Eq.~\eqref{eq:psi}\\ 
$\Omega_{\mu}(\bq)$
    & Frequency of the $({\mu},\bq)$ static approximation phonon from $\bDF$ 
    & Pag.~\columnref{staticfreq}\\     
$\bbOmega_{\mu}(\bq)$,$\Gamma_{\mu}(\bq)$    
    & Frequency and linewidth of the $({\mu},\bq)$ anharmonic phonon in the Lorentzian approximation
    & Eq.~\eqref{eq:mu_spec_lor}\\       
\end{tabular}
\end{ruledtabular}
\caption{Collection of some symbols frequently used in the main text.
First column, the symbol used. Second column, a short description. Third column,
equation or page-column of the first occurrence.}
\label{tab:symbols}
\end{table*}

In the simplest and most standard usage, the SSCHA free energy functional that is minimized depends on the centroid positions $\bRcal$ and the auxiliary force constants $\phischatrial$ as
\begin{equation}
\Fcal[\rscha,\phischatrial]=\Avgschatrial{K+V(\bR)}-T\,\Sion\left[\rhoschatrial\right].
\label{eq:fvariational}
\end{equation}
Here, we explicit that the entropy $\Sion$ only accounts for ionic degrees of freedom (not electronic). 
%
%
After the SSCHA minimization, the final estimate of the equilibrium free energy is given by
\begin{equation}
F=\min_{\rscha,\,\phischatrial}\Fcal[\rschatrial,\phischatrial]=\Fcal[\rschaeq,\phischaeq]\,.  
\label{eq:fvariationalmin}
\end{equation}
Therefore, the final result of a SSCHA free energy calculation is in general given in terms of the equilibrium configuration $\rschaeq$, the free energy $F$, and the SSCHA auxiliary force constants $\phischaeq$. The final free energy accounts for quantum and thermal ionic fluctuations without approximating the BO energy surface, valid to study thermodynamic properties, and the $\rschaeq$ positions determine the most probable atomic positions also taken into account quantum/thermal flluctuations and anharmonicity. It is important to remark, however, that the Gaussian variance $\phischatrial$ has, in principle, no relation with the experimentally observed phonon frequencies, as it is just a variable parametrizing the density matrix. The relation of it with the physical phonon frequencies is discussed in Sec. \ref{sec:method:sigma}.

%

The SSCHA can also perform the free energy minimization at fixed pressure instead. In this case, the Gibbs-Bogoliubov inequality is satisfied by the Gibbs free energy $G$, defined as
\begin{equation}
    G = F + P^*\Vol,
\end{equation}
where $P^*$ is the target pressure, $\Vol$ is the simulation box volume, and $F$ is the Helmholtz free energy. In this case, the code optimizes
\begin{equation}
    G \le {\mathcal G}[\rschatrial,\phischatrial] = \Fcal[\rschatrial,\phischatrial] + P^* \Vol, 
    \label{eq:gvariational} 
\end{equation}
which can be used, for instance, to estimate the structural changes imposed by pressure by fully accounting for fluctuations.

As made explicit in \eqname~\eqref{eq:fvariational}, only  thermal effects on the ions are taken into account so far, whereas the electrons are considered at zero temperature. However, at very high temperatures the entropy associated to electrons may be important. Within the SSCHA, it is possible to explicitly include finite-temperature effects on the electrons too. The key is to replace in \eqname~\eqref{eq:fvariational} the electronic ground state energy $V(\bR)$ with the finite-temperature electronic free energy $\Fel(\bR)=\Eel(\bR)-T\Sel(\bR)$ (if electrons have finite temperature, in the adiabatic approximation forces and equilibrium position of the ions are ruled by the electronic free energy). In this case the SSCHA method minimizes the functional
\begin{align}
\Fcal[\rscha,\phischatrial]&=\Avgschatrial{K+\Fel(\bR)}-T\,\Sion\left[\rhoschatrial\right] = \nonumber \\
& =\Avgschatrial{K+\Eel(\bR)}-T\,S\left[\rhoschatrial\right]\,,
\end{align}
where $ S\left[\rhoschatrial\right] =\Avgschatrial{\Sel(\bR)}+\Sion\left[\rhoschatrial\right]$. The same trick can be applied to the Gibbs free energy minimization as well. Therefore,  the  SSCHA  estimation  of  the  system’s  entropy can also incorporate  contributions  from  both  electrons  (averaged  through  the  ionic  distribution $\rhoschatrial$)  and  ions. In a DFT framework, for example, this simply comes down to including the electronic temperature in the  energy/forces/stress calculations for the  ensemble elements through the Fermi-Dirac occupation of the Kohn-Sham states~\cite{PhysRevLett.125.106101}.

\subsection{The implementation of the free energy minimization}
\label{sec:method:derivatives}

In the SSCHA code, the minimization of $\Fcal[\rschatrial,\phischatrial]$ 
is performed through a preconditioned gradient descent approach, which requires the calculation of the gradient of the free energy with respect to the centroid positions $\rschatrial$ and the auxiliary force constants $\phischatrial$. The partial derivatives are evaluated through the exact analytic formulas
\begin{equation}
\frac{\partial \Fcal}{\partial \Rcal_a}=-\Avgschatrial{\fBO_a(\bR) - \fscha_a(\bR)}
\label{eq:gradR}
\end{equation}
and
\begin{align}
&\frac{\partial \Fcal}{\partial \varPhi_{cd}}= \frac 12 \sum_{ab} \frac{\partial \varPsi_{ab}}{\partial \varPhi_{cd}}  \nonumber \\ 
&\times\Avgschatrial{ \left( \ft^{(BO)}_b(\bR) - \fscha_b(\bR)\right)\sum_e \varPsi^{-1}_{ae} \left(R_e -  \Rcal_e \right) }.
\label{eq:gradPhi}
\end{align}
Here $\bfBO(\bR)$ are the Born-Oppenheimer forces that act on the ions when they are in the $\bR$ positions; $\bfscha(\bR)$ is the force given by the auxiliary harmonic Hamiltonian $\Hschatrial$,
\begin{equation}
    \fscha_a(\bR) = -\sum_{b} \varPhi_{ab}(R_b - \Rcal_b);
    \label{eq:fharmonic}
\end{equation}
and $\bvarPsi$ is the displacement-displacement correlation matrix 
\begin{equation}
    \varPsi_{ab} = \Avgschatrial{u_au_b},
    \label{eq:psi0}
\end{equation}
where with $\bu=\bR-\bRcal$
we indicate the displacement from the average atomic position. Explicitly,
\begin{equation}
    \varPsi_{ab} = \frac{1}{\sqrt{M_aM_b}} \sum_\mu \frac{\hbar (2n_\mu + 1)}{2\omega_\mu} e_\mu^a e_\mu^b.
    \label{eq:psi}
\end{equation}
In Eq. \eqref{eq:psi}, $\omega_\mu$ and $\boldsymbol{e}_\mu$ are the eigenvalues and eigenvectors of the mass rescaled auxiliary force constants $\varPhi_{ab}/\sqrt{M_aM_b}$, and $n_\mu$ is the Bose-Einstein occupation number for the $\omega_\mu$ frequency. We underline again here that $\omega_\mu$ are not the phonon frequencies of the system, but just the frequencies of the auxiliary harmonic Hamiltonian $\Hschatrial$. In other words, they are only used to define the trial density matrix $\rhoschatrial$. We show how to compute the physical anharmonic phonon frequencies of the system in Sec.~\ref{sec:postprocessing}.  

It is convenient to give an explicit expression for the gradient of the free energy with respect to the auxiliary force constants in terms of the $\omega_\mu$  eigenvalues and  $\boldsymbol{e}_\mu$ eigenvectors. As shown in Ref. \cite{PhysRevB.96.014111} (see Appendix \ref{app:gradient:new}), the gradient can be rewritten as
\begin{align}
&\frac{\partial \Fcal}{\partial \varPhi_{cd}}=  \sum_{ab} \frac{\Lambda[0]^{abcd}}{\sqrt{M_aM_bM_cM_d}}  \nonumber \\ 
&\times\Avgschatrial{ \left( \ft^{(BO)}_b(\bR) - \fscha_b(\bR)\right)\sum_e \varPsi^{-1}_{ae} \left(R_e -  \Rcal_e \right)},
\label{eq:gradPhiwithLambda}
\end{align}
where
\begin{eqnarray}
    \Lambda[0]^{abcd} & = & \sum_{\mu\nu} \frac{\hbar}{4\omega_{\nu}\omega_{\mu}} e_\nu^a e_\mu^b e_\nu^c e_\mu^d \nonumber \\ & \times &
    \begin{cases}
    \frac{dn_\mu}{d\omega_\mu} - \frac{2n_\mu+1}{2\omega_\mu} &, \omega_\nu = \omega_\mu \\
    \frac{n_\mu -n_\nu}{\omega_\mu - \omega_nu} - \frac{1+n_\mu+n_\nu}{\omega_\mu + \omega_nu} &, \omega_\nu \neq \omega_\mu
    \end{cases}
    .
    \label{eq:lambda0}
\end{eqnarray}
Here, $n_{\mu}= 1/(e^{\beta\hbar\omega_{\mu}}-1)$. The reason why we have introduced the $\boldsymbol{\Lambda}[0]$ tensor will be evident in Sec. \ref{sec:postprocessing}. Even if Eq. \eqref{eq:gradPhiwithLambda} looks different to the gradient introduced in the original SSCHA work in Ref. \cite{PhysRevB.89.064302}, it can be demonstrated that both expressions are equivalent by simply playing with the permutation symmetry of $\Avgschatrial{\frac{\partial^2{V}}{\partial R_a\partial R_b}}$. However, Eq. \eqref{eq:lambda0} unambiguously determines the value taken by the $\boldsymbol{\Lambda}[0]$ tensor for the $\omega_\nu = \omega_\mu$ case, while the gradient in Ref. \cite{PhysRevB.89.064302} did not describe explicitly what to do in this degenerate limit.




At the end of the SSCHA optimization, apart from the temperature-dependent $\rschaeq$ positions and the equilibrium auxiliary force constant matrix $\phischaeq$, the code also calculates the anharmonic stress tensor $\bstress$, which includes both quantum and thermal ionic fluctuations, as derivatives of the free energy with respect to a strain tensor $\bvareps$:
\begin{align}
\stress_{\alpha\beta}
&=-\left.\frac{1}{\Vol}\frac{\partial \Fcal}{\partial {\vareps_{\alpha\beta}}}\right|_{\bvareps=0} 
= \Avgschatrial{\stressbo_{\alpha\beta}(\bR)}\nonumber \\
& -\frac{1}{2\Vol}\sum_{s = 1}^{\Nat}
\Avgschatrial{u_{s}^\alpha {\fBO}_s^\beta + u_s^\beta {\fBO}_s^\alpha}.
\label{eq:avgstress}
\end{align}
Here, we have made explicit the atomic index $s$ (lower index) and Cartesian $\alpha,\beta$ (upper index) of $\bu$ and $\bfBO$. $\bstressbo(\bR)$ is the Born-Oppenheimer stress tensor of the configuration with ions displaced in the $\bR$ coordinates. This equation is slightly different from the stress tensor equation presented in \cite{PhysRevB.98.024106}. The two equations coincide at equilibrium, but this is more general. The derivation of \eqname~\eqref{eq:avgstress} is reported in \appendixname~\ref{app:stress}. Thanks to the temperature-dependent stress, the SSCHA code can optimize also the lattice parameters and the volume. Thus, by relaxing the lattice at different temperatures, we get the thermal expansion straightforwardly.

Remarkably, the stress tensor of \eqname~\eqref{eq:avgstress} can be computed with a single SSCHA minimization at fixed volume. This is a huge advantage with respect to the standard quasi-harmonic approximation, not only because it includes quantum and anharmonic effects, but also because it is computationally much more efficient. In fact, the quasiharmonic approximation requires performing harmonic phonon calculations at different volumes (and/or internal lattice positions) to estimate the minimum of the quasi-harmonic free energy with finite differences. This process is extremely cumbersome for crystals with few symmetries and lots of internal degrees of freedom in the structure.  

In the current implementation of the code, the symmetries of the space group are imposed \emph{a posteriori} on the gradients of Eqs.~\eqref{eq:gradR} and \eqref{eq:gradPhi}, as well as on \eqref{eq:avgstress}.
This assures that the density matrix satisfies all the symmetries at each step of the minimization. Thus, during the geometry optimization, the system cannot lose any symmetry, though it can gain them. The symmetries are imposed following the methodology explained in Appendix \ref{app:symmetries}, which is different to the method originally conceived\cite{PhysRevB.89.064302}. The current SSCHA code can also work without imposing symmetries, allowing for symmetry loss, though the stochastic number of configurations needed to converge the minimization is larger (see Sec. \ref{sec:imp:stochastic}).


\subsubsection{The stochastic sampling}
\label{sec:imp:stochastic}

The stochastic nature of the SSCHA comes from the Monte Carlo evaluation of the averages in Eqs. \eqref{eq:gradR}, \eqref{eq:gradPhi}, and \eqref{eq:avgstress}. A set of random ionic configurations are created in a chosen supercell according to the Gaussian ionic probability distribution 
\begin{align}
    \rhoschatrial(\bR) & = \sqrt{\det(\psischatrial^{-1}/2\pi)} \nonumber \\
     & \times \exp \left[- \frac{1}{2}\sum_{ab} (R_a - \Rcal_a){\varPsi^{-1}}_{ab}(R_b - \Rcal_b) \right].
    \label{eq:rho:trial}
\end{align}
The Monte Carlo average of a generic observable $O(\bR)$, function only of the ionic position $\bm R$, is  calculated then as weighted sum over the created ensemble:
\begin{equation}
    \Avgschatrial{O(\bR)} = \frac{1}{\sum_{j = 1}^{\Nc} \rho_j} \sum_{j = 1}^{\Nc} \rho_j O( \bm R_{\{j\}}).
\end{equation}
Here, $\Nc$ is the total number of configurations in the ensemble, while $\bR_{\{j\}}$ is the $j$-th ionic randomly displaced configuration. Each of the $\bm{R}_{\{j\}}$ configurations is generated according to the initial trial ionic distribution $\rhoschastart(\bR)$ from which the minimization starts. To improve the stochastic accuracy, for each $\bm R_{\{j\}}$ configuration also $-\bm R_{\{j\}}$ is created, benefiting from $\rhoschatrial(\bR) = \rhoschatrial(-\bR)$ property of the Gaussian distribution. 

The $\rho_j$ weights are computed and updated along the free energy minimization as the values of $\bRcal$ and $\phischatrial$ change:
\begin{equation}
    \rho_j = \frac{\rhoschatrial\left(\bR_{\{j\}}\right)}{\rhoschastart\left(\bR_{\{j\}}\right)}.
\end{equation}
At the beginning, when the ensemble has just been generated and $\rschatrial = \rschatrial^{(0)}$ and $\phischatrial = \phischatrial^{(0)}$, all values of $\rho_j = 1$. However, as the $\bRcal$ and $\phischatrial$ are updated during the minimization, the weights change. This reweighting technique is commonly used in Monte Carlo methods\cite{Neal2001,Miotto2018} and takes the name of \emph{importance sampling}. This allows avoiding generating a new ensemble and computing \emph{ab initio} energies and forces at each step of the minimization, speeding up the SSCHA calculation.

\subsubsection{Minimization algorithm}
\label{sec:gradient}

The minimization strategy implemented in the SSCHA code for the free energy is based on a preconditioned gradient descent. At each step, the $\bRcal$ and $\phischatrial$ are updated as
\begin{equation}
    {\phischatrial}^{(n+1)} = {\phischatrial}^{(n)} - \lambda_{\phischatrial} \sum_{ab} \left(\frac{\partial^2 {\mathcal F}}{\partial {\phischatrial} \partial{\varPhi_{ab}}}\right)^{-1} \frac{\partial {\mathcal F}}{\partial{\varPhi_{ab}}}
    \label{eq:newthon:phi}
\end{equation}
\begin{equation}
    {\bRcal}^{(n+1)} = {\bRcal}^{(n)} - \lambda_{\bRcal} \sum_{a} \left(\frac{\partial^2 {\mathcal F}}{\partial {\bRcal} \partial{\Rcal_a}}\right)^{-1} \frac{\partial {\mathcal F}}{\partial{\Rcal_a}}.
    \label{eq:newthon:Rcal}
\end{equation}
In a perfectly quadratic landscape, this algorithm assures the convergence in just one step if both $\lambda_{\phischatrial}$ and $\lambda_{\bRcal}$ are set equal to one. However, in order to avoid too big steps in the minimization, often it is more  convenient to chose $\lambda_{\phischatrial|\bRcal} < 1$. This algorithm, with Hessian matrices that multiplies the gradient, is the \emph{preconditioned} steepest descent. If the \emph{preconditioning} option is set to false, a standard steepest descent minimization is followed instead, with $\lambda_{\phischatrial}$ and $\lambda_{\bRcal}$ re-scaled to the maximum eigenvalue of the $\frac{\partial^2 \mathcal F}{\partial \phischatrial^2}$ and $\frac{\partial^2 \mathcal F}{\partial \bRcal^2}$ Hessian matrices, respectively, in order to have adimensional values independent on the system. 

The preconditioning Hessian matrices that multiplies the gradients in \eqname~\eqref{eq:newthon:phi} and \eqname~\eqref{eq:newthon:Rcal} are approximated by the code. Following the procedure introduced in Ref.\cite{PhysRevB.98.024106}, we use the exact Hessian in the minimum of a perfectly harmonic oscillator with the same frequencies as the SCHA auxiliary Hamiltonian. In particular, they are:
\begin{equation}
    \frac{\partial^2 \mathcal F}{\partial \varPhi_{ab} \partial \varPhi_{cd}} \approx \frac 12 \frac{\partial\varPsi_{ab}}{\partial\varPhi_{cd}}\label{eq:precond:dyn}
\end{equation}
and
\begin{equation}
    \frac{\partial^2 \mathcal F}{\partial \bRcal \partial \bRcal} \approx \phischatrial.
\end{equation}
Eq. ~\eqref{eq:precond:dyn} is presented differently from the original work in which it was derived\cite{PhysRevB.98.024106}. We prove in \appendixname~\ref{app:gradient:new} that they are exactly the same. Considering that the Hessian preconditioner cancels out the $\frac 12 \frac{\partial\varPsi_{ab}}{\partial\varPhi_{cd}}$ term in Eq. \eqref{eq:gradPhi},  
the resulting update of the variational parameters at each step in the minimization is  performed as
\begin{align}
    &\varPhi_{ab}^{(n+1)} = \varPhi_{ab}^{(n)} + \nonumber \\
    & -\lambda_{\phischatrial} \Avgschatrial{ \left( \fBO_b(\bR) -\fscha_b(\bR)\right)\sum_c {\varPsi^{-1}}_{ac} \left(R_c -  \Rcal_c \right)}
    \label{eq:newthon:phi}
\end{align}
and
\begin{equation}
    {\Rcal}^{(n+1)}_a = {\Rcal}^{(n)}_a + \lambda_{\bRcal} \sum_{b} \varPhi^{-1}_{ab} \Avgschatrial{\fBO_b(\bR) - \fscha_b(\bR)}. 
    \label{eq:newthon:Rcal}
\end{equation}
This implementation is very efficient, especially for \eqname~\eqref{eq:newthon:phi}, as there is no need to calculate the $\LambdaR[0]$ tensor. Therefore, computing directly \eqname~\eqref{eq:newthon:phi} is much faster than calculating the gradient of \eqname~\eqref{eq:gradPhi}.






The code allows the user to select a different minimization algorithm specifically for the minimization  with respect to $\phischatrial$: the \emph{root representation}.
Since the minimization with respect to $\phischatrial$ is the more challenging, this technique aims to further improving the $\phischatrial$ optimization. In particular, the gradient has a stochastic error and the minimization is performed with a finite step size. For these reasons, $\phischatrial$ could become non positive definite during the optimization (i.e. the dynamical matrix has imaginary frequencies). If this occurs, the minimization is halted raising an error, as the density matrix of \eqname~\eqref{eq:rho:trial} diverges. In such a case, the minimization must be manually restarted, either by taking a smaller step or by stopping the minimization before reaching imaginary frequencies (fixing the maximum number of steps). This kind of halts do not occur often when using the preconditioning in the minimization. However, they may be encountered if few configurations are generated for each ensemble or the starting dynamical matrix is very far from equilibrium.

To solve these problems, we implement the root representation, in which, instead of updating $\phischatrial$ as in \eqname~\eqref{eq:newthon:phi}, it updates a root of $\phischatrial$:
\begin{equation}
    \sqrt[n]{\phischatrial}^{(i+1)} = \sqrt[n]{\phischatrial}^{(i)} - \lambda_{\phischatrial} {\bm {\mathcal G_n}}\label{eq:root:step}.
\end{equation}
The updating direction $\bm {\mathcal G_n}$ depends depends on the root order $n$:
\begin{equation}
    {\bm {\mathcal G_2}} =   \sqrt{\phischatrial} \cdot \frac{\partial\mathcal F}{\partial{\phischatrial}} +  \frac{\partial\mathcal F}{\partial{\phischatrial}}\cdot \sqrt{\phischatrial},
    \label{eq:g2:root}
\end{equation}
where with the $\cdot$ we indicate a matrix product. Similarly,
\begin{equation}
    {\bm {\mathcal G_4}} =  \sqrt[4]{\phischatrial} \cdot {\bm{\mathcal G_2}} +   {\bm{\mathcal G_2}}\cdot \sqrt[4]{\phischatrial}.
\end{equation}
We select the positive definite root matrix.
Indeed, after the step of \eqname~\eqref{eq:root:step} the original force constant matrix is obtained as
\begin{equation}
    \phischatrial^{(i+1)} = \left(\sqrt[n]{\phischatrial}^{(i+1)}\right)^n.
    \label{eq:positive}
\end{equation}
Thanks to the definition in \eqname~\eqref{eq:positive}, the dynamical matrix is always positive definite for any even value of $n$.

The \emph{root representation} is independent of the preconditioning. With preconditioning, we replace the free energy gradient in \eqname~\eqref{eq:g2:root} with the preconditioned direction in \eqname~\eqref{eq:newthon:phi} (the gradient multiplied by the approximated Hessian). This is different from what was proposed in the original work\cite{PhysRevB.98.024106}, where the Hessian matrix was computed also for the $\sqrt{\phischatrial}$ and $\sqrt[4]{\phischatrial}$ cases. However, we noticed that in systems with many atoms, the Hessian matrix calculation becomes the bottleneck as it scales with $\Nat^6$. The implementation here described allows for a much faster $\phischatrial$ update and avoids calculating the Hessian matrix.
The drawback is that the optimization step is not as optimal as it would be if the proposal in Ref. \cite{PhysRevB.98.024106} was followed.
The code offers six combinations for the minimization procedure: no root, square root ($n = 2$), and fourth-root ($n = 4$), all of them with or without the preconditioned direction. The optimal minimization step is $n=1$ with preconditioning. If the square root is employed ($n=2$), it is preferable to use the preconditioning. If fourth-root is employed ($n=4$), the best performances are without \emph{preconditioning}.

\subsubsection{The lattice geometry optimization}

The lattice degrees of freedom $\{\bm a_i\}$ are relaxed only after the minimization of the free energy with respect to $\bRcal$ and $\phischatrial$ at a constant volume stops (see Sec. \ref{sec:flowchart} for a detailed description of the stopping criteria). For this reason, the lattice geometry optimization is an ``outer'' optimization: at each step the lattice geometry optimization, we perform a full free energy minimization with respect to the centroids $\rschatrial$ and auxiliary force constants $\phischatrial$. 
This means that each step of the lattice geometry optimization is performed with a different ensemble, whose configurations are all generated with the same lattice vectors.

To update the lattice, the code calculates the stress tensor with \eqname~\eqref{eq:avgstress}, and generates a strain for the lattice as
\begin{equation}
    \varepsilon_{\alpha \beta} =\Vol \left( \stress_{\alpha \beta} - P^* \delta_{\alpha \beta}\right),
\end{equation}
where $P^*$ is the target pressure of the relaxation and $\delta_{\alpha \beta}$ is the Kronecker delta. The lattice parameters $\{ \bm a_i\}$ are updated as
\begin{equation}
    {a_i'}_\alpha = {a_i}_\alpha + \lambda_{\{\bm a_i\}}\sum_{\beta} \varepsilon_{\alpha\beta} {a_i}_\beta,
\end{equation}
where $\lambda_{\{\bm a_i\}}$ is the update step. Since each step requires a new ensemble, it is  crucial to reduce the number of steps to reach convergence by properly picking the right value for $\lambda_{\{\bm a_i\}}$.
In an isotropic material with a constant bulk modulus
\begin{equation}
    B_0 = -\Vol \left.\frac{d^2V^{(BO)}}{d\Vol^2}\right|_{\bR = \rschatrial},
    \label{eq:bulk:modulus}
\end{equation}
the optimal value of the step is
\begin{equation}
\lambda_{\{\bm a_i\}} = \frac{1}{3\Vol B_0}.
\end{equation}
$B_0$ is an input parameter given in GPa units. Good values of $B_0$ may range from \SI{10}{\giga\pascal} for crystals at ambient conditions, like ice, up to \SI{800}{\giga\pascal} in systems at \SI{}{\mega\bar} pressures (or for diamond). Remember that increasing the value of $B_0$ produces smaller steps in the cell parameters. The user can estimate the optimal value of $B_0$ to assure the fastest convergence by manually computing it from \eqname~\eqref{eq:bulk:modulus}, by taking finite differences of the pressure obtained at two uniformly strained volumes, or by looking for the experimental value of similar compounds. 

Alternatively to the fixed pressure optimization, it is also possible to perform the geometry lattice optimization at fixed volume. In this case $P^*$ is recomputed at each step so that $\Tr{{\bm \varepsilon}} = 0$. In this case, the final lattice parameters are also rescaled so that the final volume matches the one before the step. Since this algorithm has one less degree of freedom than the fixed pressure one, it usually converges faster.

\subsubsection{The code flowchart}
\label{sec:flowchart}

To start a SSCHA simulation, we need a starting guess on the trial positive definite force constants matrix $\phischatrial^{(0)}$ and on the average atomic positions $\rschatrial^{(0)}$.
Even if in principle the starting point is arbitrary, the closer to the solution we begin, the faster the minimization converges. Thus, the coordinates at the minimum of the Born-Oppenheimer energy landscape and the harmonic force constants are usually good starting points, which can be obtained from any code that computes phonons. The supercell of the simulation is given by the dimension
of the input force constants matrix, while the centroids $\bR$ are defined in the unit cell (they satisfy  translational symmetry). If the original dynamical matrix contains imaginary frequencies, it can be reverted to positive definite as
\begin{equation}
    \varPhi^{(0)}_{ab} = \sqrt{M_a M_b} \sum_\mu |\omega_\mu^2| e_\mu^a e_\mu^b.
\end{equation}
Then, the first random ensemble (that we call population in the SSCHA language) can be generated. For each configuration inside the population, its total Born-Oppenheimer energy as well as its classical atomic forces $\bfBO$ and stress tensor $\bstressbo$ must be computed. This is done with an external code, either manually (by computing externally the energies, forces, and stress tensor, and loading them back into the SSCHA code), or automatically (with an appropriate configuration discussed in Sec.~\ref{sec:python}). Once the Born-Oppenheimer energies, forces, and stress tensor of all the configurations have been computed, the minimization starts. The gradients of the free energy are computed as described in Sec.~\ref{sec:gradient} and the minimization continues either until the stochastic sampling is not good or the algorithm converges.

If $\rschatrial$ and $\phischatrial$ change a lot during the minimization of the free energy, the original ensemble no longer describes well the new probability distribution $\rhoschatrial(\bR)$, and the stochastic error increases. This occurrence is automatically checked by the SSCHA code calculating the Kong-Liu\cite{Kong1994,liu64} \emph{effective sample size} $N_{eff}$:
\begin{equation}
    N_{eff} = \frac{\sum_{j = 1}^{\Nc} \rho_j^2}{\left(\sum_{j = 1}^{\Nc} \rho_j\right)^2}.
\end{equation}
We halt the minimization when the ratio between $N_{eff}$ and the number of configurations $\Nc$ is lower than a parameter $\eta$ defined by the user:
\begin{equation}
\label{eq:cond:stochastic}
\frac{N_{eff}}{\Nc} < \eta.
\end{equation}
A standard value of $\eta$ that ensures a correct minimization is 0.5, but it can be convenient to lower it a bit to accelerate convergence in the first steps.

The convergence, on the contrary, is achieved only if the two gradients with respect to $\bRcal$ and $\phischatrial$ are lower than a given threshold:
\begin{equation}
    \left|\frac{\partial \Fcal}{\partial\phischatrial}\right| < \delta_{\phischatrial}
\end{equation}
\begin{equation}
    \left|\frac{\partial \Fcal}{\partial\bRcal}\right| < \delta_{\bRcal}.
\end{equation}
The $\delta$ threshold is provided by the user and re-scaled at each step by the estimation of the stochastic error on the corresponding gradient (\emph{meaningful\_factor}). So, at each step, $\delta$ is
\begin{equation}
\delta_{\phischatrial} = \mbox{meaningful\_factor} \cdot \left| \Delta\frac{\partial \Fcal}{\partial\phischatrial} \right|
\end{equation}
\begin{equation}
\delta_{\bRcal} = \mbox{meaningful\_factor} \cdot \left| \Delta\frac{\partial \Fcal}{\partial\bRcal} \right|.
\end{equation}
In this way, the user-provided variable \emph{meaningful\_factor} is independent on the system size or the number of configurations used. 

If the lattice parameters are free to move, then an additional condition must be fulfilled in order to end the minimization: each component of the strain per unit-cell volume $\Vol$ must be smaller than the stochastic error on the stress tensor:
\begin{equation}
    \frac{\varepsilon_{\alpha \beta}}{\Vol} \le \Delta \stress_{\alpha \beta}. 
    \label{eq:cond:stress}
\end{equation}
If \eqname~\eqref{eq:cond:stress} is not fulfilled, even if all the gradients are lower than the chosen threshold, the code generates a new ensemble and continues (until both conditions are satisfied).

\begin{figure*}
\centering 
\includegraphics[width=0.6\textwidth]{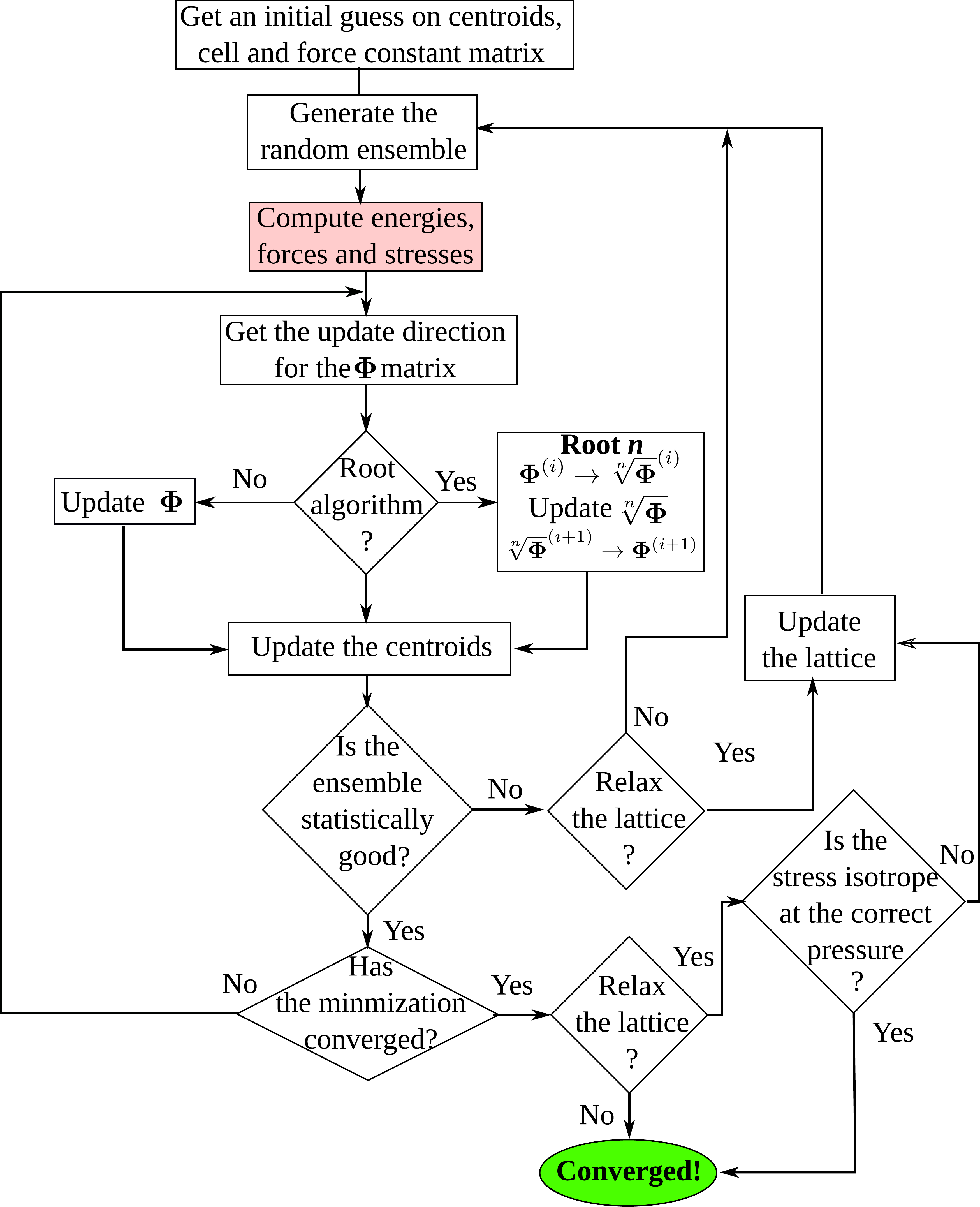}
\caption{Flowchart of the SSCHA code. The most time consuming part of the diagram is the \emph{ab initio} calculation of the Born-Oppenheimer forces, energies, and stress tensors for all the configurations inside the ensemble, and it is shaded in red. All the other steps usually take few seconds when executed on a standard workstation, even in systems that contain several hundreds of atoms. 
}
\label{fig:flowchart}
\end{figure*}

At the end of the minimization, the output of the SSCHA minimization gives the total free energy (with stochastic error), the average equilibrium ionic positions $\bRcaleq$, the equilibrium auxiliary force constant matrix $\phischaeq$, and the stress tensor $\bstress$. All output quantities are temperature-dependent, and include quantum-thermal fluctuations and anharmonicity. A flowchart that represents the whole execution of a SSCHA run is presented in \figurename~\ref{fig:flowchart}.
If the SSCHA code is coupled with an \emph{ab initio} total-energy engine, the most expensive calculation in the flowchart is by far the calculation of Born-Oppenheimer energy, forces, and stress tensors on the whole ensemble, which may contain up to several hundreds or thousands of configurations. For this reason, the pretty complex workflow we set up is aimed to pass by the calculation of a new ensemble as few times as possible. Most materials studied and presented in Sec. \ref{sec:examples} are converged within 3 populations, and the CPU time required to minimize each population is few minutes on a single CPU of modern laptops. 

\subsection{The self-consistent equation and possible alternative implementations of the SSCHA}

The preconditioned gradient descent approach sketched above offers a very efficient implementation of the SSCHA theory, in which the anharmonic free energy  is optimized by all degrees of freedom in the crystal structure, including internal coordinates as well as lattice vectors. If the centroid positions $\rscha$ are kept fixed in the minimization, the SSCHA self-consistent equation 
\begin{equation}
    \varPhi_{ab} (\bRcal) = \AvgschaRfixed{\frac{\partial^2V}{\partial R_a \partial R_b}}
    \label{eq:SSCHAself-consistent}
\end{equation}
offers an alternative way of implementing the SSCHA theory (see Ref. \cite{PhysRevB.96.014111} for a proof of Eq. \eqref{eq:SSCHAself-consistent}). It is important to underline the self-consistent condition required by the equation above, as the quantum statistical average is taken with a density matrix dependent on $\phischatrial(\rscha)$, which must equal the result of the average. As in this approach the centroid positions are not optimized, the obtained auxiliary force constant matrix depends parametrically on $\rscha$.

The self-consistent equation can be implemented stochastically, following the procedure outlined in Sec. \ref{sec:imp:stochastic}. By using integration by parts \cite{PhysRevB.96.014111}, the right-hand-side of Eq. \eqref{eq:SSCHAself-consistent} can be rewritten in terms of forces and displacements. Thus, with the importance sampling technique and reweighting, the equation can be solved by calculating forces in supercells generated with the SSCHA density matrix. An equivalent approach \cite{roekeghem2020quantum} is to extract the auxiliary force constants by fitting the obtained forces in the supercells generated with the SSCHA density matrix to Eq. \eqref{eq:fharmonic}. This approach has been followed recently \cite{roekeghem2020quantum,PhysRevB.95.014302,PhysRevLett.119.185901}, where a  least-squares technique is followed for the fitting. 

The use of the self-consistent equation is valid, thus, only for fixed centroid positions. If the centroid positions want to be optimized as well within this approach, the self-consistent procedure should be repeated for different values of $\rscha$, calculate the free energy for these positions, and see where its minimum is. Clearly this is a very cumbersome procedure unless centroid positions are fixed by symmetry. Moreover, solving the self-consistent equation fixing the centroid positions at the classical $\bR_0$ positions, which it is usually the case \cite{roekeghem2020quantum,PhysRevB.95.014302,PhysRevLett.119.185901}, neglects all the effects of quantum and thermal fluctuations on the structure. Since within our approach based on the gradient descent we can optimize the free energy not only with respect to the auxiliary force constants but also all degrees of freedom in the crystal structure, the workflow outlined in Sec. \ref{sec:method:derivatives} provides a full picture of the effect of quantum/thermal fluctuations as well as anharmonicity on crystals, much more efficient than the approaches based on Eq. \eqref{eq:SSCHAself-consistent}.   


\section{Post-minimization tools: positional free energy Hessian, phonon spectral functions, and phonon linewidths}
\label{sec:postprocessing}

In the previous section we described how to compute the free energy of a system and fully optimize its structure by taking into account the anharmonicity that arises from both thermal and quantum fluctuations. After the free energy functional minimization, additional information can be extracted from the results obtained, namely, the second derivative (Hessian) of the positional free energy with respect to the centroids, the anharmonic phonon spectral functions, and the anharmonic frequency linewidths and shifts. In the next subsections we will explain why these quantities are of physical interest and what is the strategy adopted by the code to computing them.
The theory here reviewed was introduced in Ref.~\cite{PhysRevB.96.014111} and extensively applied for the first time in \textit{ab initio} calculations to H${}_3$S in Ref.~\cite{Bianco2018}.

\subsection{Positional free energy Hessian}
As shown in Sec.~\ref{sec:overview}, for a given temperature the free energy at equilibrium $F$ of a system with Hamiltonian $H$ is obtained by minimizing the density-matrix functional 
$\Fcal[\rhotrial]=\Avg{K+V(\bR)}{\rhotrial}-TS[\rhotrial]$:
\begin{equation}
F=\min_{\tilde \rho}\Fcal[\tilde \rho]=\Fcal[\rho]\,,
\end{equation}
where $\rho$ is the equilibrium density matrix of the system obtained at the minimum. The average atomic positions at equilibrium are $\Avg{\bR}{\rho}=\bRcaleq$. 
By minimizing the functional keeping fixed the average atomic positions in a generic configuration $\bRcal$, $\Avg{\bR}{\rhotrial}=\bRcal$, 
we define the positional free energy $F(\bRcal)$:
\begin{equation}
F(\bRcal)= \min_{\substack{\tilde\rho\\ \Avg{\bR}{\tilde\rho}=\bRcal}} \Fcal[\tilde \rho]=\Fcal[\rho_{\bRcal}]\,,
\end{equation}
where $\rho_{\bRcal}$ is the density matrix giving the constrained minimum for the considered average position $\bRcal$. Since
\begin{equation}
F=F(\bRcaleq)=\min_{\bRcal}F(\bRcal)\,,
\end{equation}
$\bRcal$ and $F(\bRcal)$ can be interpreted as a multidimensional order parameter and a thermodynamic potential, respectively,
in the study of displacive phase transitions according to Landau's theory. Properly speaking, the ``order'' parameter 
would be $\bRcal-\bRcal_{\hs}$, where $\bRcal_{\hs}$ is the average position of the atoms when the system is in the high-symmetry phase. Therefore, the knowledge of the positional free energy landscape as a function of external parameters, like temperature or pressure, gives crucial information about the structural stability and evolution of a system, as it allows to determine the (meta-)stable configurations corresponding to (local) minima of the positional free energy.


The Hessian of the positional free energy $F(\bRcal)$ in the equilibrium configuration is the inverse response function to a static perturbation on the nuclei (i.e. the inverse of the static susceptibility). In presence of a second order phase transition the static response function diverges, which results in one or more eigenvalues of the positional free energy Hessian going to zero. This means that the occurrence of displacive second-order phase transitions, like CDW or ferroelectric transitions~\cite{PhysRevB.97.014306,PhysRevLett.122.075901,doi:10.1021/acs.nanolett.9b00504,PhysRevB.100.214307,PhysRevLett.125.106101}, can be characterized by analyzing the evolution with temperature of the eigenvalues of the equilibrium positional free energy Hessian. Typically, in these cases at high temperature the minimum point of the free energy, i.e. the equilibrium configuration $\rschaeq$, is a high-symmetry configuration $\rschahs$. Therefore, at high temperature the free energy Hessian in $\rschahs$ is positive definite, i.e. its eigenvalues are positive. As the temperature decreases, the minimum in $\rschahs$ becomes less and less deep, until it becomes a saddle point at the transition temperature (i.e. at least one eigenvalue is zero), so that a second-order displacive phase transition occurs and the equilibrium configuration $\rschaeq$ moves towards lower-symmetry configurations that reduce the free energy as the temperature decreases further (following the pattern indicated by the eigenvector of the vanishing eigenvalue). 
Using the same approach, it is possible  to characterize second-order displacive phase transitions driven by other external parameters, like the pressure in high-pressure superconducting hydrides~\cite{PhysRevLett.111.177002,PhysRevLett.114.157004,Errea81,Bianco2018,Errea66}.

The role played by the eigenvalues and eigenvectors of the positional free energy Hessian in tracing the system's structural stability recalls the role played by the harmonic
dynamical matrix in the standard harmonic approximation, but now including lattice thermal and quantum anharmonic effects in the dynamics of the nuclei. Therefore, the Hessian of the positional free energy, divided by the masses, \columnlabel{posHesdivM}{$\DF_{ab}=\left.\partial^2F/\partial\Rcal^a\partial\Rcal^b\right|_{\bRcaleq}/\sqrt{M_aM_b}$}, can be considered a natural generalization of the harmonic dynamical matrix that, however, includes thermal and quantum effects.

What explained hitherto about the role played by the positional free energy and its Hessian is general. In particular, the evaluation of the positional free energy within the SSCHA is pretty straightforward. Indeed, the average position for a trial harmonic density matrix $\rhoschatrial$ coincides with the centroid parameter $\bRcal$, 
\begin{equation}
\Avg{\bR}{\rhoschatrial}=\bRcal\,.
\end{equation}
Thus, within the SSCHA the positional free energy is obtained by minimizing the SSCHA free energy functional $\Fcal[\bRcal,\bvarPhi]$
with respect to the trial quadratic amplitude $\bvarPhi$ only:
\begin{equation}
F(\bRcal)=\min_{\bvarPhi}\Fcal[\bRcal,\bvarPhi]\,.
\label{eq:fmin_only_fc}
\end{equation}
The auxiliary force constants that minimize Eq. \eqref{eq:fmin_only_fc} for a given $\bRcal$ position of the centroids will be labeled in the following as \columnlabel{phiR}{$\bPhi_{\sssrcal}$}. Solving Eq. \eqref{eq:fmin_only_fc} allows to employ the SSCHA code to have direct access to $F(\bRcal)$ for any $\bRcal$ and, in principle, to compute the Hessian by finite differences. However, as discussed above, such a finite-difference approach would be extremely expensive for two main reasons. First, it would require a large number of configurations in the ensemble to reduce the stochastic error and calculate the derivatives by finite differences. Second, because the large number of degrees of freedom in $\bRcal$ prevents any realistic finite-difference approach. Luckily the SSCHA code allows to avoid any cumbersome finite-difference approach by exploiting an analytic formula for the positional free energy Hessian. 

Before describing the analytic formula, let us introduce a notation that will simplify the mathematical expressions. Given two tensors $\bX$ and $\bY$, with the single dot product $\bX\singledot\bY$ we will indicate the contraction of the last index of $\bX$ with the first index of $\bY$, $\sum_c X_{\ldots c}\,Y_{c\ldots}$. Likewise, with the double-dot product $\bX\doubledot\bY$ we will indicate the contraction of the last two indices of $\bX$ with the first two indices of $\bY$, $\sum_{cd} X_{\ldots cd}\,Y_{cd\ldots}$. Moreover, any fourth-order tensor $X_{pqlm}$ can be interpreted as a ``super" matrix $X_{AB}$, with the composite indices $A=(pq)$ and $B = (lm)$, and vice versa (through this correspondence we can define, for example, the inverse of a fourth-order tensor and the identity fourth-order tensor $\mathds{1}$). Using this notation, we can express the positional free energy Hessian, $1/\sqrt{M_aM_b}\,\,\partial^2 F/\partial\Rcal^a\partial\Rcal^b$, in component-free form as
\begin{align}
&\frac{1}{\sqrt{\bM}} \singledot \frac{\partial^2 F}{\partial \bRcal \partial \bRcal} \singledot \frac{1}{\sqrt{\bM}}= \nonumber \\
& = 
\bDR+ \bDthreeR 
\doubledot 
\bLambdaR[0]
\doubledot
\left[\mathds{1}-\bDfourR\doubledot\bLambdaR[0]\right]^{-1}
\doubledot
\bDthreeR\,,
\label{eq:Hessian}
\end{align}
where $M_{ab}=\delta_{ab}M_a$ is the mass matrix,
\begin{align}
&(\DR)_{ab}=
\frac{1}{\sqrt{M_a M_b}}\AvgschaR{\frac{\partial^2 V}{\partial R^a\partial R^b}}\mkern-20mu \nonumber\\
&\mkern70mu=\frac{(\Phi_{\sssrcal})_{ab}}{\sqrt{M_a M_b}}\label{eq:2ndD}\,,\\
&(\DthreeR)_{abc}=\frac{1}{\sqrt{M_a M_b M_c}}\AvgschaR{\frac{\partial^3 V}{\partial R^a\partial R^b\partial R^c}} \nonumber \\
&\mkern70mu=\frac{(\Phithree_{\sssrcal})_{abc}}{\sqrt{M_a M_b M_c}}\,,
\label{eq:3rdD}\\
&(\DfourR)_{abcd}= \frac{1}{\sqrt{M_a M_b M_c M_d}}\AvgschaR{\frac{\partial^4 V}{\partial R^a\partial R^b\partial R^c \partial R^d}}
\nonumber \\ 
&\mkern70mu= \frac{(\Phifour_{\sssrcal})_{abcd}}{\sqrt{M_a M_b M_c M_d}}\,,
\label{eq:4thD}
\end{align}
and $\bLambdaR[0]$ is the $z=0$ value of the fourth-order tensor $\bLambdaR[z]$, already introduced in Eq. \eqref{eq:lambda0}. In the equations above the quantum statistical averages are taken with $\rhoschaR$, which for a given $\bRcal$ position of the centroids is taken with the $\bPhi_{\sssrcal}$ auxiliary force constants that minimize the free energy. $\bLambdaR[z]$ is given in components by
\begin{equation}
\left(\LambdaR[z]\right)^{abcd}=\sum_{\mu\nu}
\mathscr{F}(z,\omega_{\mu},\omega_{\nu})\, 
e^a_{\nu}
e^b_{\mu}
e^c_{\nu}
e^d_{\mu},
\label{eq:lambda_1}
\end{equation}
where 
$\omega_\mu^2$ and ${e^a_{\nu}}$ are the eigenvalues and eigenvectors  of $\bD_{\sssrcal}$, and 
\begin{align}
\mathscr{F}(z,\omega_\nu,\omega_{\mu})=\frac{\hbar}{4\omega_{\mu}\omega_{\nu}}
& \bigg[  \frac{(\omega_{\mu}-\omega_{\nu})(n_{\mu}-n_{\nu})}{(\omega_\mu-\omega_\nu)^2-z^2}  + \nonumber \\ 
& -\frac{(\omega_{\mu}+\omega_{\nu})(1+n_{\mu}+n_{\nu})}{(\omega_\mu+\omega_\nu)^2-z^2}\bigg].
\label{eq:def_F0}
\end{align}
The only difference between $\bLambdaR[0]$ and $\boldsymbol{\Lambda}[0]$ (introduced in Eq. \eqref{eq:lambda0}) is that in the former the eigenvalues and eigenvectors entering the equation are those associated to the $\bPhi_{\sssrcal}$ auxiliary force constants at the centroid positions $\bRcal$, while in the latter this is not necessarily the case. The subindex $\bRcal$ in the equations above precisely indicates that the averages are calculated with a density matrix defined by $\bRcal$ and $\bPhi_{\sssrcal}$ (after a full SCHA relaxation at fixed nuclei position $\bRcal$). We will refer to the $\Phin_{\sssrcal}$ tensors as the $n$th-order SSCHA force constants (FCs). Note that for the second-order we drop the $(2)$ upper index.

The SSCHA code computes the free energy positional Hessian through Eq.~\eqref{eq:Hessian}. At the end of a SSCHA free energy functional minimization, the SSCHA matrix Eq.~\eqref{eq:2ndD}, with its eigenvectors and eigenvalues, is available. Thus, $\bLambdaR[0]$ is readily computable and the only quantities that need some effort to be calculated are the averages of Eqs.~\eqref{eq:3rdD} and~\eqref{eq:4thD}. The code computes them through these equivalent expressions (obtained by integrating by parts):
\begin{subequations}
\begin{align}
&\mkern-10mu(\Phithree_{\sssrcal})_{abc}=
-\sum_{pq}
\,(\Psi_{\sssrcal}^{\inv})_{ap}
\,(\Psi_{\sssrcal}^{\inv})_{bq}
\AvgschaR{\,u^pu^q\bbf_c\,}\\
&\mkern-10mu(\Phifour_{\sssrcal})_{abcd}=
-\sum_{pqr}
\,(\Psi_{\sssrcal}^{\inv})_{ap}
\,(\Psi_{\sssrcal}^{\inv})_{bq}
\,(\Psi_{\sssrcal}^{\inv})_{cr}
\AvgschaR{\,u^pu^qu^r\bbf_d\,},
\end{align}
\label{eq:ho_fc}
\end{subequations}
where 
$\bPsi_{\sssrcal}$ is the $\bvarPsi$ matrix with $\phischatrial = \phischaR$ and
\begin{equation}
\pmb{\bbf}(\bR)=\bfBO(\bR)-\AvgschaR{\bfBO(\bR)} - \bfschaR(\bR).
\label{eq:bbf}
\end{equation}
These averages are computed employing the stochastic approach already described in Sec.~\ref{sec:imp:stochastic} (indeed, as explained in Ref.~\cite{PhysRevB.96.014111}, the choice of Eq.~\eqref{eq:bbf}, among other possible alternatives, aims at reducing the statistical noise). Note that if the calculation of the free energy Hessian is performed at $\rschaeq$, $\AvgschaR{\bfBO(\bR)}$ vanishes. 

In order to minimize the number of energy-force calculations needed, it is advisable to compute these averages using the same ensemble used to minimize the free energy functional and obtain $\bPhi_{\sssrcal}$ (at most adding new elements to reduce the statistical noise, if needed). Of course, since in this case the ensemble is not generated from $\bPhi_{\sssrcal}$, an importance sampling reweighting has to be employed in order to evaluate the averages $\AvgschaR{\,\cdot\,}$. After computing the averages, the code symmetrizes the results with respect to the space group symmetries (including the lattice translation symmetries) and the index-permutation symmetry, following the approach described in Appendix~\ref{app:symmetries}.

In order to reduce the computational cost, the SSCHA code can also compute the free energy Hessian discarding the contribution coming from the higher-order terms of the geometric-series expansion in Eq.~\eqref{eq:Hessian}, i.e. discarding the terms coming from $\bDfourR$. In many cases this approximation is extremely good, but it must be checked case by case. Within this so called ``bubble" approximation, the free energy Hessian becomes 
\begin{equation}
\frac{1}{\sqrt{\bM}}
\singledot
\frac{\partial^2 F}{\partial \bRcal \partial \bRcal}
\singledot 
\frac{1}{\sqrt{\bM}}
\simeq
\bDR+\bDthreeR
\doubledot 
\bLambdaR[0]
\doubledot
\bDthreeR .
\label{eq:HessianBubble}
\end{equation}

Using Eq.~\eqref{eq:Hessian}, or its approximated expression Eq.~\eqref{eq:HessianBubble}, the SSCHA code can compute the Hessian of the free energy at any $\bRcal$. However, as said, its most significant usage is in $\rschaeq$, due to its relevance to characterize displacive second-order phase transitions. In this case,  Eq.~\eqref{eq:Hessian} can be written in a quite explanatory form. 
At the end of a full SSCHA minimization, the obtained $\rschaeq$ and $\phischaeq$ define the so-called SSCHA effective harmonic Hamiltonian 
\begin{equation}
\Hscha=K+\frac{1}{2}(\bR-\rschaeq)\cdot\phischaeq\cdot(\bR-\rschaeq)\,,
\label{eq:HSCHA}
\end{equation}
which replaces the conventional harmonic Hamiltonian to define non-interacting bosonic quasiparticles as a basis to describe the collective vibrational excitations in presence of strong anharmonic effects. In terms of the dynamical matrix  \columnlabel{DSdef}{$\DS_{ab}=(\Phi_{\eq})_{ab}/\sqrt{M_aM_b}$} of the SSCHA Hamiltonian $\Hscha$, the anharmonic generalization of the dynamical matrix $\bDF$ can be written as
\begin{equation}
\bDF=
\bDS+
\bPi(0)\,,
\label{eq:def_DF}
\end{equation}
where 
\begin{equation}
\bPi(0)=\bDthreeeq 
\doubledot 
\bLambdaeq[0]
\doubledot
\left[\mathds{1}-\bDfoureq\doubledot\bLambdaeq[0]\right]^{-1}
\doubledot
\bDthreeeq
\label{eq:static_SE}
\end{equation}
is the static SSCHA self-energy (the reason behind the use of this name will be clear 
in the Sec.~\ref{sec:method:sigma}). In particular, in the bubble approximation we have
\begin{align}
&\bDF=\bDS+\bPiB(0)\,,
\label{eq:static_bubble_DF}
\intertext{where}
&\bPiB(0)=\bDthreeeq\doubledot\bLambdaeq[0]\doubledot\bDthreeeq
\label{eq:static_bubble_SE}
\end{align}
is the so called ``bubble'' static self-energy.

In conclusion, after the SSCHA minimization, the code allows to compute the high-order SSCHA force constants, Eqs.~\eqref{eq:ho_fc}, and  
the free energy Hessian dynamical matrix
\begin{subequations}
\begin{empheq}[left={\bDF(\bq)=\bDS(\bq)+\empheqlbrace}]{align}
&\bPi(\bq,0)\label{eq:DFq}\\
&\bPiB(\bq,0)\label{eq:DFqB}
\end{empheq}
\label{eq:DSplusPi}
\end{subequations}
(depending on whether the full or only the ``bubble'' static self-energy is computed) on the $\bq$-points belonging to the reciprocal space grid commensurate with the real space supercell used to generate the ensemble. Here we are explicitly using the reciprocal-space formalism, i.e.
we are Fourier transforming the quantities with respect to the lattice vector indices (see Appendix \ref{sec:app_rec_space} for more details).
From the softening of the eigenvalues of $\bDF(\bq)$ as a function of external parameters (like temperature or pressure), it is possbile to observe the occurrence of second order displacive phase transitions, characterize the distortion patterns and compute the critical value of the external parameters driving it. Examples of the employ of this method are given for H${}_3$S in Fig.~3 of Ref.~\cite{Bianco2018}, with the softening of an optical mode driven by pressure release, and for SnSe in Fig.~2 of Ref.~\cite{Aseginolaza2019Phonon}, with the softening of the distortion mode obtained by decreasing the temperature.

\subsection{Static bubble self-energy calculation: improved free energy Hessian calculation}

The SSCHA code also allows to compute the free energy Hessian dynamical matrix $\bDF(\bq)$ on any reciprocal space $\bq$-point, allowing to analyze the structural instabilities incommensurate with the used supercell. After the free energy evaluation and the subsequent free energy Hessian calculation, the real-space $\bDS(\bl_1,\bl_2)$ and 
$\bDthree_{\eq}(\bl_1,\bl_2,\bl_3)$ are available. Here, $\DS_{ab}(\bl_1,\bl_2)$ is the real space $\bDS$ matrix in which we made explicit the dependence of the lattice vectors $\bl_1$ and $\bl_2$ that identify the unit cells in which atom $a$ and $b$ are located, respectively.  Using them, the code allows to compute the static bubble $\bPiB(\bq,0)$ in any $\bq$-point, through the formula 
\begin{align}
\PiB{}_{\mu\nu}(\bq,0)&=
\frac{1}{N_{\bk}}
\sum_{\substack{\bk_1\bk_2 \\\rho_1\rho_2}}
\sum_{\bG}\,
\delta_{\bG,\bq+\bk_1+\bk_2}
\mathscr{F}(0,\omega_{\rho_1}(\bk_1),\omega_{\rho_2}(\bk_2)) \nonumber \\
&\times 
{\Dthree}_{\mu{\rho_1}{\rho_2}}(-\bq,-\bk_1,-\bk_2)\,  
{\Dthree}_{{\rho_1}{\rho_2}\nu}(\bk_1,\bk_2,\bq)\ .
\label{eq:static_bubble_SE_formula}
\end{align}
This equation is  Eq.~\eqref{eq:static_bubble_SE} written in reciprocal space and SSCHA normal mode components, i.e. in components of the $\bDS(\bq)$'s eigenvector basis. In Eq. \eqref{eq:static_bubble_SE_formula}
 the $\bk_i$ sums are performed on a  Brillouin zone ($BZ$) mesh of $N_{\bk}$ points; ${\Dthree}_{\mu{\rho_1}{\rho_2}}(\bk_1,\bk_2,\bq)$ are the SSCHA normal components of $\bDthree_{\eq}(\bk_1,\bk_2,\bq)$, the Fourier transform of $\bDthree_{\eq}(\bl_1,\bl_2,\bl_3)$  in $(\bk_1,\bk_2,\bq)$; $\omega_{\rho}(\bk)$ are the frequencies of $\bDS(\bk)$;  the function $\mathscr{F}$ is defined by Eq.~\eqref{eq:def_F0}; $\bG$ are reciprocal lattice vectors; and $\delta_{\bG,\bq+\bk_1+\bk_2}$ preserves crystal momentum. In this formula the $\bq$ and the $\bk_i$'s are not confined to the grid commensurate with the supercell used in the SSCHA minimization, as long as the ranges of the real space $\bDS(\bl_1,\bl_2)$ and $\bDthree(\bl_1,\bl_2,\bl_3)$ are smaller than the supercell size, so as to be legitimately Fourier interpolated on any reciprocal space points (more about the Fourier interpolation in appendix~\ref{sec:app_rec_space_and_FT}).
This allows to obtain two results at once. First, the $\bk$-mesh in  Eq.~\eqref{eq:static_bubble_SE_formula} can be arbitrarly increased up to  convergence, so as to reach the thermodynamic limit in the evaluation of the bubble static self-energy. Second, from $\bPiB(\bq,0)$ and $\bDS(\bq)$, through Eq.~\eqref{eq:DFqB} the code allows to compute the free energy Hessian dynamical matrix $\bDF(\bq)$ (useful to detect and characterize the system instabilities) in $\bq$ points not necessarily commensurate with the supercell (at variance with what is obtained with the simple Hessian calculation). This can be used, for instance, to study incommensurate second-order displacive phase transitions. In particular, this is the correct way to compute the frequencies {\columnlabel{staticfreq}$\Omega_\mu(\bq)$} along a reciprocal-space path, where $\Omega^2_\mu(\bq)$ are the eigenvalues of $\bDF(\bq)$. This is, for example, the procedure followed to compute
the (static) SCHA phonon dispersions of NbS${}_2$ shown in Figs.~2,~3 of Ref.~\cite{doi:10.1021/acs.nanolett.9b00504}, and to compute the interpolation-based convergence analysis shown in Fig.~3 of Ref.~\cite{doi:10.1021/acs.nanolett.0c00597} for TiSe${}_2$ monolayer.

\subsection{Dynamic bubble self-energy calculation: spectral functions, phonon linewidth and shift}
\label{sec:method:sigma}

The anharmonic generalization of the harmonic dynamical matrix described in the previous sections is the starting point to build a quantum anharmonic ionic dynamical theory. As  shown  in  Refs.~\cite{PhysRevB.96.014111,Bianco2018},  in  the  context  of  the  SSCHA it is possible  to  formulate  an \textit{ansatz} to  give  the 
expression  of the  one-phonon  Green function $\bG(z)$ for the
variable $\sqrt{M_a}(R^a-\Rcaleq^a)$. This \emph{ansatz} has been rigorously proved within the Time Dependent Self-Consistent Harmonic Approximation (TD-SCHA)\cite{monacelli2020time,lihm2020gaussian}.
In this dynamical theory
\begin{equation}
\bG^{-1}(z)=z^2\mathds{1}-\left(\bDscha+\bPi(z)\right)\,,
\label{Eq:Gm1}
\end{equation}
where $\bDscha$ is the dynamical matrix of the SSCHA effective harmonic Hamiltonian $\Hscha$, and $\bPi(z)$ is the SSCHA self-energy, in general given by 
\begin{equation}
\bPi(z)={\bDthreeeq}\doubledot\bLambdaeq(z)\doubledot
\left[\mathds{1}-{\bDfoureq}\doubledot\bLambdaeq(z)\right]^{-1}\doubledot\,{\bDthreeeq}\,,
\label{Eq:app_Pi}
\end{equation}
and in the bubble approximation by
\begin{equation}
\bPiB(z)={\bDthreeeq}\doubledot\bLambdaeq(z)\doubledot{\bDthreeeq}\,.
\label{Eq:app_Pi_B}
\end{equation}
In the equations above we use the ``eq'' subindex to specify that the eigenvalues and eigenfunctions entering the equations are obtained from $\phischaeq$ with the centroid positions at $\bRcaleq$.

With the Green function we obtain the spectral function
$\sigma(\Omega)=- 2\,\Im\mathrm{Tr}\left[\bG(\Omega + i0^+)\right]$, which provides the information that can be obtained with inelastic scattering experiments.  Taking explicitly into account the lattice translational symmetry (i.e. Fourier transforming the quantities with respect to the lattice vector indices) we can write
\begin{align}
\sigma(\bq,\Omega)&=-\frac{\Omega}{\pi}\,\Im\mathrm{Tr}\left[\bG(\bq,\Omega + i0^+)\right]
\label{eq:spfuq}
\end{align}
or
\begin{align}
\sigma(\bq,\Omega)=-\frac{\Omega}{\pi}\Im&\mathrm{Tr}\bigg[\bigl(\Omega+i0^+\bigr)^2\mathds{1}  + \nonumber \\
& -\bigl(\bDS(\bq)+\bPi(\bq,\Omega + i0^+)\bigr)\bigg]^{-1}\,,
\label{eq:def_spec}
\end{align}
where the multiplicative factor $\Omega/2\pi$ has been included to have, for each $\bq$, a function that integrated on the real axis gives the total number of modes $3\Natu$
($\Natu$ is the number of atoms in the unit cell, that may be different from the total number of atoms in the supercell $\Nat$). 

In the so-called ``static approximation'',  we replace the full self-energy $\bPi(z)$ with the  static self-energy 
$\bPi(0)$, where $z$ is blocked at zero. In this case the spectral function is 
\begin{align}
\overset{\static}{\sigma}(\bq,\Omega)
&=-\frac{\Omega}{\pi}\Im\mathrm{Tr}\bigg[\bigl(\Omega+i0^+\bigr)^2\mathds{1}  -\bigl(\bDS(\bq)+\bPi(\bq,0)\bigr)\bigg]^{-1} \nonumber \\
&=-\frac{\Omega}{\pi}\Im\mathrm{Tr}\left[\bigl(\Omega+i0^+\bigr)^2\mathds{1}-\bDF(\bq)\right]^{-1},
\label{eq:spec_static_tot_b}
\end{align}
where in the last line we have used Eq.~\eqref{eq:DSplusPi}. Therefore,
\begin{equation}
\overset{\static}{\sigma}(\bq,\Omega)=\sum_{\mu}
\overset{\static}{\sigma}{}_{\mkern-10mu\mu}(\bq,\Omega)\,
\label{eq:spec_static_tot}
\end{equation}
with
\begin{equation}
\overset{\static}{\sigma}{}_{\mkern-10mu\mu}(\bq,\Omega)=\frac{1}{2}
\left[\delta(\Omega-\Omega_\mu(\bq))
+\delta(\Omega+\Omega_\mu(\bq))
\right]\,,
 \label{eq:spec_static_analy}
\end{equation}
where $\Omega^2_\mu(\bq)$ are the eigenvalues of the free energy Hessian matrix $\bDF(\bq)$. In other words, the spectral function in the static limit is formed with delta peaks at the eigenvalues of $\bDF(\bq)$.

In the current version, the SSCHA code computes the full dynamical SSCHA self-energy ($z\neq 0$) only in the bubble approximation with the equation (see Eqs.~\eqref{eq:static_bubble_SE},~\eqref{eq:static_bubble_SE_formula}, and~\eqref{Eq:app_Pi_B}))
\begin{align}
\PiB{}_{\mu\nu}(\bq,\Omega+i\delta_{\se})=
&\frac{1}{N_c}
\sum_{\substack{\bk_1\bk_2 \\\rho_1\rho_2}}
\sum_{\bG}\,
\delta_{\bG,\bq+\bk_1+\bk_2}\nonumber\\
&\mkern-50mu\times\,\mathscr{F}(\Omega+i\delta_{\se},\omega_{\rho_1}(\bk_1),\omega_{\rho_2}(\bk_2))\nonumber\\
&\mkern-50mu\times\,{\Dthree}_{\mu{\rho_1}{\rho_2}}(-\bq,-\bk_1,-\bk_2)\,\,  
{\Dthree}_{{\rho_1}{\rho_2}\nu}(\bk_1,\bk_2,\bq)\,,
\label{eq:dyn_slfrg}
\end{align} 
where the summation $\bk$-grid can be arbitrarily fine as long as the interpolation
of the third-order SSCHA FCs can be performed (as in the static case Eq.~\eqref{eq:static_bubble_SE_formula}), and $\delta_{\se}$ is an arbitrary small, but finite, positive smearing value used to obtain converged results in the computation. In fact, the exact result corresponds 
to the limiting value obtained with an infinite $\bk$-grid and a zero $\delta_{\se}$ smearing.  In actual, finite-time calculations, the converged value of the dynamic self-energy is therefore estimated in this way. For a given summation $\bk$-grid, the corresponding self-energy converged value is estimated by analyzing the result given by Eq.~\eqref{eq:dyn_slfrg} for smaller and smaller $\delta_{\se}$ values (for the used $\bk$-grid, there will be a minimum value of $\delta_{\se}$ under which the result shows numerical instability). This analysis is performed with finer and finer summation grids until the converged value in the thermodynamic limit is obtained.
In principle, a dedicated convergence study of this kind has to be performed for all the specific observables of interest.

With  the  dynamical  SSCHA  bubble  self-energy, the code allows to compute the spectral function by the equation (see Eq.~\eqref{eq:def_spec})
\begin{align}
\sigma(\bq,\Omega)=-\frac{\Omega}{\pi}\Im& \mathrm{Tr}
\bigg[\bigl(\Omega+i\delta_{\id}\bigr)^2\mathds{1}+ \nonumber \\ 
& -\bigl(\bDS(\bq)+\bPiB(\bq,\Omega+i\delta_{\se})\bigr)\bigg]^{-1}\,,
\label{eq:spec_bub}
\end{align}
with $\delta_{\id}$ another arbitrary small, but finite, positive smearing value. The role of $\delta_{\id}$ is significant when the imaginary part of the self-energy is small. A prominent example where this happens is when the spectral function is calculated in the static approximation, i.e. when the bubble self-energy is kept fixed at the static value $\PiB{}_{\mu\nu}(\bq,0)$ (see Eq. \eqref{eq:spec_static_tot_b}).
Indeed, in this case the self-energy is real (Hermitian) and the computed spectral function becomes
\begin{align}
\overset{\static}{\sigma}(\bq,\Omega) =\sum_\mu \frac{1}{2} & \Im
 \bigg[\frac{1}{\pi}\frac{1}{\Omega-\Omega_\mu(\bq)+i\delta_{\id}} + \nonumber \\
 & +\frac{1}{\pi}\frac{1}{\Omega+\Omega_\mu(\bq)+i\delta_{\id}}\bigg]\,,
\label{eq:spec_static_comp}
\end{align}
where $\Omega^2_\mu(\bq)$ are the eigenvalues of $\bDF(\bq)$ in the bubble approximation. Therefore, for the numerical computation of the static spectral function, a finite $\delta_{\id}$ value is necessary to recover the analytical result, Eq.~\eqref{eq:spec_static_analy},  but with smeared Dirac delta functions. Actually, this is not just an extreme example, since the code really gives the opportunity to compute the spectral function in the static approximation, replacing
in Eq.~\eqref{eq:spec_bub} the full bubble self-energy with its static value computed through~Eq.~\eqref{eq:static_bubble_SE_formula}. This can be used to double-check that, as expected from Eqs.~\eqref{eq:spec_static_analy} and~\eqref{eq:spec_static_comp}, the obtained spectral function is given by spikes around the frequencies of the Hessian free energy matrix $\bDF$ (computed in the bubble approximation). However, the role played by $\delta_{\id}$ is not as critical as $\delta_{\se}$ since it is not typically system-dependent and it does not require a convergence study: in the code its default value is automatically set depending on the spacing of the energy $\Omega$-grid used to compute the spectral function.

Given a $\bq$,  the calculation of the full spectral function $\sigma(\bq,\Omega)$ through Eq.~\eqref{eq:spec_bub} turns out to be quite a heavy task due to the inversion of a different $3\Natu\times3\Natu$ matrix for each $\Omega$ value. 
The code also allows to employ a much less computational demanding approach by discarding the off-diagonal elements of the computed dynamical self-energy in the SSCHA normal modes components (i.e. the components in the $\bDS(\bq)$'s eigenvector basis). Within this ``no mode-mixing" approximation, which usually proves to be extremely good, the SSCHA modes keep their individuality even after the renormalization due to anharmonic effects. Indeed in this case, as in the static approximation, Eq.~\eqref{eq:spec_static_tot}, the total spectral function is given by the superposition of individual mode spectral functions:
\begin{equation}
\sigma(\bq,\Omega)=\sum_{\mu}\sigma_{\mu}(\bq,\Omega),
\label{eq:mu_spec_0}
\end{equation}
where now the $(\bq,\mu)$-mode spectral function $\sigma_{\mu}(\bq,\Omega)$ is computed with 
\begin{align}
\sigma_{\mu}(\bq,\Omega)  =
&\frac{1}{2}
\bigg[\frac{1}{\pi}\frac{-\Im\Zcal_{\mu}(\bq,\Omega)}{[\Omega-\Re\Zcal_{\mu}(\bq,\Omega)]^2+[\Im\Zcal_{\mu}(\bq,\Omega)]^2}
\vphantom{\frac{\biggl(\biggr)}{\biggl(\biggr)}}+ \nonumber \\
& + 
\frac{1}{\pi}\frac{\Im\Zcal_{\mu}(\bq,\Omega)}{[\Omega+\Re\Zcal_{\mu}(\bq,\Omega)]^2+[\Im\Zcal_{\mu}(\bq,\Omega)]^2}\bigg]\,
\label{eq:mu_spec}
\end{align}
and
\begin{equation}
\Zcal_{\mu}(\bq,\Omega)=\sqrt{\omega^2_{\mu}(\bq)+\Pi_{\mu\mu}(\bq,\Omega+i\delta_{\se})}\,.
\label{eq:mu_spec_z}
\end{equation}
Therefore, computing the spectral function in the ``no mode-mixing" approximation, by measuring the deviation of $\sigma_{\mu}(\bq,\Omega)$ from a Dirac delta function around $\Omega_\mu(\bq)$, it is possible to asses the impact that anharmonicity has on the different SSCHA modes $(\bq,\mu)$, separately.

The form of the $(\bq,\mu)$-mode spectral function $\sigma_{\mu}(\bq,\Omega)$ in Eq.~\eqref{eq:mu_spec} resembles a Lorentzian, but with frequency-dependent center and width, meaning that the actual form of the spectral function $\sigma(\bq,\Omega)$ can be quite different from the superposition of true Lorentzian functions. However, in some cases the $\sigma_{\mu}(\bq,\Omega)$ can be expressed with good approximation as a true Lorentzian with a certain half width at half maximum (HWHM) $\Gamma_{\mu}(\bq)$ and center $\bbOmega_{\mu}(\bq)$,
\begin{align}
\sigma_{\mu}(\bq,\Omega)&=
\frac{1}{2}
\left[\frac{1}{\pi}\frac{\Gamma_{\mu}(\bq)}{[\Omega-\bbOmega_\mu(\bq)]^2+[\Gamma_{\mu}(\bq)]^2}\right.\nonumber \\
&\vphantom{\frac{\biggl(\biggr)}{\biggl(\biggr)}}+
\left.\frac{1}{\pi}\frac{\Gamma_{\mu}(\bq)}{[\Omega+\bbOmega_\mu(\bq)]^2+[\Gamma_{\mu}(\bq)]^2}\right],
\label{eq:mu_spec_lor}
\end{align}
meaning that the quasiparticle picture is still valid, even after the inclusion of anharmonicity, with the $(\mu,\bq)$ 
quasiparticle having frequency (energy) $\bbOmega_{\mu}(\bq)$  and lifetime $\tau_{\mu}(\bq)=1/(2\Gamma_{\mu}(\bq))$. The difference 
between the renormalized and the ``bare'' SSCHA frequency, 
$\Delta_{\mu}(\bq)=\bbOmega_{\mu}(\bq)-\omega_{\mu}(\bq)$,
is called the frequency
shift of the $(\mu,\bq)$ mode. 

The SSCHA code offers several tools to perform such a ``Lorentzian analysis''. In general, the best Lorentzian approximation is obtained with
\begin{align}
&\vphantom{\frac{1}{2}}
\bbOmega_{\mu}(\bq)=\Re\Zcal_{\mu}(\bq,\bbOmega_\mu(\bq))
\label{eq:CENTER}\\
&\Gamma_{\mu}(\bq)=-\Im\Zcal_{\mu}(\bq,\bbOmega_\mu(\bq)).
\label{eq:HWHM}
\end{align}
Once the dynamical self-energy and the $\Zcal_{\mu}(\bq,\Omega)$ are computed, the SSCHA code allows to compute the single-mode spectral functions in the Lorentzian approximation, estimating the frequency $\bbOmega_{\mu}(\bq)$ and HWHMs $\Gamma_{\mu}(\bq)$ in different ways. One, optional, possibility is to solve self-consistently Eq.~\eqref{eq:CENTER} to estimate $\bbOmega_{\mu}(\bq)$, and then $\Gamma_{\mu}(\bq)$, through Eq.~\eqref{eq:HWHM}. However, by default, the ``one-shot'' approximation is employed with
\begin{align}
&\vphantom{\frac{1}{2}}
\overset{\OS}{\bbOmega}{}_{\mkern-4mu\mu}(\bq)=\Re\Zcal_{\mu}(\bq,\omega_\mu(\bq))
\label{eq:CENTER_os}\\
&\overset{\OS}{\Gamma}_{\mkern-8mu\mu}(\bq)=-\Im\Zcal_{\mu}(\bq,\omega_\mu(\bq))\,.
\label{eq:HWHM_os}
\end{align}
If the SSCHA self-energy $\bPi$ is a (small) perturbation on the SSCHA free propagator
(not  meaning  that  we are  in  a  perturbative  regime  with  respect  to  the  harmonic approximation),
then perturbation theory can be employed to evaluate the spectral function. If we keep the first order in the self-consistent equations \eqname~\eqref{eq:HWHM_os}, we get:
\begin{align}
&\vphantom{\frac{1}{2}}
\overset{\pert}{\bbOmega}{}_{\mkern-11mu\mu}(\bq)=\frac{1}{2\omega_{\mu}(\bq)}\Re\Pi_{\mu\mu}(\bq,\omega_\mu(\bq))
\label{eq:SHIFT_pert}\\
&\overset{\pert}{\Gamma}{}_{\mkern-12mu\mu}(\bq)=-\frac{1}{2\omega_{\mu}(\bq)}\Im\Pi_{\mu\mu}(\bq,\omega_\mu(\bq)+i\delta_{\se})\,.
\label{eq:HWHM_pert}
\end{align}
This perturbative approach is also employed by the SSCHA code to evaluate the quasiparticles' energies and lifetimes.
Examples of spectral function calculations done with Eq.~\eqref{eq:spec_bub}, Eqs.~\eqref{eq:mu_spec_0}-~\eqref{eq:mu_spec_z}
and Eq.~\eqref{eq:mu_spec_lor} can be found in Fig.~4 of Ref.~\cite{Bianco2018}. In Fig.~5 of the same reference, the anaharmonic phonon frequencies and linewidths along a path, computed using the Lorenztian approximation through Eqs.~\eqref{eq:CENTER} and~\eqref{eq:HWHM}, are shown. The spectral function computed with~Eq.~\eqref{eq:mu_spec} along a path is shown with a colorpolot in Fig.~3 of Ref.~\cite{PhysRevB.97.014306} for PbTe, and in Fig.~4 of Ref.~\cite{Aseginolaza2019Phonon} for SnSe.

In conclusion, with the SSCHA code we can calculate three frequencies for a mode $(\bq,\mu)$: $\omega_{\mu}(\bq), \Omega_{\mu}(\bq)$, and $\bbOmega_{\mu}(\bq)$, which are the frequency of the SSCHA auxiliary boson, the frequency coming from the SSCHA free energy Hessian (i.e. from the static approximation), and the  frequency of the SSCHA quasiparticle in the Lorentzian approximation. Only the last one is a true physical quantity as it can be measured in experiments. However, the static $\Omega_{\mu}(\bq)$ is also a physical meaningful quantity, as its zero value corresponds to a structural instability driving a second-order phase transition along the pattern characterized by the mode $(\bq,\mu)$. The SSCHA provides a specific physical meaning of each of these frequencies, in contrast to other approaches used to estimate anharmonic phonons, where no distinction is usually done.

\section{The Python code}
\label{sec:python}

Two different Python libraries are provided with the SSCHA code: CellConstructor and Python-sscha.  The latter is the library that performs the SSCHA minimization itself, while the former is a library that deals with the dynamical matrix, the crystal structure, the symmetrization, and performs the calculation of phonon spectral functions and linewidths as a post-processing tool.

The SSCHA code allows to set up the calculations with a simple Python script. In the standard calculation, the script loads the starting dynamical matrices; sets up the ensemble and the parameters for the SSCHA run; performs the calculations of the Born-Oppenheimer energies, forces, and stress tensors on the configurations in the ensemble by calling to a external total-energy-force engine; and starts the minimization of the free energy. A simple input script that performs all these steps requires less than 20 lines. Examples are provided within the code, as well as step-by-step tutorials to perform a full SSCHA calculation starting just with the structure in a \emph{cif} file. Python scripting the SSCHA run makes it versatile, as it can be interfaced with other Python libraries to facilitate the analysis of the results. As an alternative, it is also possible to write an input file and run the SSCHA code as a stand-alone command-line program. 

\subsection{Code structure}

Most of the program is written in python with an object-oriented style.
The system status (density matrix) is described by a class defined in CellConstructor (\emph{Phonons}), that contains all the information about the system, including lattice parameters, atomic positions, and the auxiliary force-constant matrix (plus eventual extra data, as effective charges used for post-processing purposes). Methods of this class allow the user to impose symmetries on the system, constrain the auxiliary force to be positive definite (\eqname~\ref{eq:positive}), extract auxiliary phonon frequencies and polarization vectors, or interpolate them to other points in the Brillouin zone.

All the calculations related to the SSCHA averages are performed by the \emph{Ensemble} class (inside Python-sscha). This class generates and stores all the randomly displaced ionic configurations, and can submit or load the results of the energy, forces, and stress tensors calculations. It also computes the quantities related to averages on the ensemble, as the free energy, the gradients, the stress tensor, and the free energy Hessian.

Finally there are other classes, which employ the ensemble and perform the minimization of the free energy, take care of communicating with a remote cluster to run the calculation of forces and energies (see next section), and manage the post-processing to compute the spectral function (the full description of them is provided within the documentation of the code).

Most of the code is written in Python, however, the heaviest CPU-intensive calculation is written in Fortran and interfaced with python through the f2py utility provided by numpy\cite{Harris2020}. In particular, the calculation of the free energy gradient, the free energy hessian, the spectral functions, the interpolation, and the symmetrization are performed by a Fortran module compiled with the code. For this reason, in order to compile and use the code, a Fortran compiler as well as LAPACK and BLAS libraries are required.

\subsection{Parallelization}

The nature of the algorithm makes it very simple to exploit massive parallelization strategies available in high performance computing (HPC) facilities. In particular, the most expensive part of the code is the calculation of Born-Oppenheimer energies, forces, and stress tensors of the generated ionic configurations in each population (the red shaded cell in the code flowchart in \figurename~\ref{fig:flowchart}). Each of this calculation is independent from the others, so they can be trivially run in parallel on different computing nodes. This is a huge advantage with respect to other methods based on AIMD or PIMD, which mimic a time evolution of the system and thus require to calculate atomic forces on one configuration after the other. 

The SSCHA code does not include a particular engine for computing energies, forces and stresses, but relies on external software. For this reason, it is possible to exploit the efficient parallelization already implemented by the chosen software. For example, the widely used Quantum ESPRESSO package recently implemented also a hybrid parallelization that exploits together multi-threading (OpenMP), multiprocessing (MPI), and GPU (CUDA) parallelization\cite{Giannozzi2020}. In this way the SSCHA code stands on the shoulders of giants, exploiting the most efficient parallelization available today. 

All other steps of the code are generally computationally very cheap compared to the energy and force calculation, especially when an \emph{ab initio} approach is followed. The SSCHA minimization cannot exploit so well the possibilities offered by parallelization, since each step of the main cycle depends on the previous one. Most of the computations executed in the cycle are linear algebra calculations carried out with the numpy library\cite{Harris2020}, some of them speeded up with an explicit Fortran implementation. Thanks to the numpy implementation\cite{5725236}, if this library is correctly compiled, the linear algebra calculations will exploit multi-threading. For this reason, the best performances of the SSCHA are obtained by executing the \emph{ab initio} calculations on an HPC facility, while the SSCHA minimization on a commercial workstation in which the minimization can take few seconds.

Post-processing calculations, like the free energy Hessian and the phonon dynamical spectral functions, may be executed with additional Python scripts after the end of the SSCHA run. The calculation of the free energy Hessian has been parallelized with OpenMP (multi-threading), while the calculation of the spectral functions, which may require a dense $k$-point grid for the interpolation, exploits multiprocessor parallelization through MPI (both mpi4py and pypar can be used\cite{Dalcn2005,Dalcn2008,Dalcin2011}).

\subsection{Execution modes}

Since the best performances of the code are obtained by running it in different computers, we introduced three different execution modes: manual, automatic local, and automatic remote submission.

In the manual mode, the code stops after generating the ensemble and printing on files the structures of the randomly distributed ionic configurations. At this point the user must feed these structures to a total-energy engine, e.g. a DFT code, to calculate their Born-Oppenheimer energies, atomic forces, and stress tensors. The user should prepare later specific files with the output of these calculations. Then, the SSCHA code should be manually restarted; it reads the output of energies, atomic forces, and stress tensors and runs the minimization until the exit criteria is fulfilled. After, it is up to the user to decide whether to start a new population or not. In this sense, the manual mode does not require direct interaction between the SSCHA code and any other external software, and consequently this execution mode does not require the installation of the SSCHA code on an HPC facility.

In the local automatic mode that can be scripted in Python, the code has to be supplied with an interface to an external code that is able to compute atomic forces and energies. This can be done through the Atomic Simulation Environment (ASE) library\cite{ASE}, which already implements interfaces with most common \emph{ab initio} codes like Quantum ESPRESSO\cite{Giannozzi2009,Giannozzi2017}, VASP\cite{VASP}, SIESTA\cite{SIESTA}, CP2K\cite{CP2K}, and many more. Force-field codes like LAMMPS\cite{LAMMPS} may also be used. In this execution mode, the code will proceed automatically to perform the calculations locally, and the full flowchart in \figurename~\ref{fig:flowchart} is executed without requiring any direct interaction with the user. While this execution mode is very useful, as it does not waste human time to manually restart the code at each population, it requires the most expensive part of the code, the calculation of total energies and forces, to be executed on the same machine as the SSCHA algorithm. This has a drawback when running the whole process in an HPC facility: the overall cost in terms of hours and parallel resources that needs to be allocated for the ensemble computation could be very expensive, and the SSCHA code will not exploit this amount of resources during the minimization. For this reason, the automatic local mode is indicated only when the calculation of energies and forces is fast and the requested resources are not so expensive, for example when force fields are used, such as in the SnTe example provided below.

Lastly, a remote automatic mode is also implemented. In this case the software will submit the energy and force calculations into a server through a queue job manager, and retrieve the results when finished. 
This last mode is the most suited for standard calculations as it exploits the HPC parallelization when computing the total energies and forces of the configurations, but runs the SSCHA minimization on a local computer, which benefits from the high speed multi-thread processors of commercial workstations. Moreover, as in the manual mode, there is no need to install the SSCHA code on a HPC cluster. Thanks to the complete automatic workflow, the only effort required by the user is to setup the communication with the clusters, which is mostly system independent.

\subsection{Distribution}

The packages is distributed as a standard Python application, and can be installed with a setup.py script. Since parts of the code are written in Fortran and C, it requires the appropriate compilers with LAPACK and BLAS libraries to be installed. Part of the Fortran subroutines are modified versions of Quantum ESPRESSO subroutines from the PHonon package, especially those regarding the symmetries. Together with the github page, we also provide the stable release in the pip repository, to facilitate installation. A different setup.py script is provided to facilitate the installation of the package on clusters to fully exploit MPI parallelization for the post-processing. The code is documented with Sphinx. We release the package and the source code under the GPLv3 license.

\section{A model calculation on Tin Telluride}
\label{sec:model}

\begin{figure*}
    \centering
    \includegraphics[width=\textwidth]{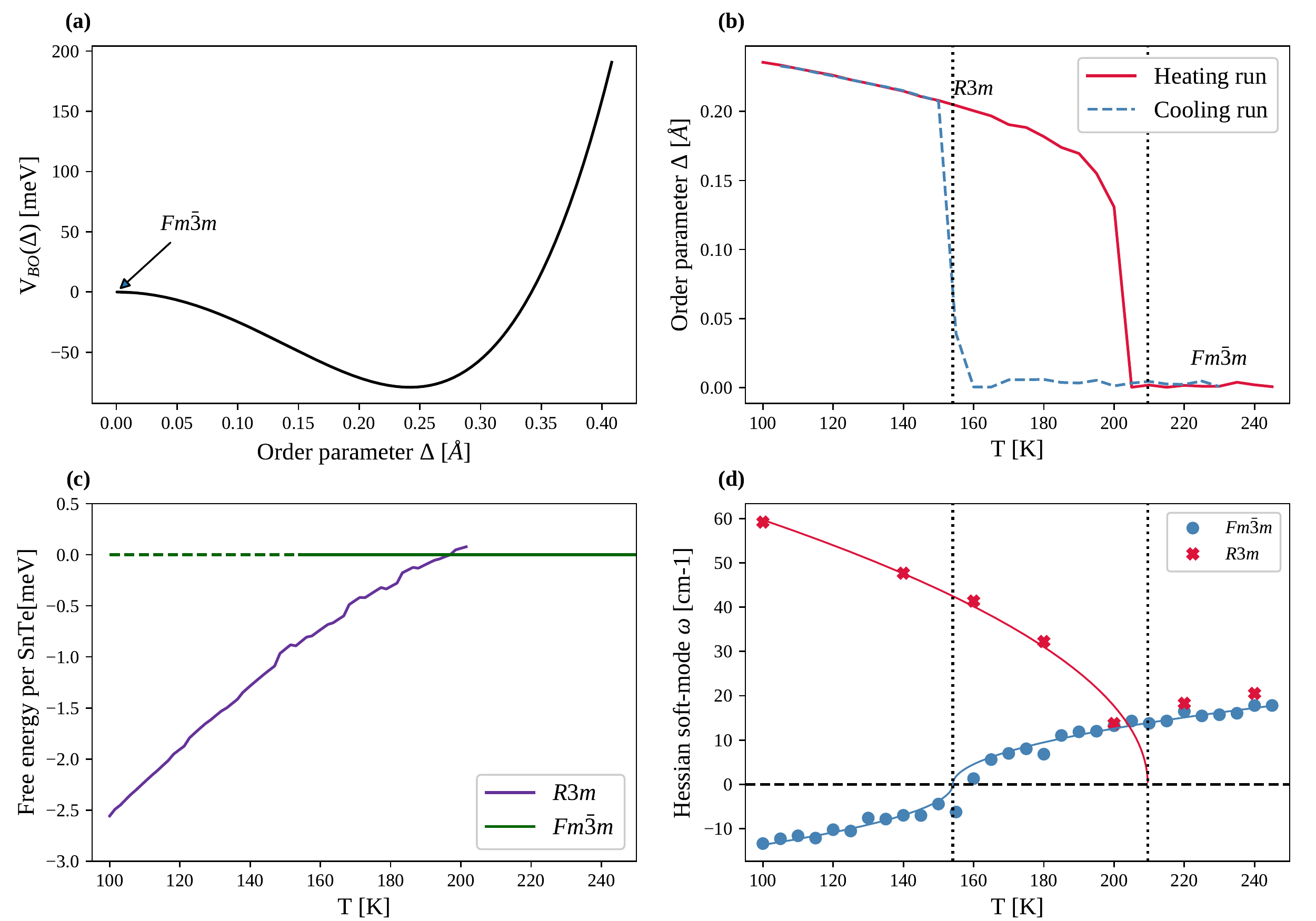}
    \caption{(a) Born-Oppenheimer energy as a function of the order parameter of the ferroelectric phase transition of \ch{SnTe} obtained with a model force field. (b) Hysteresis cycle between the ferroelectric phase $R3m$ and the cubic paraelectric phase $Fm\bar 3m$. In both heating and cooling we constrained the SSCHA simulation to the $R3m$ symmetry, which is a subgroup of $Fm\bar 3 m$. (c) Free energy of the two phases. The high-symmetry phase becomes more stable around \SI{200}{\kelvin}, slightly before the $R3m$ falls into the high-symmetry phase in the heating cycle. The dashed line indicates a dynamical instability. (d) The free energy curvature around the order parameter in the high-symmetry $Fm\bar 3m$ phase. Positive values mean (meta)stability; negative values indicate a dynamical instability. The transition occurs at $T = \SI{154}{\kelvin}$, and it coincides with the lower bound for the $Fm\bar 3 m$ in the hysteresis cycle.}
    \label{fig:SnTe:toy}
\end{figure*}

To display the potentiality of the code, we provide an example calculation on a \ch{SnTe} toy model force field, where the lattice has been artificially stretched to enhance the anharmonicity. More details on this force field can be found in Ref. \cite{PhysRevB.96.014111}. We provide it as a separate package under GPL license.

\ch{SnTe}, as other ferroelectric materials\cite{PhysRevLett.122.075901,PhysRevB.100.214307}, undergoes a displacive phase-transition, where a phonon mode at $\Gamma$ softens with temperature lowering and provokes a cell distortion from the high-temperature high-symmetry $Fm\bar 3 m$ phase to the low-temperature $R3m$ phase. The toy model is able to reproduce this behavior, although it does not pretend to accurately describe the real \ch{SnTe} transition and it is just provided as an artificial example.

\begin{figure*}
    \centering
    \includegraphics[width=\textwidth]{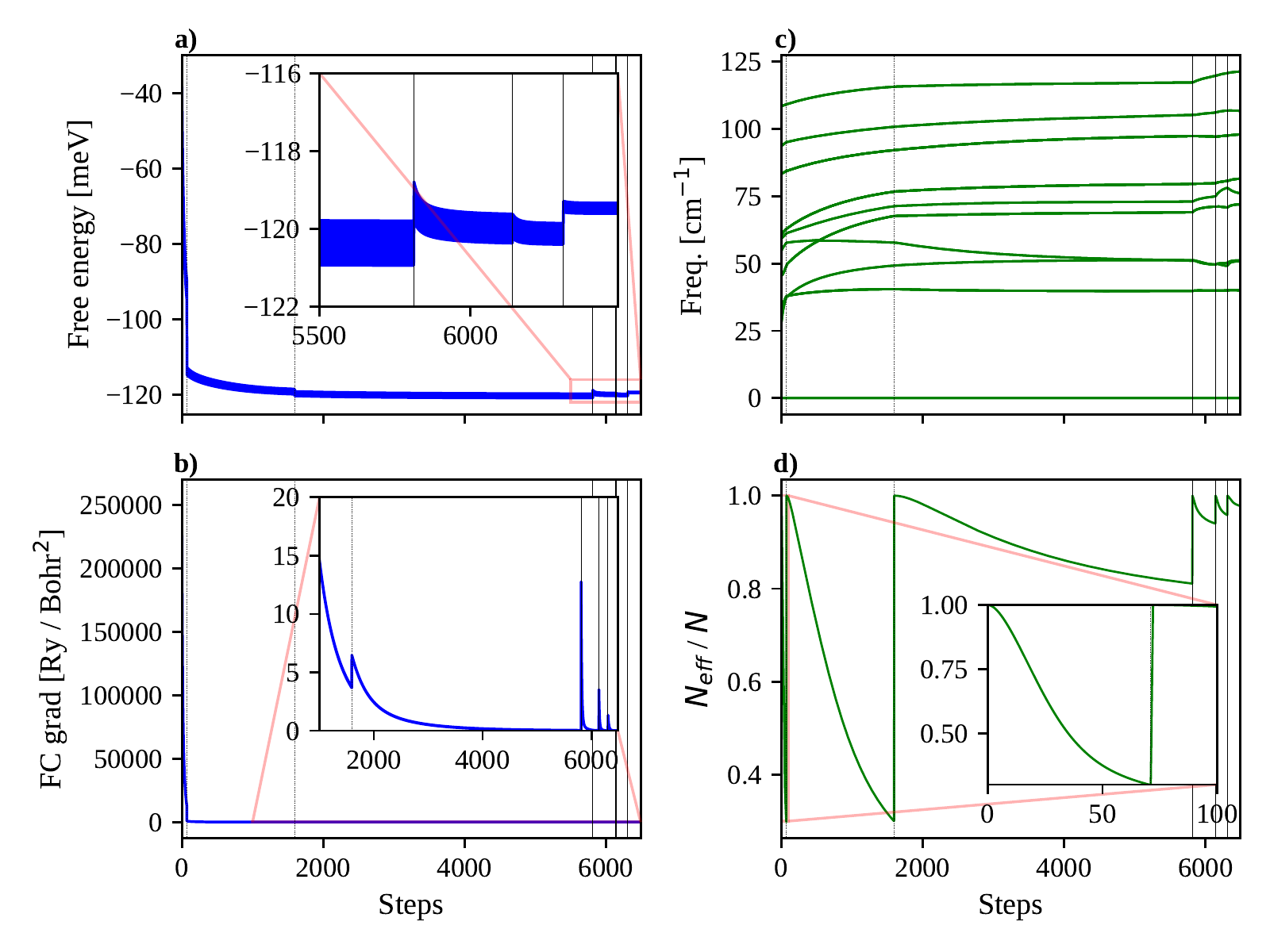}
    \caption{Convergence of the SSCHA minimization in  the \ch{SnTe} system in a 2$\times$2$\times$2 supercell at $T = \SI{250}{\kelvin}$. \textbf{a}: Evolution of the free energy per unit cell during the minimization. The width of the line denotes the stochastic error. \textbf{b}: Evolution of the modulus of the gradient of the auxiliary force constants. \textbf{c}: Evolution of the frequencies of the auxiliary force constants. \textbf{d}: The Kong-Liu effective sample size ratio (we used 0.4 as stochastic criterion for restarting) during the minimization.
    Vertical dotted lines indicate the new population after the simulation was out of the stochastic criterion. Vertical solid lines indicate a new population after the simulation converged by increasing the number of configurations to improve the accuracy. The simulation is performed starting with 50 configurations and increasing after successful convergence to 100 and 200. The overall number of total energy and forces calculations required to converge this example is 450. One last step is shown for demonstrative purposes, where we increased the number of configurations to 1000 to show how well the result converges already with few configurations (in particular the auxiliary frequencies of panel \textbf{c}).}
    \label{fig:minimization}
\end{figure*}

The system has an incipient ferroelectric instability, marked by the negative curvature of the energy in the high-symmetry position. This means that an optical vibrational mode has an imaginary frequency at $\Gamma$ within the harmonic approximation. The BO energy of the toy model as a function of the atomic displacements projected onto the eigenvecotrs of the imaginary mode (the order parameter $\Delta$) is reported in \figurename~\ref{fig:SnTe:toy}(a), where it is clear that the high-symmetry $Fm\bar 3 m$ is not at the minimum of $V(\bR)$. However, as extensively discussed above, the stability of a structure is determined by the temperature-dependent free energy,
\begin{equation}
F = E - TS,
\label{eq:free:energy:t}
\end{equation}
and not the BO potential. Notably, $E$ is not the energy profile reported in \figurename~\ref{fig:SnTe:toy}(a), as it also includes the vibrational contribution to the energy. For this reason, the energy profile (and the harmonic approximation) does not correctly describe even the behavior at $T = 0$, where there is no entropy contribution. Since entropy usually is higher in high-symmetric positions ($\Delta = 0$), the $Fm\bar 3m$ high symmetry phase will become progressively more stable as temperature increases.

In \figurename~\ref{fig:minimization} we show the evolution of the free energy, its gradient, and the frequencies of the auxiliary force constants during a typical SSCHA minimization at $T = \SI{250}{\kelvin}$ for this system. We start the minimization from the harmonic solution of the high-symmetry phase $Fm\bar 3 m$, which has imaginary frequencies. Since the system is strongly anharmonic, the starting solution is very far from the solution. To approach the minimum quickly and with low computational effort, we start the minimization with a small number of configurations (here 50). \figurename~\ref{fig:minimization}(\textbf{d}) reports the stochastic condition to stop the minimization (and extract a new ensemble), as defined in \eqname~\eqref{eq:cond:stochastic} (we chose $\eta = 0.4$). Here, we need only three ensembles to converge to the minimum, as a zero gradient is obtained with a reasonably large value of $\eta$. To improve the quality of the calculation (and decrease the stochastic error) we further run two more populations with 100 and 200 configurations, both converging in one population. 
Both the gradient and the free energy rapidly decrease and the result converges. An extra population with 1000 configurations is included to see that the result is converged. \figurename~\ref{fig:minimization}(\textbf{c}) presents the evolution of the auxiliary phonon frequencies associated to the auxiliary force constant matrix. The small change in these frequencies when the number of configurations is increased means that a small number of configurations is sufficient to have a good estimate of the auxiliary frequencies. Indeed, a good check for a well-converged result is to verify that these frequencies are stationary and do not change more in the minimization. The SSCHA code prints in output, if requested, this information at each run. We provide the code scripts that produce this kind of graphs from the raw data generated by the code, which facilitates the user to control if the minimization is working correctly.

In \figurename~\ref{fig:SnTe:toy}(b) we report the order parameter obtained at the end of the SSCHA minimizations at different temperatures. The starting structure at low temperatures is the low-symmetry $R3m$. When temperature is increased, the structure obtained at the previous lower temperature is used as input. At low temperatures the output structure reamins the $R3m$, with $\Delta \neq 0$, but at $T = \SI{205}{\kelvin}$, the low-symmetry phase jumps into the high-symmetry phase, marking a first-order phase transition. We can confirm it is a first-order phase transition as we can start cooling down from the high-symmetry phase (without constraining the new symmetries acquired) and the system remains stable up to $T=\SI{160}{\kelvin}$, when it transforms back to the low-symmetry phase. This is the hysteresis cycle of the material.

We can further analyze the thermodynamic properties. The SSCHA provides also the free energies of the two phases. We compare them in \figurename~\ref{fig:SnTe:toy}(c). As clearly shown, the low-symmetry phase is more stable up to \SI{200}{\kelvin}, so that the phase diagram in this model is formed by the $R3m$ phase below \SI{200}{\kelvin} and the $Fm\bar 3m$ above. We can also see whether the $Fm\bar 3 m$ becomes dynamically unstable by calculating the hessian matrix of the free energy. We plot the second derivative of the free energy with respect to the order parameter in \figurename~\ref{fig:SnTe:toy}(d). 
The free energy curvature becomes negative below $T = \SI{154}{\kelvin}$. This is a threshold below which the $Fm\bar 3m$ phase is no longer stable, and cannot exist or be observed. Consequently, it coincides with the lower bound of the hysteresis cycle. On the other side, looking at the free energy Hessian of the $R3m$ phase, we see that the frequency of the mode along the order parameter softens to zero at $T = \SI{210}{\kelvin}$, marking an upper bound to the stability of the low symmetry phase. Interestingly, while in the high symmetry phase the ``bubble approximation'' (\eqname~\ref{eq:HessianBubble}) is very accurate, the correct estimation of the free energy Hessian in the $R3m$ phase requires the full expression of the Hessian (\eqname~\eqref{eq:Hessian}).

This example shows that the SSCHA can fully characterize a complex first-order phase transition, and thanks to the possibility of exploiting symmetries, we can even study a phase that is dynamically unstable, i.e., the $Fm\bar 3m$ below the critical point. We can do simulations directly in the high-symmetry phase, with a considerable gain in the computational cost, and spot instabilities by the Hessian matrix calculation, as in \figurename~\ref{fig:SnTe:toy}(d).

\begin{figure}
    \centering
    \includegraphics[width=\columnwidth]{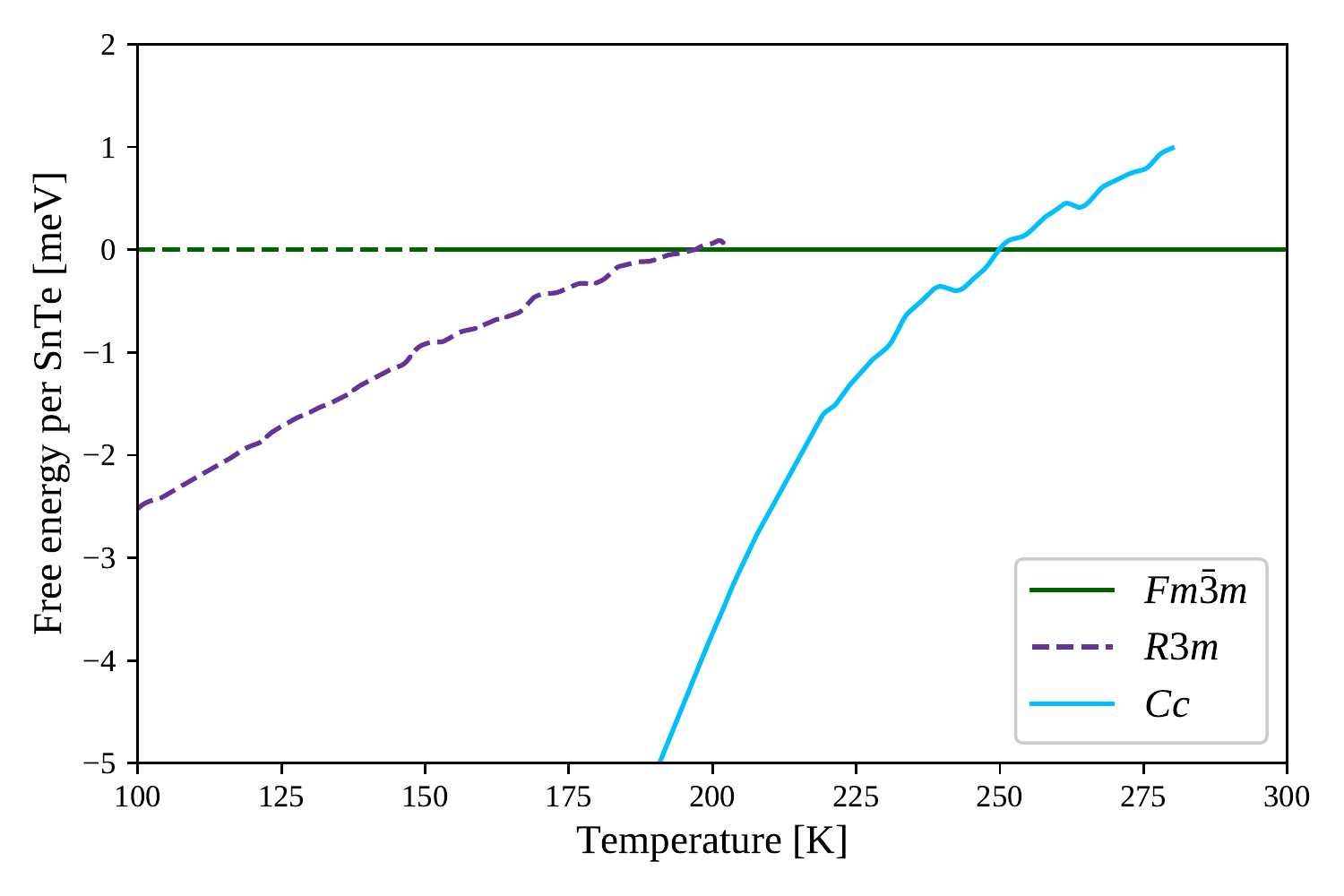}
    \caption{Full phase diagram of the SnTe toy model. In dashed lines we report the unstable phases (whose free energy Hessian has an imaginary mode).}
    \label{fig:PD}
\end{figure}

However, we can do even more: {\it finite temperature structure search}. To investigate whether the $R3m$ is the actual ground state within the toy model or a lower symmetry phase is energetically favored, we calculate the Hessian also in the $R3m$ phase. We find that in the whole region of the simulation, the $R3m$ phase is dynamically unstable and the system wants to break the symmetry once again. To find the real ground state, we release all the symmetry constrains in our simulation and perform a full relaxation with the SSCHA at $T = \SI{100}{\kelvin}$. We discovered a new phase of $Cc$ symmetry defined in a 1$\times$2$\times$1 supercell of the original cubic cell. In \figurename~\ref{fig:PD} we report the final phase diagram for the SnTe toy model. The new $Cc$ phase is found to be the ground state up to $\SI{250}{\kelvin}$, where the cubic $Fm\bar 3m$ becomes again energetically favorable. The $Cc$ continues to exists until $\SI{280}{\kelvin}$, where it transforms into the $Fm\bar 3 m$ phase.

We want to remark that the particular temperature, phase transitions, as well as the real existence of phase $Cc$ are just features of the toy model and do not pretend to represent the physics of this system. The real \ch{SnTe} has a ferroelectric $R3m$ ground state at low temperatures and the phase transition to the $Fm\bar 3m$ phase is of second-order type (the order parameter does not jump, and the free energy lines of the two phases touch when the $Fm\bar 3m$ mode becomes imaginary). This system has been already studied with the SSCHA in Ref. \cite{PhysRevB.97.014306} with \emph{ab initio} energies and forces. However, even if it is just a toy model, this example shows how the code can reach high-symmetry phases in an unsupervised way starting from the low-temperature structure.

For this reason, the SSCHA is also attractive for a structure search perspective: it is able to perform a search of saddle-point structures in the classical Born-Oppenheimer energy landscape that become the ground state due to ionic quantum and/or thermal fluctuations. This can be a great advantage to find saddle-point structures in complex systems with many atoms in the unit cell, such as in molecular crystals, where symmetry constrains may be inefficient\cite{Monserrat_2018}.

The other post-processing utility the code provides is the calculation of spectral functions and dynamical phonon spectra. We remark that the auxiliary phonons, i.e. the eigenvalues of $\bDscha$, are just an auxiliary quantity used to define the density matrix. For this reason, they only describe quantum flucuations around the centroid positions. The eigenvalues of the Hessian matrix $\bDF$, instead, are the response to a static external  perturbation,  and  describe  the  stability  of  the  structure  with  respect  to a spontaneous  symmetry breaking. Last, physical phonons, those observed by experimental probes like vibrational spectroscopy and inelastic scattering, must be computed from the dynamical interacting Green function. 
While all these definitions of phonon frequencies coincide in perfectly harmonic crystals, when anharmonicity is involved, they can differ significantly.
The SSCHA code offers a tool to easily compute the dynamical Green functions as a post-processing utility as discussed in Sec. \ref{sec:method:sigma}.

\begin{figure*}
    \centering
    \includegraphics[width=0.85\textwidth]{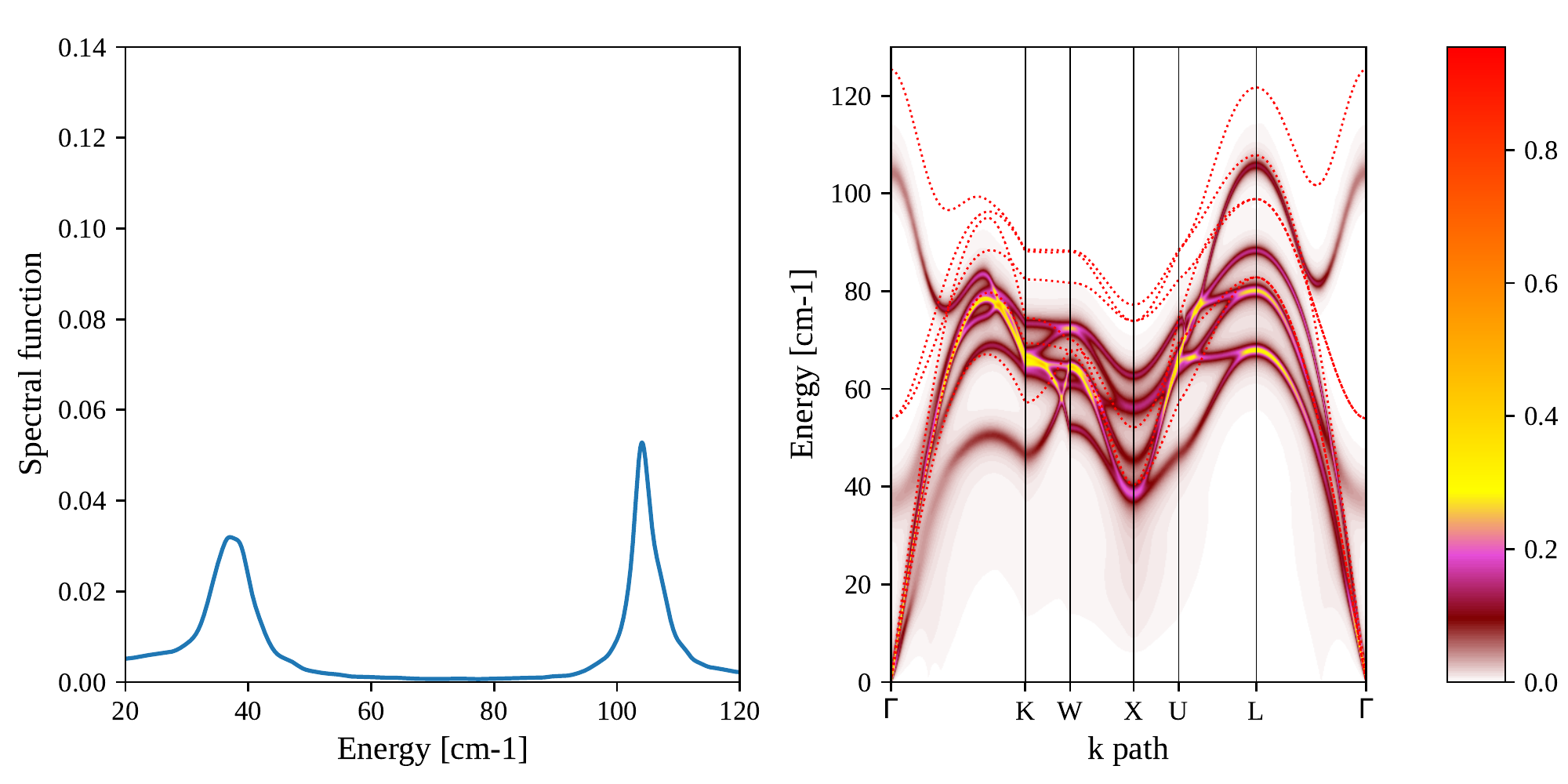}
    \caption{Phonon spectrum of \ch{SnTe} at $T = \SI{280}{\kelvin}$. Left panel: spectral function at $\Gamma$. The two main peaks are the LO-TO splitting. Right panel: the full spectral function along a path in the Brillouin zone. The red dashed line is the dispersion of the auxiliary phonons (eigenvalues of $\bDscha$).}
    \label{fig:SnTe:spectral}
\end{figure*}

\begin{figure}
    \centering
    \includegraphics[width=\columnwidth]{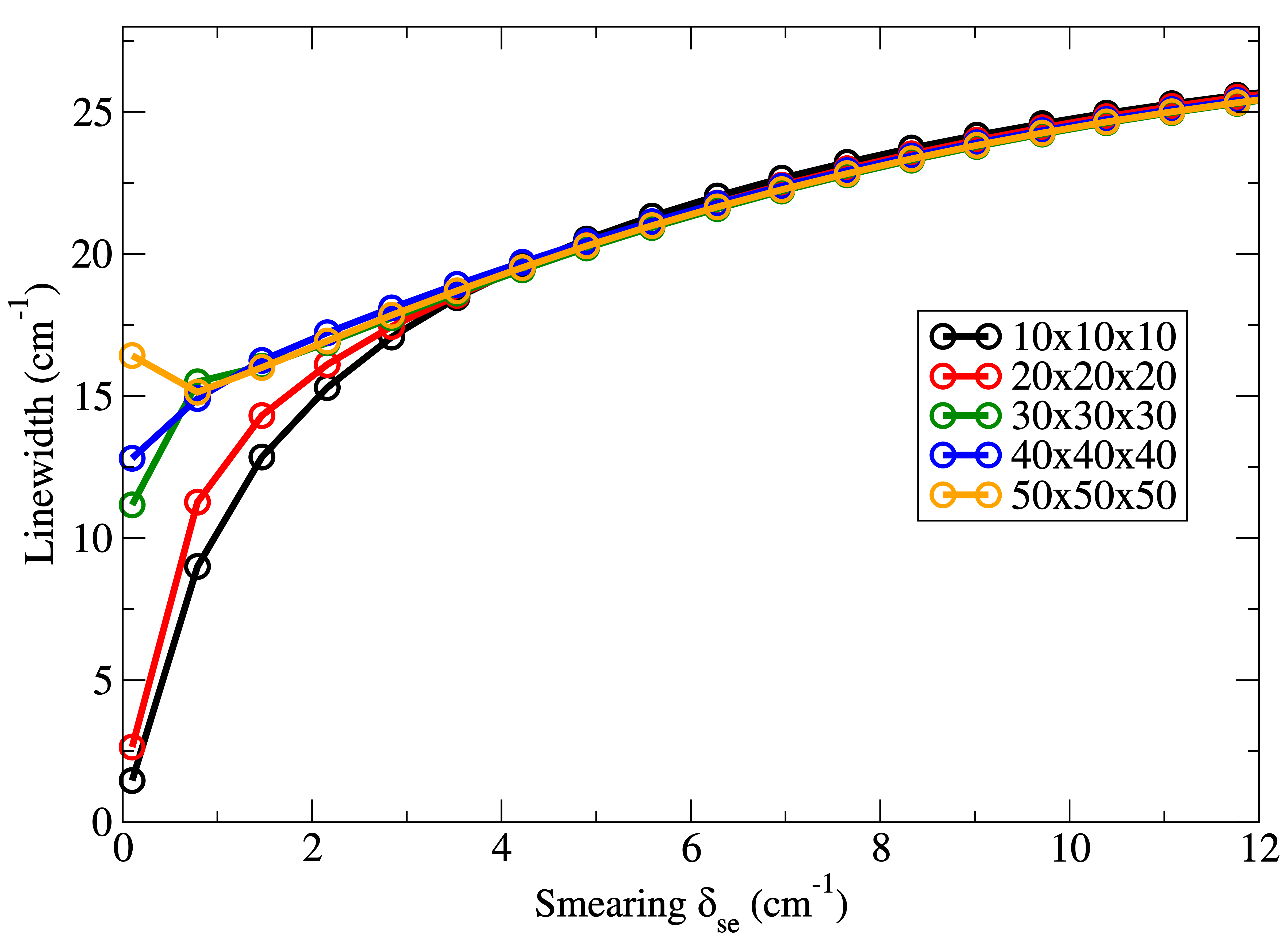}
    \caption{Convergence study of the linewidth (full width at half maximum) of the SnTe highest optical phonon frequency in $\Gamma$. For increasing size of the used \textbf{k}-mesh summation grid, the linewidth as a function of the smearing parameter $\delta_{\se}$ is studied (see Sec.~\ref{sec:method:sigma} for details about these quantities). The result shows that the converged value of the linewidth (15 cm${}^{-1}$) is obtained with a $30\times 30\times 30$ \textbf{k}-mesh summation grid, at least, and smearing $\delta_{\se}$ around 0.8 cm${}^{-1}$.}
    \label{fig:convlinewidth}
\end{figure}

In \figurename~\ref{fig:SnTe:spectral}, we plot the phonon spectrum, computed as the spectral function obtained from the dynamical Green functions. As anticipated, the peaks of the spectral function in \figurename~\ref{fig:SnTe:spectral} do not coincide with the dispersion obtained from the auxiliary dynamical matrix $\bDscha$, and show a rather anomalous behavior. It is worth mentioning that effective charges are considered in the calculation of the spectral functions. The effective charges are considered following the procedure outlined in Appendix \ref{sec:app_EC}. In Fig.~\ref{fig:convlinewidth} we illustrate the convergence for a phonon linewidth $2\Gamma_{\mu}(\bq)$ with respect to the $\delta_{\se}$ parameter and the \textbf{k}-mesh summation grid in Eq. \eqref{eq:dyn_slfrg}.

All the data of this simulation has been obtained in less than one hour, using a single processor on a laptop, proving the high-efficiency of the SSCHA package, which is beyond standard molecular dynamics software. We provide in the additional materials the Python scripts to run and analyze all the simulations here reported in this example.

\section{Applications of the SSCHA method}
\label{sec:examples}    

In order to illustrate some physical problems and materials that have already been efficiently tackled with the SSCHA, in this section we briefly overview some of the systems studied with this method. One should not consider that the applications are limited to these examples, the SSCHA provides a general utility to treat accurately and efficiently all materials where ionic vibrations play a crucial role both in the thermodynamic and transport properties.

\subsection{Hydrogen-based compounds}

\begin{figure*}
    \centering
    \includegraphics[width=0.75\textwidth]{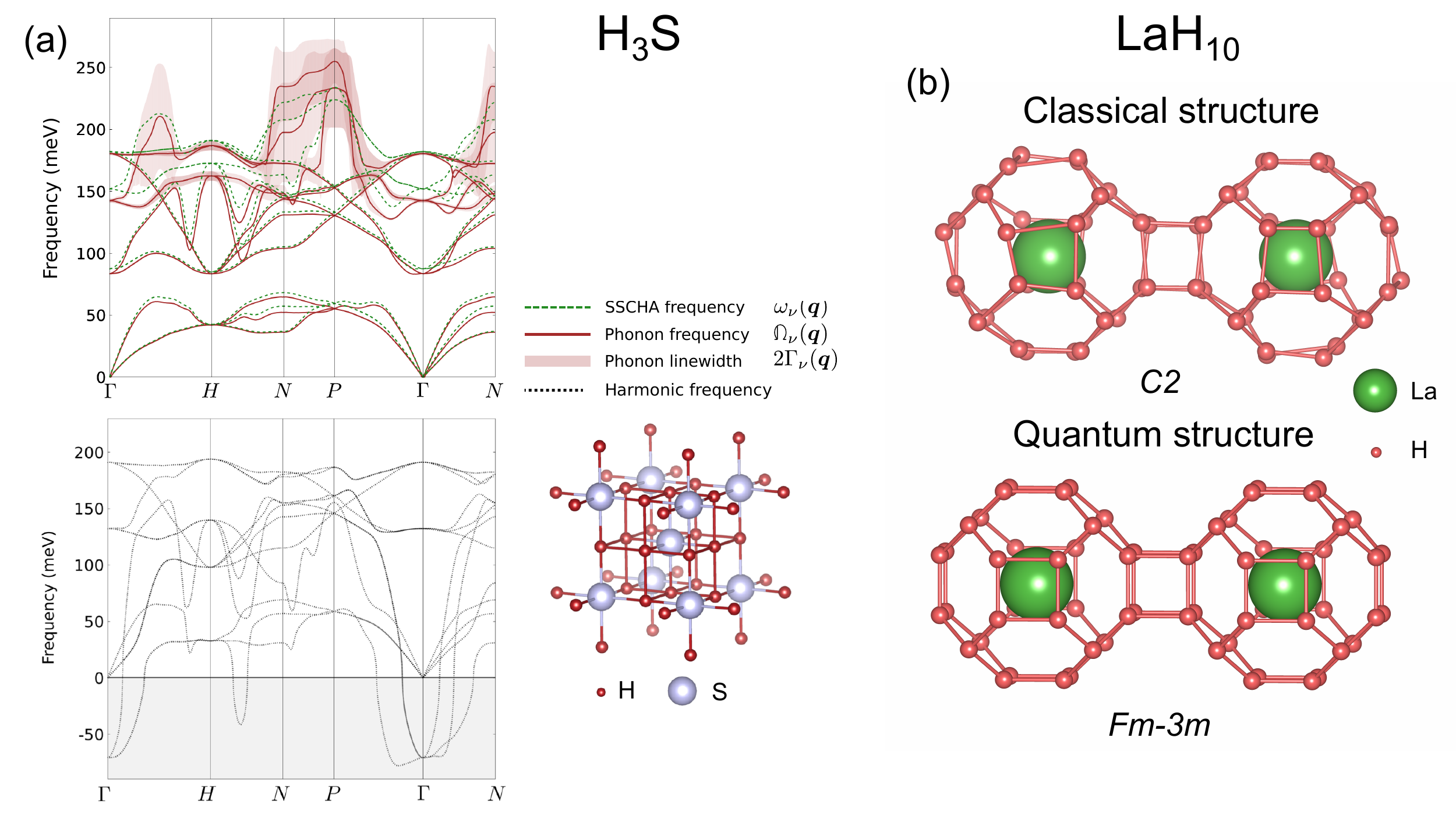}
    \caption{(a) Anharmonic phonon spectra obtained with the SSCHA for $H_3S$ at 158 GPa in the $Im\bar{3}m$ phase (top panel). The SSCHA auxiliary phonon frequencies are given, those obtained diagonalizing $\phischaeq$, together with the phonon frequencies obtained from the spectal function in the Lorentzian approximation. The linewidth obtained in the latter case is also given. The harmonic phonons are also shown (bottom panel). We also provide the structure of $Im\bar{3}m$ $H_3S$. Data taken from Ref. \cite{Bianco2018}. (b) Crystal structures for LaH$_{10}$ obtained from the minimum of the classical Born-Oppenheimer energy landscape, $C2$, and from the SSCHA quantum energy landscape, $Fm\bar{3}m$.}
    \label{fig:hydrides}
\end{figure*}

Hydrogen is the lightest atom in the periodic table, and, consequently, it is subject to high amplitude fluctuations even at zero Kelvin. Hydrogen atoms thus sample the $V(\bR)$ potential far from its minima. Not surprisingly, it has been shown with the SSCHA that the phonons of many hydrogen-based compounds and hydrogen itself are characterized by a huge anharmonic renormalization, impossible to capture within perturbative approaches\cite{PhysRevLett.111.177002,PhysRevB.89.064302,PhysRevLett.114.157004,Errea81,PhysRevB.93.174308,0953-8984-28-49-494001,Bianco2018,PhysRevB.99.024108,Errea66}. The anharmonic renormalization of phonons in these compounds has been crucial to explain the superconducting properties of many hydrogen-based superconductors. For instance, the anomalous inverse isotope effect on palladium hydrides, which makes the deuterium compound acquire a larger superconducting critical temperature $T_c$ than the protium compound\cite{stritzker1972superconductivity,PhysRevB.10.3818}, is a consequence of a huge anharmonic renormalization of the phonons\cite{PhysRevLett.111.177002}. Also, the experimentally found high-temperature superconductivity in H$_3$S around 200 K\cite{Drozdov2015} and in LaH$_{10}$ around 250 K\cite{Somayazulu_2019,Nature_LaH_Eremets_2019} at high pressures can only be explained if phonon frequencies renormalized by anharmonicity are considered in the superconductivity equations\cite{Errea81,Errea66}. In Fig. \ref{fig:hydrides} we show the huge anharmonic renormalization of the phonon frequencies for H$_3$S\cite{Errea81,Bianco2018}. Superconductivity in hydrogen compounds can be both largely suppressed but also enhanced by anharmonicity depending on the system\cite{doi:10.1146/annurev-conmatphys-031218-013413}.

The quantum effects and anharmonicity that the SSCHA captures go beyond the renormalization of phonon frequencies. For crystals with Wyckoff positions not fixed by symmetry, quantum or thermal fluctuations may strongly modify the atomic positions, resulting in a structure with atoms far from the positions that minimize the $V(\bR)$ potential,  occupying, instead, those that minimize the quantum $F(\bRcal)$ free energy. The change in the structure can eventually be so large that changes the crystal symmetry. For instance, the experimental crystal structure of both H$_3$S and LaH$_{10}$ compounds is stable thanks to quantum effects in the pressure range where they highest superconducting critical temperatures have been experimentally observed\cite{Errea81,Errea66}. A large modification of the structure of molecular phases of hydrogen has also been predicted within the SSCHA, which is crucial to understand the experimental Raman and infrared spectra\cite{0953-8984-28-49-494001,monacelli2019black}. The change in the crystal structure that the SSCHA captures goes beyond the internal degrees of freedom and can largely impact also the cell parameters. In Fig. \ref{fig:hydrides} we illustrate the apparent difference between the structure found classically from the minimum of $V(\bR)$ and the one obtained from the quantum energy landscape for LaH$_{10}$.  

It has been recently argued\cite{Errea66} that the large impact of quantum effects and anharmonicity on hydrogen-based compounds is precisely due to the large electron-phonon coupling of these compounds. This means that quantum effects will lower the pressure needed to synthesize these compounds with superconducting $T_c$'s approaching room temperature. The SSCHA method will be of great importance in the quest of new high-$T_c$ compounds at low pressures as it can be used for crystal structure predictions in the quantum energy landscape thanks to its capacity to relax crystal structures including quantum and anharmonic effects at any target pressure.  

\subsection{Charge density wave materials}

A CDW is a structural phase transition that induces a static modulation of the electronic density. CDW transitions are often second-order phase transitions in which the frequency of the phonon mode that drives the CDW instability rapidly softens as temperature is lowered and vanishes exactly at the CDW temperature $T_{cdw}$\cite{PhysRevLett.107.266401,PhysRevLett.107.107403,PhysRevB.92.140303}. As the temperature dependence of phonon frequencies is a purely anharmonic property, the SSCHA has been used to predict from first principles $T_{cdw}$ in several transition metal dichalcogenides (TMDs) both in the bulk and the monolayer\cite{doi:10.1021/acs.nanolett.9b00504,doi:10.1021/acs.nanolett.0c00597,ZhouTiTe2,PhysRevLett.125.106101,diego2020phonon}. 

\begin{figure*}
    \centering
    \includegraphics[width=0.9\textwidth]{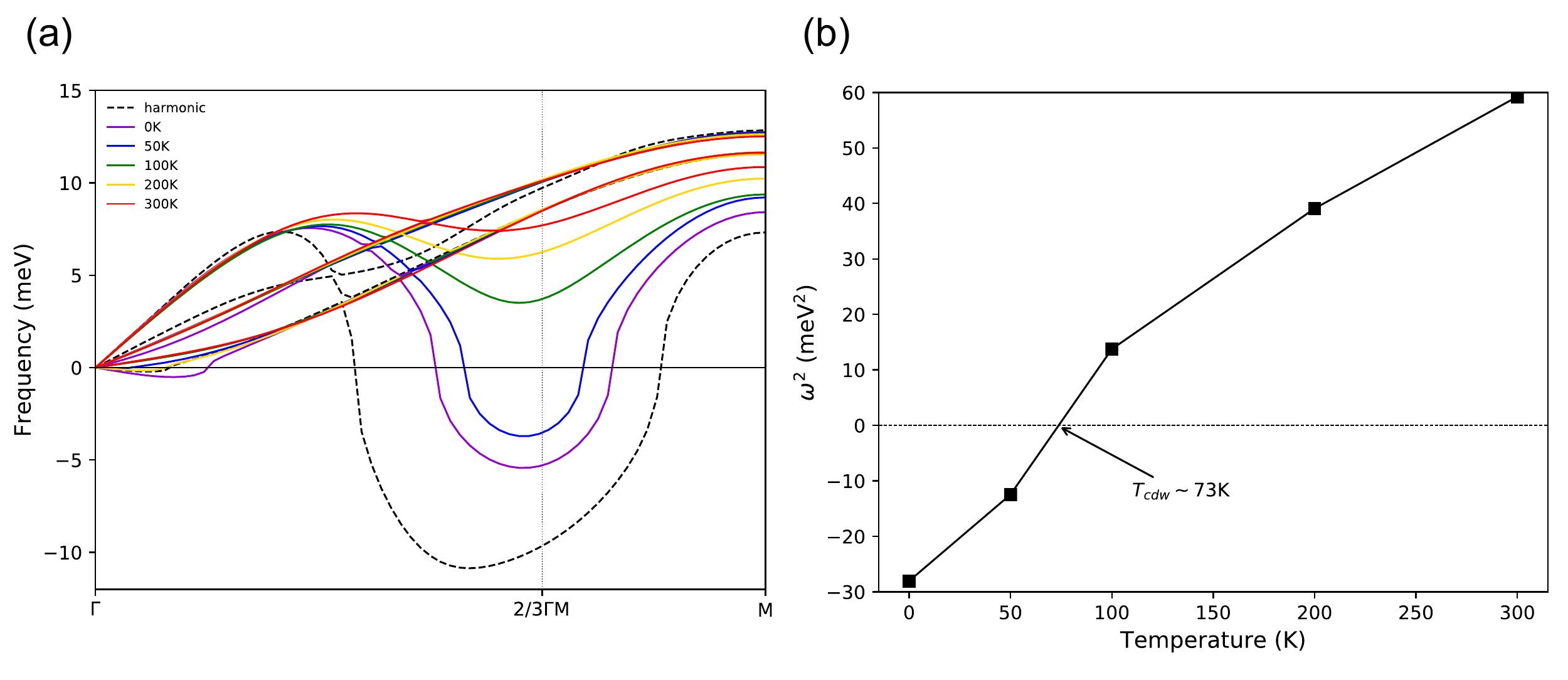}
    \caption{(a) Phonon spectra of monolayer NbSe$_2$ derived from the free energy Hessian as a function of temperature. (b) Squared phonon frequency of the lowest energy mode at $\bq=1/3\Gamma$M as a function of temperature and the determination of the CDW temperature. Data taken from Ref. \cite{PhysRevLett.125.106101}.}
    \label{fig:cdw}
\end{figure*}

The standard procedure in these calculations is to apply the SSCHA for the high-symmetry phase at different temperatures and calculate the spectra associated to the free energy Hessian $\bDF$. These phonons represent the static limit of the physical phonons observed experimentally, which can be accessed with the SSCHA by calculating instead the spectral function as described in Sec. \ref{sec:postprocessing}. At the temperature at which $\bDF$ develops a null eigenvalue, the high-symmetry structure is no longer a minimum of the free energy and the CDW distortion occurs leading the structure into a phase modulated by the wave vector at which the phonon collapse occurs. In Fig. \ref{fig:cdw} we show as an example the temperature dependence of the phonon spectra derived from the free energy Hessian in monolayer NbSe$_2$ and the consequent theoretical determination of the CDW temperature\cite{PhysRevLett.125.106101}. 
In most of the cases the calculation of $\bDF$ within the ``bubble'' approximation yields good results for the calculation of $T_{cdw}$, and setting $\bDfoureq=0$ in Eq. \eqref{eq:static_SE} seems in general a good approximation. However, converging $T_{cdw}$ is rather sensitive to the SSCHA supercell and rather large supercells may be needed to converge the CDW transition temepratures\cite{doi:10.1021/acs.nanolett.0c00597,PhysRevLett.125.106101}.
 
The capacity of the SSCHA of predicting $T_{cdw}$ purely {\it ab initio} without empirical fitting parameters offers a fantastic tool to determining the physics behind CDW transitions. The force calculations needed for the SSCHA variational minimization can be performed at different theoretical levels or at different thermodynamic conditions, disentangling the driving forces of the instability. For instance, calculations within the SSCHA have enlightened the sensitivity of CDW transitions in monolayer TMDs to strain\cite{doi:10.1021/acs.nanolett.9b00504} and doping \cite{doi:10.1021/acs.nanolett.0c00597}, the difference (or similarities) between the CDW transitions in bulk and the corresponding two-dimensional structures\cite{doi:10.1021/acs.nanolett.9b00504,doi:10.1021/acs.nanolett.0c00597}, as well as the importance of Van der Waals forces in the melting of CDW transitions\cite{diego2020phonon}. Consequently the SSCHA program is expected to have a large impact on theoretical studies of CDW transitions for many type of materials, not just TMDs.   

\subsection{Phase transitions, spectral functions, and thermal conductivity in semiconducting materials}

\begin{figure}
    \centering
    \includegraphics[width=\columnwidth]{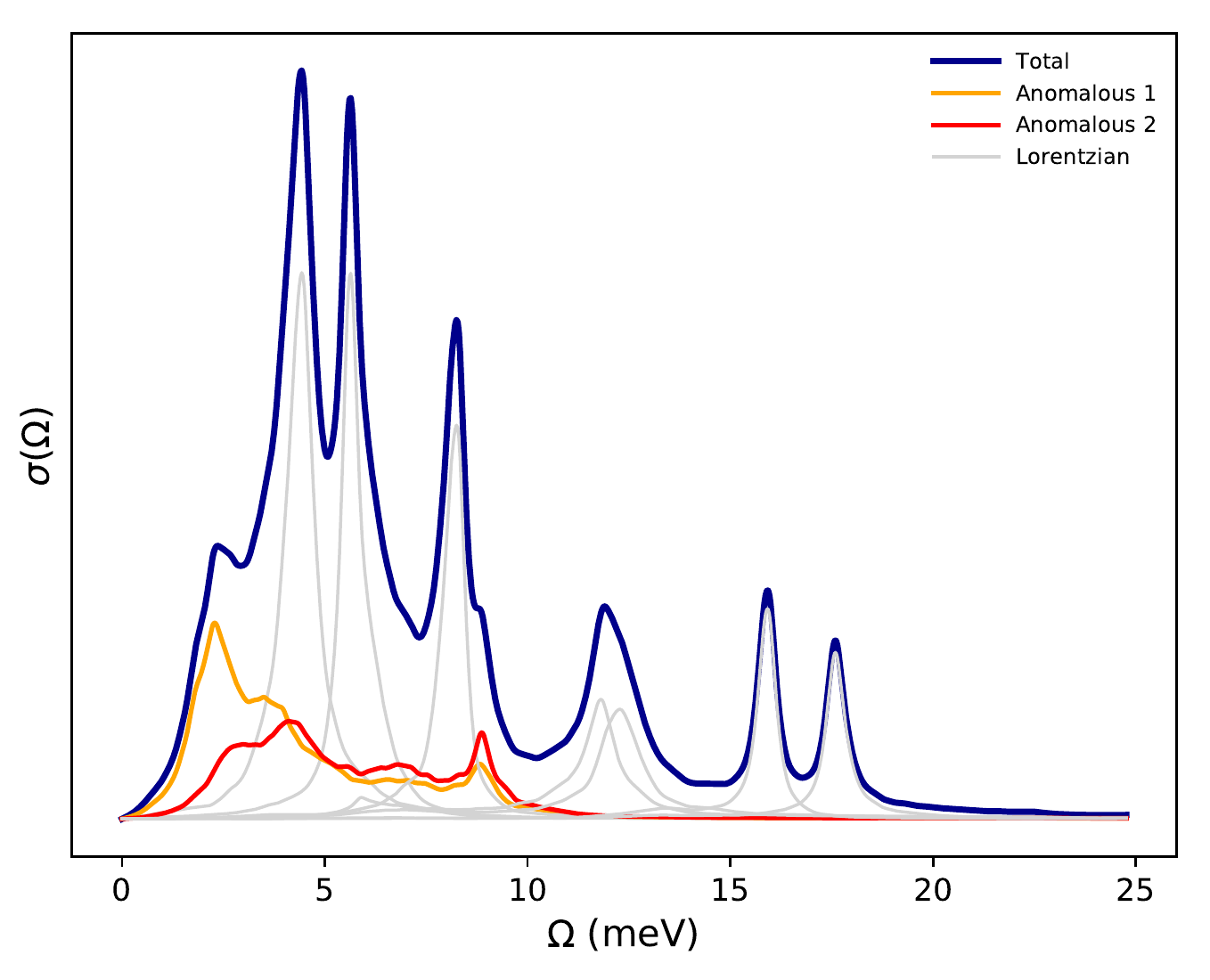}
    \caption{Spectral function $\sigma(\bq=\Gamma,\Omega)$ of SnSe at 800 K in the $Cmcm$ phase calculated within the SSCHA without assuming the Lorentzian approximation. The partial contribution  $\sigma_{\mu}(\bq=\Gamma,\Omega)$ of different modes is shown. Those anomalous modes that do not have a Lorentzian line-shape are highlighted. Data taken from Ref. \cite{PhysRevLett.122.075901}.}
    \label{fig:sigma}
\end{figure}

Phase transitions related to soft phonons are also very common in ferroelectric, thermoelectric, and other functional materials. The SSCHA is again a perfect method to study these phase transitions considering that many of the high-temperature phases of these compounds are not a minimum of the Born-Oppenheimer potential $V(\bR)$, but saddle points. Thus, it becomes imperative to adopt a non-perturbative treatment of anharmonicity in order to study their thermodynamic and transport properties such as the thermal conductivity. Whether these phase transitions are purely second-order or first order it is not always evident experimentally, unless a clear softening to zero of a phonon mode at the transition temperature  is observed. The SSCHA can distinguish between continuous and discontinuous transitions as discussed in the practical example provided in Sec. \ref{sec:model}. For instance, in order to convincingly show that the transition between the high-temperature $Cmcm$ phase of SnSe and the low-temperature $Pnma$ is second order, at the temperature at which the free energy Hessian developed a negative eigenvalue a SSCHA relaxation was performed starting from the low-temperature phase. It was shown that the $Pnma$ phase relaxed at this temperature into the $Cmcm$, showing that the $Pnma$ phase is no longer a minimum of the free energy \cite{PhysRevLett.122.075901}. The SSCHA has also been used to study phase transitions in the similar SnS\cite{PhysRevB.100.214307} and the ferroelectric SnTe\cite{PhysRevB.97.014306}.

Many of these semiconducting calchogenides are among the most efficient thermoelectric materials due to their very low thermal conductivity. The low value of the thermal conductivity of these materials is  linked to the very large linewidth of its phonon modes. The anharmonic interaction is the main responsible for the large linewidths of the phonons and, consequently, their low lifetimes. Thanks to the strong anharmonic coupling, many of these compounds develop very anomalous spectral functions with satellite peaks and a clear departure from the Lorentzian-like behavior. Such anomalies can be very misleading for the interpretation of experiments, since the emergence of extra peaks can be misinterpreted with phase transitions. The SSCHA is a perfect method for capturing these subtleties as it provides the spectral function $\sigma(\bq,\Omega)$ without the Lorentzian approximation. It has been used to understand the complex $\sigma(\bq,\Omega)$ in PbTe, SnTe, SnSe, and SnS\cite{PhysRevB.97.014306,PhysRevLett.122.075901,PhysRevB.100.214307}. In Fig. \ref{fig:sigma} we show the spectral function calculated within the SSCHA for $Cmcm$ SnSe at 800 K, where the  $\sigma_{\mu}(\bq,\Omega)$  contribution of some particular modes is clearly anomalous and deviates from the standard Lorentzian picture. 

With the phonon frequencies and the phonon linewidths obtained with the SSCHA, transport properties such as the thermal conductivity can be calculated with an external code, for instance, within Boltzmann transport equations\cite{PhysRevB.88.045430,ShengBTE_2014}. It has been shown that employing the SSCHA phonon scattering tensor $\Phithree_{\sssrcal}$ in the thermal transport calculations leads to a very good agreement with experimental results, in contrast with what obtained by employing the standard third-order derivatives of the Born-Oppenheimer total energy. The difference is that the former includes higher-order anharmonic terms coming from the average over the thermal ensemble (\eqname~\ref{eq:3rdD}). Both in good thermoelectric SnSe and SnS compounds, higher order terms captured by $\Phithree_{\sssrcal}$ reduce considerably the thermal conductivity, bringing it closer to the experimentally observed values\cite{PhysRevLett.122.075901,PhysRevB.100.214307}. Therefore, the SSCHA also provides a fantastic platform to calculate the basic ingredients for transport properties when high-order terms of the Born-Oppenheimer potential are important both for the renormalization of the phonon frequencies and the anharmonic scattering terms. Due to the large effort devoted currently to the quest of more efficient thermolectric materials, the SSCHA may become a reference method to understand the thermoelectric properties of materials and predict the efficiency of new promising compounds, which overcomes the limits of standard harmonic and perturbative approaches. 

\subsection{Other type of materials}

As mentioned above, beyond those examples listed above, the SSCHA code provides an efficient platform to calculate any property affected by ionic fluctuations, specially when it is affected by strong anharmonicity. For instance, it has been used to determine the muon implantation sites in metallic systems and to understand the effect of the large muon quantum fluctuations on the contact hyperfine field\cite{Onuorah_2019}. The SSCHA has also been employed to understand the thermal expansion and the behavior of low-energy acoustic modes of graphene\cite{aseginolaza2020bending}, finally explaining the origin of the sound propagation and the non-diverging bending rigidity of graphene as well as any other strictly two-dimensional membrane. Many other exciting applications of the SSCHA code to interesting physical problems are expected in the coming years.

\section{Conclusions}
\label{sec:conclusions}

We present here the implementation of the SSCHA theory into an efficient modular Python code, which can be run in conjunction with other Python modules and interfaced with HPC clusters for the Born-Oppenheimer total energy, force, and stress tensor calculations needed. The SSCHA provides an efficient way of calculating the effect of ionic quantum and/or thermal fluctuations on the free energy, as well as their impact on the atomic positions. It is a unique feature of the SSCHA to optimize the atomic positions, including the lattice degrees of freedom, by considering quantum and finite temperature fluctuations and without any approximation on the Born-Oppenheimer energy landscape. As a postprocessing, it calculates the free energy Hessian, which allows to infer the thermodynamic conditions at which second-order phase transitions occur. Furthermore, it enables the evaluation the interacting phonon spectral functions, predicting the outcome of most common experimental techniques (IR and Raman spectroscopies, x-ray and neutron inelastic scattering). It can also extract the phonon linewidths from the Lorentzian approximation of the spectral functions, which can be later interfaced with any code that calculates the thermal conductivity. 

In conclusion the SSCHA code provides a complete and efficient software for studying vibrational properties of materials, particularly suitable to study systems with prominent quantum and/or thermal fluctuations that are thus largely affected by anharmonicity, which can be applied to study many relevant problems in physics, chemistry, and material science.    

\section*{Acknowledgements}

R.B. and I.E. acknowledge funding from the European Research Council (ERC) under the European Union’s Horizon 2020 research and innovation programme (grant no. 802533). M. C. acknowledges support from Agence Nationale de la Recherche (Grant No. ANR-19-CE24-0028).
R.B. thanks L. Paulatto for illuminating discussions.

\appendix

\section{Stress tensor}
\label{app:stress}

Here we derive the new equation for the stress tensor reported in the main text (\eqname~\eqref{eq:avgstress}). 

First we note that the quantum statistical averages taken with the trial density matrix can be written as
\begin{equation}
\Avgschatrial{O} = \int O(\rschatrial + \mathbf J {\bm y}, \{ \bm a_i\}) \left[dy\right],
\label{eq:av_transformed}
\end{equation}
where
\begin{equation}
\left[dy\right]= \prod_{\mu} \frac{\exp\left(\frac{-y_\mu^2}{2}\right)}{\sqrt{2\pi}} dy_\mu.
\label{eq:change:variable}
\end{equation}
This is obtained rewriting Eq. \eqref{eq:rho:trial} applying the
\begin{equation}
u_a = \sum_\mu J_\mu^a y_\mu 
\end{equation}
change of variables, with
\begin{equation}
\qquad J_\mu^a = \frac{e_\mu^a}{\sqrt {M_a}}\sqrt{\frac{\hbar(1 + n_\mu)}{2\omega_\mu}}.
\end{equation}
Let's note that 
\begin{equation}
    \varPsi_{ab} = \sum_\mu J_\mu^a J_\mu^b
\end{equation}
after this change of variables. Note that in Eq. \eqref{eq:av_transformed} we explicitly indicate the dependence of the operator $O$ on the lattice parameters $\{ \bm a_i\}$. Thus, in that equation, centroid positions $\bRcal$ refer only to the internal degrees of freedom of the crystal structure.

When calculating the stress tensor from Eq. \eqref{eq:avgstress} we are deriving the free energy functional in the minimum of the SSCHA free energy with respect to the auxiliary force constants $\phischatrial$ for given centroid positions $\bRcal$. Thus, the stress tensor should be calculated considering the derivatives
\begin{equation}
\frac{\partial\Fcal(\bRcal)}{\partial\varepsilon_{\alpha\beta}} = 
\sum_{i = 1}^3 \left[ \frac{\partial\Fcal(\bRcal)}{\partial \bm a_i} \cdot \frac{\partial\bm a_i}{\partial\varepsilon_{\alpha\beta}} + \frac{\partial\Fcal(\bRcal)}{\partial \bRcal} \cdot \frac{\partial \bRcal}{\partial\varepsilon_{\alpha\beta}} \right].
\label{eq:deriv:cryst}
\end{equation}
The final equation of the strain should however be calculated for the minimum of the free energy with respect to the centroid positions, in which case the second addend above vanishes. Therefore we will just give the expression for the equilibrium situation:
\begin{equation}
\frac{\partial\Fcal}{\partial\varepsilon_{\alpha\beta}} = 
\sum_{i = 1}^3 \frac{\partial\Fcal}{\partial \bm a_i} \cdot \frac{\partial\bm a_i}{\partial\varepsilon_{\alpha\beta}}.
\label{eq:deriv:cryst}
\end{equation}
This expression coincides with the one usually employed to compute the stress tensor from the BO energy surface, but with the $V(\bR)$ potential substituted by the anharonic free energy.

Let us write the free energy at the minimum as 
\begin{equation}
\Fcal (\rscha) = F_{\phischaR} + \AvgschaR{V(\bR) - \mathcal V_{\rscha, \phischaR}},
\label{eq:fcalR}
\end{equation}
where $\phischaR$ is the dynamical matrix that minimizes $\Fcal$ fixing the average atomic positions. $F_{\phischaR}=\AvgschaR{K+\mathcal V_{\rscha, \phischaR}}$ and
\begin{equation}
    \mathcal V_{\rscha, \phischaR} = \frac 1 2  \left( \bR - \rscha\right) \cdot
\phischaR  \cdot \left(\bR -\rscha \right).
\end{equation}
The first term in Eq. \eqref{eq:fcalR} does not give any contribution to the derivative (as it depends on $\rscha$ through $\phischaR$, which minimizes already the free energy). Therefore, the only therm that survives in the stress tensor is
\begin{equation}
\frac{\partial\Fcal(\bRcal)}{\partial\varepsilon_{\alpha\beta}} = 
\sum_{i =1}^3 \frac{\partial \AvgschaR{V(\bR) - \mathcal V_{\rscha, \phischaR}}}{\partial\bm a_i} \cdot \frac{\partial\bm a_i}{\partial\varepsilon_{\alpha\beta}}.
\label{eq:first:stress:good} 
\end{equation}
Joining \eqname~\eqref{eq:first:stress:good} with \eqname~\eqref{eq:change:variable} we
can compute the derivative of an average in the SSCHA ensemble with respect to the strain:
\begin{align}
\frac{\partial \AvgschaR{O}}{\partial\varepsilon_{\alpha\beta}} & = \frac{\partial}{\partial\varepsilon_{\alpha\beta}} \int O\left(\rscha (\bm\varepsilon) + \bm J{\bm y}, \left\{\bm a_i(\varepsilon)\right\}\right)\left[ dy\right] =  \nonumber \\
& =
\int  \sum_{i=1}^3 \frac{\partial O}{\partial \bm a_i} \cdot \frac{\partial \bm a_i}{\partial\varepsilon_{\alpha\beta}} \left[dy\right],
\end{align}
\begin{align}
\frac{\partial \AvgschaR{O}}{\partial\varepsilon_{\alpha\beta}}  = 
\AvgschaR{\sum_{i=1}^3 \frac{\partial O}{\partial \bm a_i} \cdot \frac{\partial \bm a_i}{\partial\varepsilon_{\alpha\beta}}}.
\label{eq:average:stress}
\end{align}
Replacing $O$ by the BO energy landscape $V(\bR)$ we get
\begin{align}
\frac{\partial \AvgschaR{V}(\bR)}{\partial \varepsilon_{\alpha\beta}}& =
\AvgschaR{\sum_{i=1}^3 \frac{\partial V(\bR)}{\partial \bm a_i} \cdot \frac{\partial \bm a_i}{\partial\varepsilon_{\alpha\beta}}}
= \nonumber \\
& = -\Vol \AvgschaR{\stressbo_{\alpha\beta}(\bR)}.
\label{eq:pbo}
\end{align}

The term with the harmonic potential $\mathcal V$ can be derived writing its explicit dependence on the strain tensor $\bm\varepsilon$:
\begin{equation}
\mathcal V_{\rscha, \phischaR}(\bm\varepsilon) = \sum_{st}
\frac 1 2 \left[\left(\mathbb{1} + \bm\varepsilon\right) \cdot \left( \bR - \rscha\right)_s \right] \cdot
{\phischaR}_{st}  \left[\left(\mathbb{1} + \bm\varepsilon\right) \cdot \left(\bR -\rscha \right)_t\right],
\end{equation}
where the dot product is assumed in this equation only in the Cartesian indexes and $st$ are atomic labels.
From this equation we immediately can write the derivative
\begin{equation}
\left.\frac{\partial \mathcal V_{\rscha, \phischaR}}{\partial \varepsilon_{\alpha\beta}}\right|_{\bvareps=0} = - \frac 12 \sum_{s} \left(
u_s^\alpha ({\bfschaR})_s^\beta + u_s^\beta ({\bfschaR})_s^\alpha\right).
\end{equation}
From which we obtain
\begin{align}
   & \AvgschaR{\left.\frac{\partial \mathcal V_{\rscha, \phischaR}}{\partial \varepsilon_{\alpha\beta}}\right|_{\bvareps=0}} = \nonumber \\ &
-\frac 12 \sum_s \AvgschaR{\left(
u_s^\alpha ({\bfschaR})_s^\beta + u_s^\beta ({\bfschaR})_s^\alpha\right)}.
\label{eq:auxstress}
\end{align}

Combining Eqs. \eqref{eq:pbo} and \eqref{eq:auxstress} with the definition of the stress tensor, it is trivial to get Eq. \eqref{eq:avgstress}.

\section{Gradient equation}
\label{app:gradient:new}

The gradient equation presented here in \eqname~\eqref{eq:gradPhi} can be obtained starting from the 
\begin{equation}
    \frac{\partial \Fcal}{\partial \varPhi_{cd}} = \frac 12 \sum_{ab} \frac{\partial \varPsi_{ab}}{\partial\varPhi_{cd}} \left[\Avgschatrial{\frac{\partial^2 V(\bR)}{\partial R_a \partial R_b}} - \varPhi_{ab}\right]\label{eq:raf:grad}
\end{equation}
equation obtained in Ref. \cite{PhysRevB.96.014111}. By using the 
\begin{equation}
    \Avgschatrial{\frac{\partial O(\bR)}{\partial R_a}} = \sum_b \varPsi^{-1}_{ab}\Avgschatrial{u^b O(\bR)}
\end{equation}
result proved in the same reference, we have
\begin{align}
    \Avgschatrial{\frac{\partial^2 V(\bR)}{\partial R_a \partial R_b}} &= \sum_e \varPsi^{-1}_{ae}\Avgschatrial{u^e \frac{\partial V(\bR)}{\partial R_b}(\bR)} = \nonumber \\
    & = -
    \sum_e \varPsi^{-1}_{ae}\Avgschatrial{u^e \fBO_b(\bR)}\label{eq:V2bo}.
\end{align}
Anologously,
\begin{equation}
    \varPhi_{ab} =  -
    \sum_e \varPsi^{-1}_{ae}\Avgschatrial{u^e \fscha_b(\bR)}\label{eq:V2scha}.
\end{equation}
Substituting \eqname~\eqref{eq:V2bo} and \eqref{eq:V2scha} into \eqname~\eqref{eq:raf:grad} we get \eqname~\eqref{eq:gradPhi}.

In Ref. \cite{PhysRevB.96.014111} it was also shown that 
\begin{equation}
    \frac 12 \sum_{ab} \frac{\partial \varPsi_{ab}}{\partial\varPhi_{cd}} A_{ab} = \sum_{ab} \frac{\Lambda_{abcd}[0]}{\sqrt{M_aM_bM_cM_d}}  A_{ab},
    \label{eq:lambda_psiphi}
\end{equation}
where $\mathbf{A}$ is a symmetric matrix. This also proves Eq. \eqref{eq:gradPhiwithLambda}.

\section{The Hessian preconditioner}

In Ref. \cite{PhysRevB.98.024106} it was shown that 
\begin{equation}
    \frac{\partial^2 \Fcal}{\partial \varPhi_{ab}\partial\varPhi_{cd}} = \frac{\Lambda_{abcd}[0]}{\sqrt{M_aM_bM_cM_d}}.
\end{equation}
However, due to the relationship in Eq. \eqref{eq:lambda_psiphi}, we can see that we can effectively extend this equality to 
\begin{equation}
    \frac{\partial^2 \Fcal}{\partial \varPhi_{ab}\partial\varPhi_{cd}} = \frac{\Lambda_{abcd}[0]}{\sqrt{M_aM_bM_cM_d}} = \frac 12 \frac{\partial \varPsi_{ab}}{\partial\varPhi_{cd}}.
\end{equation}
With the latter result, it is trivial to see how the preconditioned gradient that is used along the minimization can be written as in Eq. \eqref{eq:newthon:phi}.

\section{Symmetries}
\label{app:symmetries}

The original algorithm proposed to account for symmetry in Ref. \cite{PhysRevB.89.064302} was based on the Gram-Schmidt orthonormalization of the symmetry generators. This algorithm is suited for systems with a reduced number of atoms in the unit cell, but scales as $n_a^6$, with $n_a$ the number of atoms in the unit cell. This symmetrization procedure becomes thus a real bottleneck of the SSCHA code for systems with more than 30 atoms in the unit cell. In the version of the code we describe here, the orthonormalized generators are not calculated and, instead, the starting dynamical matrix and the gradient are directly symmetrized. The symmetrization of the dynamical matrix, or its gradient, is made in $\mathbf{q}$ space, which allows for a very fast implementation even for big supercells.

The code enforces all the symmetries in the auxiliary force constant matrix as
\begin{equation}
    \bvarPhi(\mathbf{q}) = \frac {1}{N_S} \sum_{i = 1}^{N_S} \mathbf{T}_{\hat S_i}(S_i^{-1} \mathbf{q}) \bvarPhi(S_i^{-1} \mathbf{q}) \mathbf{T}_{\hat S_i}^\dagger (S_i^{-1} \mathbf{ q}), 
\end{equation}
where $S_i$ are the 3$\times$3 point group matrices of the space group, $N_S$ the number of symmetries of the crystal, $\mathbf{T}_{\hat S}(\mathbf{q})$ are unitary matrices that represent the $S$ symmetry in the $\mathbf{q}$ point. These matrices are reported in Refs. \cite{RevModPhys.40.1,RevModPhys.40.38,Hendrikse1995297}. To find the symmetries given the structure, we wrapped into the SSCHA the symmetry module of Quantum ESPRESSO\cite{Giannozzi2009,Giannozzi2017}.

This operation is performed also on the gradient of the dynamical matrix each time it is computed. Since the dynamical matrices satisfying the symmetries define a linear subspace, if both the gradient and the original dynamical matrix belong to this subspace, any linear combination of them will also satisfy the symmetry constrains. Thus, it is necessary to symmetrize the dynamical matrix once at the beginning, and then apply the symmetry constrains only to the gradient to preserve the symmetries in the whole simulation. 

The symmetry module from Quantum ESPRESSO only recognizes symmetries when the unit cell is the primitive one. Sometimes, it could be convenient to choose a different unit cell. Therefore, we also interfaced the SSCHA code with the \emph{spglib} package\cite{togo2018spglib} to improve the identification of symmetries. Instead of working in the unit cell in $\mathbf{q}$ space, spglib provides the symmetry operations in real space. In this case, the SSCHA code divides the symmetry matrices identified by spglib into pure translations and point group operations. Then, symmetries are enforced in real space by first imposing pure translations, followed by point group operations as
\begin{equation}
    \bvarPhi = \frac {1}{N_S} \sum_{i = 1}^{N_S} \mathbf{T}_{\hat S_i} \bvarPhi \mathbf{T}_{\hat S_i}^\dagger .
\end{equation}
Then, the permutation symmetry on the indices is imposed.
Finally, the code transforms back the real space dynamical matrix (or the gradient) in $\mathbf{q}$ space.

This operation takes more time than the symmetrization in $\mathbf{q}$ space, as it is performed in the supercell. However, due to its simplicity and to avoid the cumbersome $\mathbf{q}$-space symmetrization of higher-order force constants, the same supercell approach is used to symmetrize the third- and fourth-order force constant matrices, namely $\Phithree$ and $\Phifour$ introduced in Sec.~\ref{sec:method:derivatives}.

Symmetries are also enforced on the average positions of the nuclei (and the forces). After computing the SCHA forces on  atom $t$ along direction $\alpha$, $\ft^\alpha_t$, we impose symmetry as
\begin{equation}
    \ft_t^\alpha = \frac{1}{N_S} \sum_{i = 1}^{N_S} \sum_{t = 1}^{n_{a}}\sum_{\beta = 1}^3 {S_i}_{\alpha\beta} \ft_{S^{-1}(t)}^\beta,
\end{equation}
where $S^{-1}(t)$ is the atom in which the $S$ symmetry maps the $t$ one to. In this way, the forces are correctly directed only along the Wyckoff coordinates, and the atomic positions relax subsequently keeping the correct Wyckoff positions.

\subsection{Acoustic sum rule on the auxiliary force constants}

Besides space group symmetries, also the acoustic sum rule (ASR) must be imposed. The acoustic sum rule is a condition that arises from the momentum conservation (the center of mass of the system is fixed). The energy must not change after a rigid translation of the whole system. This can be translated in a trivial condition for the force constant matrix in the supercell:
\begin{equation}
    \sum_{t} \varPhi_{st}^{\alpha\beta} = \sum_s \varPhi_{st}^{\alpha\beta} = 0.
    \label{eq:asr}
\end{equation}

In general, the SSCHA gradient computed from a finite ensemble violates this condition due to the stochastic noise. 
We enforce the sum rule on the gradient at each step. As for the symmetries, also matrices that satisfy the acoustic sum rule define a linear subspace. Thus we define the orthogonal projector operator that enforces the acoustic sum rule as
\begin{equation}
    \phischatrial^{(asr)} = {\bm P} \phischatrial {\bm P}^\dagger.
\end{equation}
The projection matrix in real space is
\begin{equation}
P_{st}^{\alpha\beta} = \delta_{st}\delta_{\alpha\beta} - \frac{\delta_{\alpha\beta}}{n_a}\sum_{u = 1}^{n_a}\delta_{t u}.
\end{equation}
This operation only affects the dynamical matrix at $\Gamma$.
The same projector is employed to impose the ASR on the forces:
\begin{equation}
    {\ft^{(asr)}}_s^\alpha = ({\bm P} \bft)_s^\alpha = \ft_s^\alpha -  \frac{1}{n_{a}}\sum_{k=1}^{n_a} \ft_{k}^\alpha
\end{equation}
Notably, it can be proved that this procedure does not spoil the symmetrization described above.

This ASR imposition procedure analytically cancels out the frequencies of acoustic modes at $\Gamma$ and any rigid translation of the atomic positions, thus it is the most indicated for the SSCHA minimization. A different approach, implemented for the Fourier interpolation, is described in Appendix~\ref{sec:app_FT}. The latter affects not just phonons at $\Gamma$, and, thus, it is more suited for interpolating dynamical matrices close to the Brillouin zone center.

\section{Reciprocal space formalism and Fourier interpolation}
\label{sec:app_rec_space_and_FT}

\subsection{Reciprocal space formalism}
\label{sec:app_rec_space}

The SSCHA code is designed to be used with crystals, thus it takes advantage of lattice periodicity and Fourier transforms the relevant quantities with respect to the lattice vectors.
That allows to make independent analysis for each $\bq$ point in reciprocal space.
When we need to stress this aspect, we will modify the notation adopted, partitioning the supercell atomic index 
into a unit-cell index plus a lattice index $(s,\bl)$, with $s$ now ranging from 1 to $\Natu$ (the number of atoms in the unit cell), and $\bl$ being a 3 dimensional integer vector assuming $\Nc$ total values (the numer of unit cells forming the supercell). Thus, in this notation, in general we will have $n$-th order tensors in a $3\Natu$ dimensional space (indicated with bold symbols, in free-component notation), which in real space depend on $n$ lattice-vector parameters, $\overset{\ss{(n)}}{\bD}(\bl_1,\ldots\bl_n)$ (to be precise, due to the translation symmetry, this real-space tensor actually depends only on $n-1$ independent values $\bl_i$). The reciprocal-space 
expression of such a tensor, $\overset{\ss{(n)}}{\bD}(\bq_1,\ldots,\bq_n)$, is obtained through the Fourier transform
\begin{equation}
\overset{\ss{(n)}}{\bD}(\bq_1,\ldots,\bq_n)=\frac{1}{\Nc}\sum_{\bl_1\ldots\bl_n}
e^{i\sum_h\bq_h\cdot\bl_h}
\overset{\ss{(n)}}{\bD}(\bl_1,\ldots,\bl_n).
\end{equation}
Notice that, due to the lattice translation symmetry, $\bDS(\bq_1,\ldots,\bq_n)$ is zero unless
$\sum_h\bq_h$ is a reciprocal lattice vector, thus we have again only $n-1$ independent parameters $\bq_i$. 
In particular, after the calculation performed on a real-space supercell, for each $\bq$ point of the commensurate grid of the reciprocal-space unit cell, the SSCHA code computes the Fourier transformed matrices $\bDS(-\bq,\bq)$, which we will shortly indicate as $\bDS(\bq)$,
and the relative eigenvalues $\omega_\mu(\bq)$ and eigenvectors $\be_{\mu}(\bq)$. Similarly, the Hessian calculation provides
the matrix $\bDF(-\bq,\bq)$, which we indicate as $\bDF(\bq)$, where (see Eq.~\eqref{eq:def_DF})
\begin{equation}
\bDF(\bq)=\bDS(\bq)+\bPi(\bq,0)\,,
\label{eq:DFq}
\end{equation}
and its eigenvalues $\Omega_\mu(\bq)$ and eigenvectors $\bf_{\mu}(\bq)$.

\subsection{Fourier interpolation: centering and acoustic sum rule}
\label{sec:app_FT}
The SSCHA code computes the FCs in real space supercells with periodic boundary conditions (PBCs). As shown in the previous section, a crucial feature of the SSCHA code is the use of the Fourier
interpolation technique in order to extrapolate the results to the thermodynamic limit (infinite supercell results) without recurring to expensive large supercell calculations. In order to Fourier interpolate the computed FCs on arbitrary points of the reciprocal space, as a first thing it is necessary to reconstruct the real-space infinite-crystal FCs from them. Roughly speaking, this is done by removing the PBCs, i.e. superlattice equivalent atoms are not considered identical anymore, 
and assuming that only the FCs between atoms in the same supercell are different from zero. 
Of course, this gives correct results as long as the supercell used in the calculations is large enough to consider negligible the FCs between atoms at distances comparable with the
distances between the periodic boundary replica. However, an intrinsic arbtrariness is present in this recipe, due to the fact that the supercell is not univocally defined and the choice of different supercells leads to different interpolation results (i.e. as long as the reciprocal-space point in which we are interpolating does not belong to the original commensurate grid, different - yet superlattice equivalent - lattice points give different contributions to the Fourier transform). This problem is solved by wisely selecting the supercell according to a prescription based on a physical principle: among equivalent superlattice points, the ones closest to each other must be selected. This procedure defines the so called ``centering'' of the FCs and, as explained, it is a necessary step to be done before Fourier interpolating the real space FCs. 
The SSCHA code centers 2nd and 3rd order FCs (with a procedure that can be generalized to any $n$th-order FCs. In particular, the next release of the code will apply the same procedure to center and interpolate the 4th order FCs). Here we explictly describe the 3rd FCs centering algorithm~\cite{PhysRevB.87.214303}. 

The PBCs are defined on a superlattice $\RSlat$ of the original lattice $\Rlat$.
The lattice vectors set $\Rlat$ can be equivalently described as the superlattice $\RSlat$ plus the basis given
by the lattice vectors in a superlattice unit cell $SC$. 
In other words, a lattice vector $\bl\in\Rlat$ indentifies a set of 
superlattice-equivalent lattice vectors $\{\bl+\bT \,\,\text{ with } \bT\in\RSlat\}$,
and we have $\Rlat=\{\bl+\bT \,\,\text{ with } \bl\in SC\,,\bT\in\RSlat\}$.
Given three atoms $s_1,s_2,s_3$ in the unit cells $\bnot,\bl_2,\bl_3$, respectively (due to the lattice translation symmetry we can confine the first atom to the origin unit cell), they indentify a triangle with vertices in $\btau_{s_1},\btau_{s_2}+\bl_{2},\btau_{s_3}+\bl_3$ ($\btau_{s_i}$ is the position vector of atom $s_i$ in the original unit cell). 
For these three points we define the weight $\Wcal_{s_1s_2s_3}(\bnot,\bl_2,\bl_3)$ in this way: it is zero
if there is at least another ``equivalent-vertices" triangle having as vertices points 
$\btau_{s_1}$, $\btau_{s_2}+\bl_{2}+\bT_2$, $\btau_{s_3}+\bl_3+\bT_3$ with $\bT_2,\bT_3\in\RSlat$ (i.e. points that are superlattice-equivalent to $\btau_{s_1},\btau_{s_2}+\bl_{2},\btau_{s_3}+\bl_3$) with smaller perimeter,
otherwise it is the inverse of the number of equivalent-vertices triangles having the same (minimal) perimeter.
In formulas, indicated with $\Pcal_{s_1s_2s_3}(\bnot,\bl_2+\bT_2,\bl_3+\bT_3)$ the perimeter of the triangle with vertices $\btau_{s_1}$, $\btau_{s_2}+\bl_{2}+\bT_2$, $\btau_{s_3}+\bl_3+\bT_3$, this amounts to
\begin{widetext}
\begin{equation}
\scaleto{
\Wcal_{s_1s_2s_3}(\bnot,\bl_2,\bl_3)=
\left\{
\begin{aligned}
&\mkern180mu \scaleto{0}{8pt} 
\vphantom{\frac{\Biggl[}{\Biggl[}}
&& \text{ if }&&
\scaleto{
\begin{gathered}
\exists\,\bT_2,\bT_3\in\RSlat\,:\\
\Pcal_{s_1s_2s_3}(\bnot,\bl_2+\bT_2,\bl_3+\bT_3) < \Pcal_{s_1s_2s_3}(\bnot,\bl_2,\bl_3)
\end{gathered}}{25pt}
\\
&
\left[
\begin{gathered}
\# (\bT_2,\bT_3)\in\RSlat\,:\\
\Pcal_{s_1s_2s_3}(\bnot,\bl_2+\bT_2,\bl_3+\bT_3) = \Pcal_{s_1s_2s_3}(\bnot,\bl_2,\bl_3)
\end{gathered}
\right]^{-1}
\vphantom{\frac{\Biggl[}{\Biggl[}}
&& \text{ if }&&
\scaleto{
\begin{gathered}
\nexists\,\bT_2,\bT_3\in\RSlat\,:\\
\Pcal_{s_1s_2s_3}(\bnot,\bl_2+\bT_2,\bl_3+\bT_3) < \Pcal_{s_1s_2s_3}(\bnot,\bl_2,\bl_3)
\end{gathered}}{25pt}
\end{aligned}
\right.
}{118pt}
\end{equation}
The weights $\Wcal_{s_1s_2s_3}(\bnot,\bl_2,\bl_3)$ are pure geometrical factors,
different from zero for ``compact'' three-atom clusters, and they
satisfy the normalization
\begin{equation}
\sum_{\bT_2,\bT_3\in\RSlat}\Wcal_{s_1s_2s_3}(\bnot,\bl_2+\bT_2,\bl_3+\bT_3)=1\mkern80mu
\begin{aligned}
&\forall\,\bl_2,\bl_3\in\Rlat\\
&\forall\,s_1,s_2,s_3\in\{1,\ldots,\Natu\}\,.
\end{aligned}
\label{eq:norm_W}
\end{equation}

The weights are used to define the centering. Given a 3rd-order FCs, 
$\Phi^{\alpha_1 \alpha_2 \alpha_3}_{s_1,s_2,s_3}(\bnot,\bl_2,\bl_3)$, 
its ``centered" version 
$\Phicent{}^{\mkern-8mu\alpha_1 \alpha_2 \alpha_3}_{\mkern-8mu s_1 s_2 s_3}(\bnot,\bl_2,\bl_3)$ is given by
\begin{equation}
\Phicent{}^{\mkern-8mu\alpha_1 \alpha_2 \alpha_3}_{\mkern-8mu s_1 s_2 s_3}(\bnot,\bl_2,\bl_3)
=
\Phi{}^{\alpha_1 \alpha_2 \alpha_3}_{s_1 s_2 s_3}(\bnot,\bl_2,\bl_3)
\,\times\,
\Wcal_{s_1s_2s_3}(\bnot,\bl_2,\bl_3)\,,
\label{eq:def_centering}
\end{equation}
where we have separately indicated cartesian ($\alpha_h$) and atomic ($s_h$) indices.
The idea behind this definition is pretty simple: 
once the PBCs are discarded, of the infinite set of superlattice-equivalent atoms only the``closest one''
are characterized by a force constant different from zero. If there are several equivalent triplets at the minimal reciprocal distance, all of them are considered (to preserve the symmetry) and the force constants are consequently scaled (to avoid a wrong multiple counting effect). The centering definition has some degree of arbitrariness, though, due to the arbitrariness of the criterion employed to evaluate the ``size'' of a three atoms cluster. We took the perimeter of the triangle, a criterion that is a direct generalization of the distance between atoms, which is the one used in the 2nd order FCs centering. More in general, for an $n$-atoms cluster this size measure is readily 
generalized as the sum of the distances between all the $n(n-1)/2$ couples of atoms. However, even if in principle other choices could be done, this arbitrariness is immaterial as long as the supercell calculation is large enough (in the thermodynamic limit all the possible choices, if reasonable, are expected to be equivalent).

A delicate issue is associated with the centering: the spoiling of the acoustic sum rule (ASR). 
For a $n$th-order FC, the acoustic sum rule is
\begin{equation}
\sum_{\bl_i}\sum_{s_i}\Phi^{\alpha_1\ldots\alpha_i\ldots\alpha_n}_{s_1\ldots s_i\ldots s_n}(\bl_1,\ldots,\bl_i,\ldots,\bl_n)=0
\begin{aligned}
&\mkern40mu\forall\,\alpha_h\in\{x,y,z\}&&\\
&\mkern40mu\forall\, s_h\in\{1\ldots\Natu\} &&\text{ with }h\neq i\\
&\mkern40mu\forall\,\bl_h\in\Rlat&&\text{ with }h\neq i
\end{aligned}
\quad.
\label{eq:ASR}
\end{equation}
The ASR comes is crucial, among other things, to have the correct acoustic phonon dispersion at and close to $\Gamma$. The FCs computed with SSCHA (in supercells with PBCs) fulfill the acoustic sum rule but, in general, the centering spoils it (except for the $n=2$ case). 
In fact, in general the centered FCs fulfill a ``weaker'' version of the ASR
in~Eq.~\eqref{eq:ASR}, since only the simultaneous sum on $n-1$ indices is zero:
\begin{equation}
\sum_{\bl_{i_1},\ldots,\bl_{i_{n-1}}}\sum_{s_{i_1},\ldots,s_{i_{n-1}}}
\Phicent{}^{\mkern-8mu \alpha_1\ldots\alpha_n}_{\mkern-8mu s_1\ldots s_n}(\bl_1,\ldots,\bl_n)=0\,.
\label{eq:ASRweak}
\end{equation}
In particular, this explains why the centering of 2nd-order FCs does not spoil the ASR 
(for $n=2$ the weak ASR is nothing but the proper ASR, as the sum over  $n-1$ indices 
coincides with the sum over one index).

In order to see why this happens let us consider, as an example, the 3rd-order FCs case 
and the sum over the third index.
It is:
\begin{align}
&\sum_{s_3}\,\sum_{\bl_3\in\Rlat}\,
{\Phicent}{}^{\mkern-8mu \alpha_1\alpha_2\alpha_3}_{\mkern-8mu s_1s_2s_3}(\bnot,\bl_2,\bl_3)=\\
&\mkern80mu=
\sum_{s_3}\,
\sum_{\bl_3\in SC} \,
\sum_{\bT_3\in\RSlat}
\Phicent{}^{\mkern-8mu \alpha_1\alpha_2\alpha_3}_{\mkern-8mu s_1s_2s_3}(\bnot,\bl_2,\bl_3+\bT_3)\\
&\mkern80mu=
\sum_{s_3}\,
\sum_{\bl_3\in SC} \,
\sum_{\bT_3\in\RSlat}
\Phi^{ \alpha_1\alpha_2\alpha_3}_{s_1s_2s_3}(\bnot,\bl_2,\bl_3+\bT_3)\,\,
\Wcal_{s_1s_2s_3}(\bnot,\bl_2,\bl_3+\bT_3)
\\
&\mkern80mu=
\sum_{s_3}\,
\sum_{\bl_3\in SC} \,
\sum_{\bT_3\in\RSlat}
\Phi^{ \alpha_1\alpha_2\alpha_3}_{s_1s_2s_3}(\bnot,\bl_2,\bl_3)\,\,
\Wcal_{s_1s_2s_3}(\bnot,\bl_2,\bl_3+\bT_3)
\\
&\mkern80mu=
\sum_{s_3}\,
\sum_{\bl_3\in SC} \,
\Phi^{ \alpha_1\alpha_2\alpha_3}_{s_1s_2s_3}(\bnot,\bl_2,\bl_3)\,\,
\underbrace{\sum_{\bT_3\in\RSlat}\Wcal_{s_1s_2s_3}(\bnot,\bl_2,\bl_3+\bT_3)}_{\text{nonconstant w.r.t. } s_3\text{ and }\bl_3}\,.
\end{align}
Since the last factor, highlighted with a brace under, in general is not constant with respect to $s_3$
and $\bl_3$, it cannot be factored out from the sums, so that the ASR for the original 
$\Phi^{ \alpha_1\alpha_2\alpha_3}_{s_1s_2s_3}(\bnot,\bl_2,\bl_3)$ 
\begin{equation}
\sum_{s_3}\,\sum_{\bl_3\in SC} \,\Phi^{ \alpha_1\alpha_2\alpha_3}_{s_1s_2s_3}(\bnot,\bl_2,\bl_3)=0 
\end{equation}
cannot be used to obtain the ASR for the centered
$\Phicent{}^{\mkern-8mu \alpha_1\alpha_2\alpha_3}_{\mkern-8mu s_1s_2s_3}(\bnot,\bl_2,\bl_3)$. 
However, using Eq.~\eqref{eq:norm_W}, with similar passages
we can show that the sum over the last two indices is zero:
\begin{align}
&\sum_{s_2,s_3}\,\,\sum_{\bl_2,\bl_3\in\Rlat}\,
{\Phicent}{}^{\mkern-8mu \alpha_1\alpha_2\alpha_3}_{\mkern-8mu s_1s_2s_3}(\bnot,\bl_2,\bl_3)=
\vphantom{\underbrace{\sum_{\bT_2\bT_3\in\RSlat}\Wcal_{s_1s_2s_3}(\bnot,\bl_2+\bT_2,\bl_3+\bT_3)}_{=1}}\\
&\mkern80mu=
\sum_{s_2,s_3}\,\,
\sum_{\bl_2,\bl_3\in SC} \,
\Phi^{ \alpha_1\alpha_2\alpha_3}_{s_1s_2s_3}(\bnot,\bl_2,\bl_3)\,\,
\underbrace{\sum_{\bT_2,\bT_3\in\RSlat}\Wcal_{s_1s_2s_3}(\bnot,\bl_2+\bT_2,\bl_3+\bT_3)}_{=1}\\
&\mkern80mu=
\sum_{s_2,s_3}\,\,
\sum_{\bl_2,\bl_3\in SC} \,
\Phi^{ \alpha_1\alpha_2\alpha_3}_{s_1s_2s_3}(\bnot,\bl_2,\bl_3)
\vphantom{\underbrace{\sum_{\bT_2\bT_3\in\RSlat}\Wcal_{s_1s_2s_3}
(\bnot,\bl_2+\bT_2,\bl_3+\bT_3)}_{=1}}\\
&\mkern80mu=0\,,
\end{align}
thus the weak ASR is fulfilled.

The spoling of the ASR after the centering dictates to impose it. 
In principle, there is not a unique way of doing it, as 
imposing the ASR on FCs simply consists in finding new FCs 
that fullfil the ASR and differ the least from the original FCs 
(according to some reasonable but arbitrary metric). 
In this release of the SSCHA code we impose the ASR by employing an iterative procedure,
consisting of two steps~\cite{PhysRevB.87.214303}. 
First, the ASR is imposed on one index (the last one, for example).
This spoils the permutation symmetry, which is consequently imposed. 
In general, the resulting permutation-symmetric FCs do not fulfill the ASR yet, thus this procedure is repeatedly applied until the permutation-symmetric FCs fulfill the ASR within a certain tolerance. The imposition of the permutation symmetry is a straightforward task.  The ASR is imposed on the third index of a centered 
3rd-order FCs by updating its values on the compact three-atom clusters that defined the centering
(in order to preserve the short-sightdness of the centered FCs even after the ASR imposition). Given a centered $\Phi^{ \alpha_1\alpha_2\alpha_3}_{s_1s_2s_3}(\bnot,\bl_2,\bl_3)$, the $\widetilde{\Phi}{}^{ \alpha_1\alpha_2\alpha_3}_{s_1s_2s_3}(\bnot,\bl_2,\bl_3)$ that fulfills the ASR on the third index is computed with
\begin{equation}
\widetilde{\Phi}{}^{ \alpha_1\alpha_2\alpha_3}_{s_1s_2s_3}(\bnot,\bl_2,\bl_3)=\Phi^{ \alpha_1\alpha_2\alpha_3}_{s_1s_2s_3}(\bnot,\bl_2,\bl_3)-
\mathcal{K}^{\alpha_1\alpha_2\alpha_3}_{s_1s_2s_3}(\bnot,\bl_2,\bl_3|p)\times \sum_{\ols_3,\olbl_3}\Phi^{ \alpha_1\alpha_2\alpha_3}_{s_1s_2\ols_3}(\bnot,\bl_2,\olbl_3)\,,
\label{eq:ASR_imp}
\end{equation}
where $\mathcal{K}^{\alpha_1\alpha_2\alpha_3}_{s_1s_2s_3}(\bnot,\bl_2,\bl_3|p)$ is the scaling factor
\begin{equation}
\mathcal{K}^{\alpha_1\alpha_2\alpha_3}_{s_1s_2s_3}(\bnot,\bl_2,\bl_3|p)
=
\left\{
\begin{aligned}
&\frac{ \left|\Phi^{ \alpha_1\alpha_2\alpha_3}_{s_1s_2s_3}(\bnot,\bl_2,\bl_3) \right|^p}
{\sum_{\ols_3,\olbl_3}  \left|\Phi^{ \alpha_1\alpha_2\alpha_3}_{s_1s_2\ols_3}(\bnot,\bl_2,\olbl_3)\right|^{p}}
&&\text{if  }\mkern20mu
&& p = 0 
\mkern40mu\text{or}\mkern40mu  
p > 0 \mkern10mu  \text{ and } \mkern10mu\sum_{\ols_3,\olbl_3}  \left|\Phi^{ \alpha_1\alpha_2\alpha_3}_{s_1s_2\ols_3}(\bnot,\bl_2,\olbl_3)\right|\neq0
\\
&\mkern90mu 0 \vphantom{\frac{\sum_{\ols_3,\olbl_3}  \Phi^{ \alpha_1\alpha_2\alpha_3}_{s_1s_2\ols_3}(\bnot,\bl_2,\olbl_3) }{\sum_{s_3,\olbl_3}  \left|\Phi^{ \alpha_1\alpha_2\alpha_3}_{s_1s_2\ols_3}(\bnot,\bl_2,\olbl_3)\right|^{p}}}
&&\text{if }\mkern20mu
&& p > 0
\mkern10mu\text{ and }\mkern10mu 
\sum_{\ols_3,\olbl_3}  \left|\Phi^{ \alpha_1\alpha_2\alpha_3}_{s_1s_2\ols_3}(\bnot,\bl_2,\olbl_3)\right|=0
\end{aligned}
\right.\,,
\end{equation}
with $p$ a non-negative real number which can be arbitrarily fixed to optimize
the calculation performances (in the equation above the convention $0^0=1$ has been adopted). 
The $\widetilde{\Phi}{}^{ \alpha_1\alpha_2\alpha_3}_{s_1s_2s_3}(\bnot,\bl_2,\bl_3)$ defined through Eq.~\eqref{eq:ASR_imp} fulfills the ASR on the third index, since for any $p\geq 0$ the scaling factor fulfills the normalization condition
\begin{equation}
\sum_{s_3,\bl_3}\mathcal{K}^{\alpha_1\alpha_2\alpha_3}_{s_1s_2s_3}(\bnot,\bl_2,\bl_3|p)=1\,.
\end{equation}
The value of $p$ has effects on the way the different terms of ${\Phi}{}^{ \alpha_1\alpha_2\alpha_3}_{s_1s_2s_3}(\bnot,\bl_2,\bl_3)$  are scaled. For $p=0$ the scaling factor is a pure geometric quantity related to the three atoms clusters. Indeed, given $s_1,s_2, \bl_2$, the scaling factor $\mathcal{K}^{\alpha_1\alpha_2\alpha_3}_{s_1s_2s_3}(\bnot,\bl_2,\bl_3|p=0)$ is fully determined (it is the same for all the $\alpha_h, s_3,\bl_3$) and, in particular,
it does not depend on the FCs value. On the contrary, for $p\neq0$, given $\alpha_h,s_1,s_2, \bl_2$
we have
\begin{equation}
\frac{\mathcal{K}^{\alpha_1\alpha_2\alpha_3}_{s_1s_2s'_3}(\bnot,\bl_2,\bl'_3|p)}{\mathcal{K}^{\alpha_1\alpha_2\alpha_3}_{s_1s_2s''_3}(\bnot,\bl_2,\bl''_3|p)}
=\left|\frac{{\Phi}{}^{ \alpha_1\alpha_2\alpha_3}_{s_1s_2s''_3}(\bnot,\bl_2,\bl''_3)}{{\Phi}{}^{ \alpha_1\alpha_2\alpha_3}_{s_1s_2s'_3}(\bnot,\bl_2,\bl'_3)}\right|^p
\end{equation}
so that if $p>1$ the scaling factor is higher (lower) for FCs have lower (higher) absolute value, otherwise the opposite.
\end{widetext}

\subsection{Effective charges}
\label{sec:app_EC}
In ionic crystals the nuclei displacement induces dipoles (proportional to the Born effective charge tensors),
and this adds a dipole-dipole interaction term to the interatomic forces. This contribution, because of its long-range character (it goes as the inverse of the third power of the nuclei distances),  is not suited to be Fourier interpolated and it is at the origin of the nonanalytic behavior of the dynamical matrix at $\Gamma$, with (in general anisotropic) LO-TO splitting of the phonon frequencies at $BZ$ center.  The long-range dipole-dipole contribution to the force constants can be calculated analytically since it is fully determined by the Born effective charges $(Z^*_s)^{\alpha\beta}$ (effective charge tensor of atom $s$) and the electronic dielectric permittivity tensor $(\epsel)^{\alpha\beta}$, which can both be calculated from first principles.
For a given $\bq\in BZ$, this dipole-dipole contribution is given by~\cite{PhysRevB.55.10355,PhysRevB.43.7231}  
\begin{equation}
\boldsymbol{\Phi}^{\sss{(dd)}}_{st}(\bq)=\widehat{\boldsymbol{\Phi}}{}^{\sss{(dd)}}_{st}(\bq)-\delta_{st}\sum_{\overline{t}}\widehat{\boldsymbol{\Phi}}{}^{\sss{(dd)}}_{s\overline{t}}(\bq=\bnot)
\label{eq:dd_corr1}
\end{equation}
with
\begin{widetext}
\begin{equation}
\widehat{\boldsymbol{\Phi}}{}^{\sss{(dd)}}_{st}(\bq)=
\frac{4\pi}{\Vol}\sum_{\bG}{}^{\!'}
\frac{\left[(\bG+\bq)\cdot\bZ^*_s\right] \otimes \left[(\bG+\bq)\cdot \bZ^*_t\right]}{(\bG+\bq)\cdot\bepsel\cdot(\bG+\bq)}
\,
e^{\textstyle -\frac{(\bG+\bq)\cdot\bepsel\cdot(\bG+\bq)}{4\eta^2}}
\,
e^{i(\bG+\bq)\cdot(\btau_s-\btau_t)} \,,
\label{eq:dd_corr2}
\end{equation}
where we have explicitly indicated only the atomic indices (i.e. we are using component-free notation for the cartesian indices), $\eta$ is a parameter whose value has to be large enough to allow to include only the reciprocal space terms in the Ewald sum, and $\sum_{\bG}'$ is the sum over reciprocal lattice vectors such that $\bG+\bq\neq\bnot$ (the sum includes as many $\bG$'s as it is necessary to reach the convergence for the considered $\eta$)~\cite{Giannozzi2009}.

Once $\bZ^*_s$ and $\bepsel$ are available, the problem caused to the Fourier interpolation by the long-range dipole-dipole interaction  is thus bypassed in the SSCHA code in two steps. First, from the $\bPhi(\bq)$ calculated on a (coarse) grid of $\bq$ point of the Brillouin zone, the corresponding dipole-dipole terms $\boldsymbol{\Phi}^{\sss{(dd)}}(\bq)$ are subtracted and the resulting short range FCs is Fourier transformed to the real space. Subsequently, this real space short-range FCs, $\boldsymbol{\Phi}^{\sss{(sr)}}(\bl)$, can be Fourier transformed back to any $\bk\in BZ$ and the corresponding long-range dipole-dipole analytical contribution $\boldsymbol{\Phi}^{\sss{(dd)}}(\bk)$ is added~\cite{Giannozzi2009}:
\begin{equation}
\begin{gathered}
\bPhi(\bq) \\
\textup{ on $BZ$ $\bq$-grid}
\end{gathered}
\xrightarrow{
\substack{
    \textup{Subtract dipole-dipole interaction terms}\\ 
    \textup{$\bPhi^{\sss(dd)}(\bq)$} \\  
    $+$ \\
    \textup{Fourier transform}\\
    \textup{to real space}
         }
}
\quad
\bPhi^{\sss{(sr)}}(\bl)
\quad
\xrightarrow{
\substack{
    \textup{Fourier transform}\\ 
    \textup{back to $\bk\in BZ$}\\
    $+$ \\
    \textup{Add dipole-dipole interaction term}\\
    \textup{$\bPhi^{\sss(dd)}(\bk)$}
         }
}
\bPhi(\bk)
\end{equation}
\end{widetext}
The dipole-dipole correction to the FCs given by Eqs.~\eqref{eq:dd_corr1},~\eqref{eq:dd_corr2} 
is nonanalytic at zone center and its $\bq\rightarrow\bnot$ limit
depends on the direction $\widehat{\bq}=\bq/\|\bq\|$ along which the limit is performed:
\begin{equation}
\lim_{\delta\rightarrow 0^+}\boldsymbol{\Phi}{}^{\sss{(dd)}}_{st}(\delta\widehat{\bq})=
\boldsymbol{\Phi}^{\sss{(dd)}}_{st}(\bnot)+
\boldsymbol{\Phi}^{\sss{(dd-na)}}_{st}(\widehat{\bq})\,,
\end{equation}
where
\begin{equation}
\boldsymbol{\Phi}{}^{\sss{(dd-na)}}_{st}(\widehat{\bq})=
\frac{4\pi}{\Vol}\frac{ \left[\widehat{\bq} \cdot\bZ^*_s\right] \otimes  \left[ \widehat{\bq} \cdot\bZ^*_t\right]    }{ \widehat{\bq}\cdot\bepsel\cdot \widehat{\bq}}
\end{equation}
is the nonanalytic zone-center correction term.  
When a phonon dispersion through $\Gamma$ is calculated, 
the SSCHA code includes the nonanalytic correction term in the zone center, with the direction given by the followed path~\cite{Giannozzi2009}. When the SSCHA code calculate the spectral properties (static or dynamic), it adds the nonanalytic correction term in the zone center dynamical matrix (necessary for the integral over the BZ) from a random direction.


\end{document}